\documentclass{aa}  
\usepackage{linenoaa}
\usepackage{rotating}
\usepackage{graphicx}
\usepackage{booktabs}
\usepackage{txfonts}
\usepackage{natbib} 
\bibpunct{(}{)}{;}{a}{}{,}
\usepackage{graphicx}
\usepackage{caption}
\usepackage{subcaption}
\usepackage{footnote}
\makesavenoteenv{table*}
\usepackage{threeparttable}
\usepackage{txfonts}
\usepackage{tikz}
\usepackage{amsmath}
\usepackage{soul}
\usepackage{hyperref}
\hypersetup{
    colorlinks=true,
    linkcolor=blue,
    citecolor=blue,
    urlcolor=blue
}
\begin{document} 
   \title{Classical Cepheids in the Galactic thin disk 
   }

 \subtitle{I. Abundance gradients via non-local thermodynamic equilibrium spectral analysis}

   \author{A.~Nunnari\inst{1}\fnmsep\inst{2},
          V.~D'Orazi\inst{1,2,3},
          G.~Fiorentino\inst{2},
          V.~F.~Braga\inst{2},
          G.~Bono\inst{1,2},
          M.~Fabrizio\inst{2,4},
          H.~J\"{o}nsson\inst{5},          
          R.-P.~Kudritzki\inst{6,7},
          R.~da Silva\inst{2,4},
          M.~Bergemann\inst{8},
          E.~Poggio\inst{9},
          J.~M.~Otto\inst{10},
          K.~Baeza-Villagra\inst{1,2},
          A.~Bragaglia\inst{11},
          G.~Ceci\inst{12,1,2},
          M.~Dall'Ora\inst{13},
          L.~Inno\inst{14,13},
          C.~Lardo\inst{15},
          N.~Matsunaga\inst{16},
          M.~Monelli\inst{1,17,18},
          M.~S\'anchez-Benavente\inst{17,18},
          C.~Sneden\inst{3},
          M.~Tantalo\inst{1},
          F.~Th\'ev\'enin\inst{19},
          V.~Kovtyukh\inst{20},
          M.~Di~Criscienzo\inst{2},
          and G.~B\"{o}cek~Topcu\inst{1,2}
          }

   \institute{Dipartimento di Fisica, Università di Roma Tor Vergata, via della Ricerca Scientifica 1, 00133 Rome, Italy 
         \and
          INAF – Osservatorio Astronomico di Roma, Via Frascati 33, 00078 Monte Porzio Catone, Italy.\\
          \email{antonino.nunnari@inaf.it}
          \and
          Department of Astronomy \& McDonald Observatory, The University of Texas at Austin, 2515 Speedway, Austin, TX 78712, USA
          \and
          ASI - Space Science Data Center, via del Politecnico snc, I-00133 Roma, Italy
          \and
        Materials Science and Applied Mathematics, Malm\"{o} University, SE-205 06 Malm\"{o}, Sweden
          \and
         Institute for Astronomy, University of Hawai’i at Manoa, Honolulu, HI 96822, USA
          \and
          LMU M\"{u}nchen, Universit\"{a}tssternwarte, Scheinerstr. 1, 81679 M\"{u}nchen, Germany
          \and
          Max-Planck-Institut f\"{u}r Astronomie, K\"{o}nigstuhl 17, D-69117 Heidelberg, Germany
          \and
          INAF – Osservatorio Astrofisico di Torino, via Osservatorio 20,10025 Pino Torinese (TO), Italy
          \and
          Department of Physics and Astronomy, Texas Christian University, TCU Box 298840 Fort Worth, TX 76129, USA
          \and
          INAF -- Osservatorio di Astrofisica e Scienza dello Spazio di Bologna, via Piero Gobetti 93/3, 40129 Bologna, Italy
          \and
          Dipartimento di Fisica, Sapienza Università di Roma, P.le A. Moro 5, I-00185 Roma, Italy
          \and
          INAF - Osservatorio Astronomico di Capodimonte, salita Moiariello 16, 80131, Naples, Italy
          \and
          Department of Science and Technology, Parthenope University of Naples, CDN-IC4, Naples, I-80143, Italy
          \and
          Dipartimento di Fisica e Astronomia, Università degli Studi di Bologna, Via Gobetti 93/2, 40129 Bologna, Italy
          \and
          Department of Astronomy, School of Science, The University of Tokyo, 7-3-1 Hongo, Bunkyo-ku, Tokyo 113-0033, Japan
          \and
          IAC- Instituto de Astrof\'isica de Canarias, Calle V\'ia Lactea s/n, E-38205 La Laguna, Tenerife, Spain
          \and
          Departmento de Astrof\'isica, Universidad de La Laguna, E-38206 La Laguna, Tenerife, Spain
          \and
          Universit\'e Côte d’Azur, Observatoire de la Côte d’Azur, CNRS, Laboratoire Lagrange, France
          \and
          Astronomical Observatory, Odessa National University, Shevchenko Park, 65014, Odessa, Ukraine
          }

   \date{Received Month DD, YYYY; accepted Month DD, YYYY}

    \abstract{Classical Cepheids (CCs) have long been considered excellent tracers of the chemical evolution of the Milky Way's young disk. We present a homogeneous, non-local thermodynamical equilibrium (NLTE) spectroscopic analysis of 401 Galactic CCs, based on 1,351 high-resolution optical spectra, spanning Galactocentric distances from 4.6 to 29.3 kpc. Using PySME with MARCS atmospheres and state-of-the-art grids of NLTE departure coefficients, we derived the atmospheric parameters and abundances for key species tracing multiple nucleosynthetic channels (O, Na, Mg, Al, Si, S, Ca, Ti, Mn, Fe, and Cu). Our sample is the largest CC NLTE dataset to date and it achieves high internal precision, enabling the robust modelling of present-day thin-disk abundance patterns and radial gradients. 
  
  We estimate abundance gradients using three analytic prescriptions (linear, logarithmic, bilinear with a break) within a Bayesian, outlier-robust framework. We also applied Gaussian process (GP) regression to capture non-parametric variations. We find that NLTE atmospheric parameters differ systematically from LTE determinations. Moreover, iron and most elemental abundance profiles are better described by non-linear behaviour rather than by single-slope linear models: logarithmic fits generally outperform simple linear models, while bilinear fits yield inconsistent break radii across elements.  GP models reveal a consistent outer-disk flattening of [X/H] for nearly all studied elements. The [X/Fe] ratios are largely flat with Galactocentric radius, indicating coherent chemical scaling with iron across the thin disk, with modest positive offsets for Na and Al and mild declines for Mn and Cu. 
  
  Finally, Cepheid kinematics confirm thin-disk orbits for the great majority of the sample. Comparisons with recent literature shows an overall agreement, while also  highlighting NLTE-driven differences, especially in outer-disk abundances. These results provide tighter empirical constraints for chemo-dynamical models of the Milky Way and set the stage for future NLTE mapping with upcoming large spectroscopic surveys.}

\authorrunning{A. Nunnari et al.}
   \maketitle

\defcitealias{da2023oxygen}{dS23}
\nolinenumbers
\section{Introduction}
Radial abundance gradients play a key role in constraining Galactic 
chemo-dynamical models (\citealt{palla20}; \citealt{spitoni23}; \citealt{lemasle22}). 
The radial trends observed in the present-day chemical abundances of the thin disk, 
when compared with theoretical predictions across different Galactocentric distances 
and stellar ages, allow us to constrain its chemical enrichment history (\citealt{chiappini01}). 
Furthermore, radial gradients and their steady flattening as a function of time soundly 
support stellar radial migrations in shaping the current metallicity distribution function 
of the thin disk (\citealt{hou00}; \citealt{prantzos23}).  Moreover, they also provide 
an independent constraint on the star formation efficiency when moving from the innermost to 
the outermost disk regions (\citealt{spitoni23}).

Radial gradients across the thin disk have been investigated by using different stellar tracers: 
open clusters (\citealt{friel02}; \citealt{magrini23}; \citealt{myers22}; \citealt{occaso24}; \citealt{otto25}; \citealt{dalponte25}), 
planetary nebulae (\citealt{Maciel13}; \citealt{stanghellini18}), 
HII regions (\citealt{fernandez17}), and 
OB stars (\citealt{daflon04}; \citealt{nieva12}; \citealt{braganca19}). 
Classical Cepheids (CCs) have also been extensively used to investigate metallicity distributions  
and chemical gradients in the Galactic thin disk (\citealt{yong06}; \citealt{szildi07}; \citealt{romaniello2008influence}; \citealt{pedicelli2010new}; \citealt{luck11}; \citealt{lucklambert11}; \citealt{lemasle2013galactic},  \citeyear{lemasle2017detailed}; \citealt{genovali2013metallicity}, \citeyear{genovali2014fine}, \citeyear{genovali15}; \citealt{korotin14}; \citealt{martin15}; \citealt{andrievsky16}; \citealt{da2016neutron}, \citeyear{da2022new}, \citeyear{da2023oxygen}; \citealt{proxauf2018new}; \citealt{luck18}; \citealt{inno2019first}; \citealt{kovtyukh22}; \citealt{ripepi22}; \citealt{trentin23}, \citeyear{Trentin24B}, \citeyear{trentin24A}, \citeyear{trentin25}). Cepheids are radially pulsating variable stars that serve as excellent distance indicators, since their individual distances can be estimated with an accuracy of 1-3\%. They also trace young stellar populations \citep{bono2024cepheids}. CCs are central helium-burning,  intermediate-mass stars younger than a few hundred million years, with pulsation periods ranging from days to roughly one year. They are ubiquitous across the thin disk and the Galactic centre (\citealt{matsunaga11}; \citealt{bono2024cepheids}).

Since CCs are typically distributed across the thin disk, their intrinsic colours are 
affected by uncertainties due to reddening corrections. This is the main reason why atmospheric 
parameters are estimated directly from the spectra. Several approaches have been proposed to 
address this issue.
In particular, 
\citet{proxauf2018new} and \citet{da2022new} adopted the line depth 
ratio method (\citealt{kovtyukh07}) based on empirical calibrations of different 
element pairs to derive effective temperature \citep{kovtyukh07, elgueta24}. 
A new temperature scale for Galactic CCs based on a data-driven, machine-learning technique 
applied to observed spectra was recently suggested by \citet{lemasle20}. Specifically, the flux 
ratios of different spectral features are tied to the effective temperatures derived using the 
infrared surface-brightness method (\citealt{hanke18}).

A significant fraction of abundance analysis of CCs is based on local thermodynamic 
equilibrium (LTE) atmosphere models, such as those of \citet[ATLAS]{castelli2004new} and 
\citet[MARCS]{gustafsson2008grid}. However, both theoretical predictions and empirical 
evidence show that LTE modelling can be prone to possible systematics (\citealt{lind24}; \citealt{bergemann25}). These systematics 
become more relevant  when moving from dwarf stars to giants,  and from metal-rich to 
metal-poor populations (e.g., \citealt{thevenin1999stellar}; \citealt{idiart2000non}; \citealt{bergemann2012non}; \citealt{hansen2013lte}; \citealt{fabrizio2021use}). 
Moreover, NLTE corrections are element- and line-dependent, with lines of neutral atoms being more affected than those of ionized species \citep{kiselman2001nlte, collet2005effects, merle2011grid, tautvaivsiene2015gaia, duffau2017gaia}. 

Chemical abundances for CCs based on a NLTE approach have been provided by \citet{andrievsky16} 
for all their analysed elements, as well as by \citet{luck11} for CNO elements, by \citet{martin15} for O, S and Ba, 
and by \citet{korotin14} for O. However, in these investigations the stellar atmospheric parameters were derived 
by using an LTE approach. 

Recently, \citet{amarsi2020galah} provided NLTE departure coefficients for 
H, Li, C, N, O, Na, Mg, Al, Si, K, Ca, Mn, and Ba. Additionally, NLTE departure coefficients are provided by \citet{mallinson24} for Ti, by \citet{amarsi22} for Fe, by  \citet{caliskan25} for Cu and by \citet{amarsi25} for S. These departure coefficient are computed on a 
grid of 3756 1D MARCS model atmospheres (\citealt{gustafsson2008grid}) covering $\mathrm{3000 < T_{eff}/K < 8000}$, $\mathrm{-0.5 < \log(g/cm s^{-2}) < 5.5}$, and $\mathrm{-5 < [Fe/H] < +1}$ dex. These comprehensive grids allow proper spectral modelling and NLTE analysis. 

In this study, we apply a complete NLTE approach--deriving both atmospheric parameters and 
chemical abundances--for Galactic CCs by using high-resolution optical spectra.
Although the statistical sampling in the inner (Galactocentric distance, $\mathrm{R_{GC}\lesssim 5kpc}$) 
and in the outer ($\mathrm{R_{GC}\gtrsim 20kpc }$) disk is sparse, there is solid empirical evidence of radial gradients for most of the elements that have been investigated. 
More specifically, our dataset spans a wide range in 
Galactocentric distances ($\mathrm{R_{GC}}$=5--29~kpc) and provides homogeneous NLTE abundances for light (O),  
odd-Z (Na, Al, Cu), $\alpha$ (Mg, Si, S, Ca, Ti), and iron peak (Mn, Fe) elements. 
The current analysis is a stepping stone for exploiting thousands of new high-resolution 
spectra for CCs that will be collected from upcoming large spectroscopic surveys 
in optical (\citealt{dejong16}, 4MOST) and near-infrared (\citealt{cirasuolo12}, MOONS). 
This paper is organized as follows: in Sect.~\ref{dataset_and_method}, we present the spectral dataset and the 
line list we adopted for the spectral analysis. In Sect.~\ref{abundance_analysis} we show the results concerning 
the atmospheric parameters and the chemical abundances, including the gradients, and the 
kinematic properties. The summary of the results, the discussion and a brief outline of the near 
future developments of this project are presented in Sect.~\ref{summary}. The paper also includes several appendices. Appendix~\ref{tables} presents the tables included in the paper, while Appendix~\ref{distances} discusses the new distances of two CCs. 
Appendix~\ref{details_synthesis} provides details of the adopted spectral synthesis, and Appendix~\ref{test_nlte} presents  
the validation of NLTE atmospheric parameters and chemical abundances.

\section{Dataset and method}
\label{dataset_and_method}
   \begin{figure}[!ht]
   \centering
   \includegraphics[width=9cm]{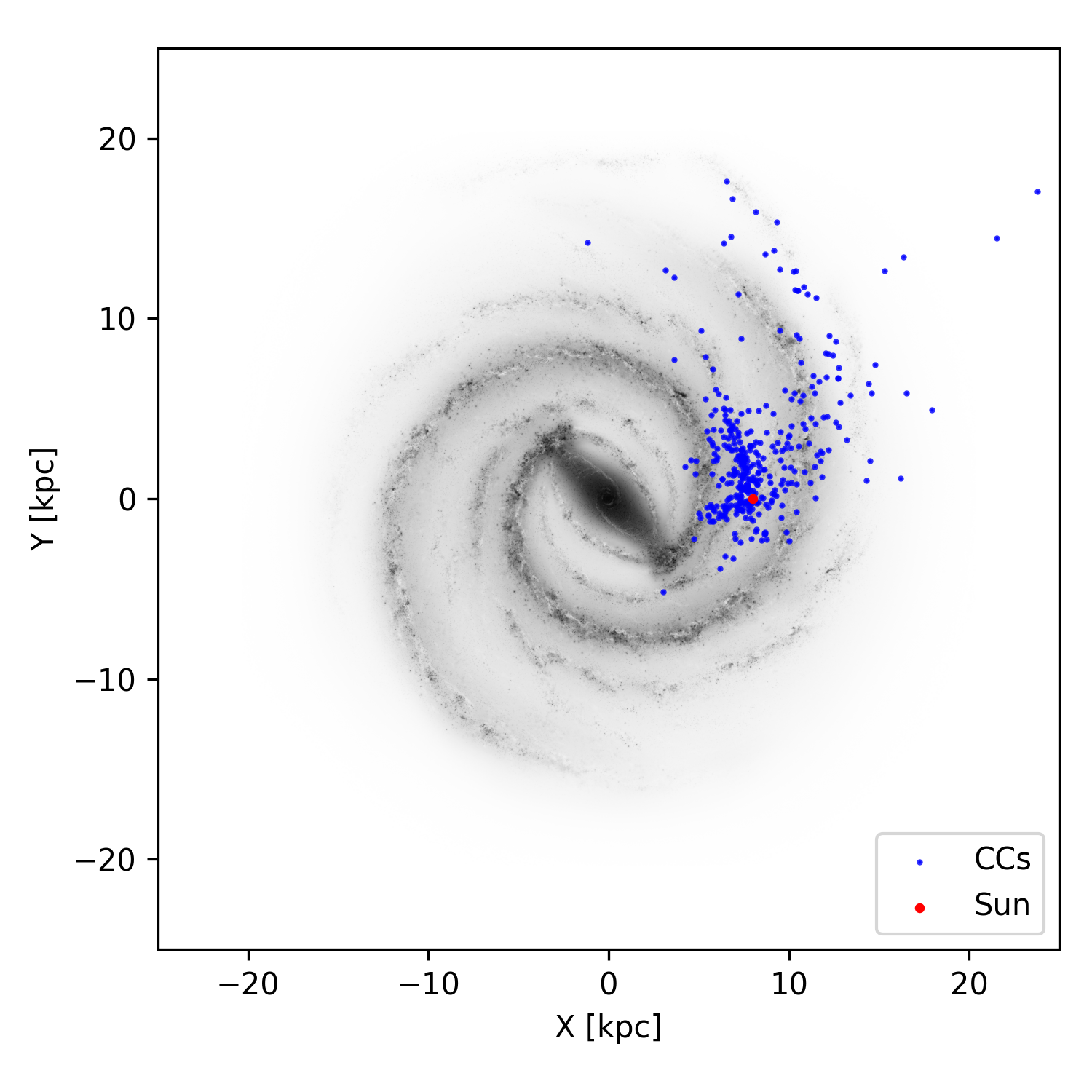}
      \caption{Distribution of the sample of CCs across the thin disk 
      in a Galactocentric 
      reference frame. The face-on map of the Milky Way was made with the Python library \texttt{mw-plot}.}
         \label{CC_distribution}
   \end{figure}
The current spectroscopic sample includes 401 CCs distributed across the thin disk, as shown in Fig.~
\ref{CC_distribution}\footnote{NASA/JPL-Caltech/R. Hurt (SSC/Caltech), https://milkyway-plot.readthedocs.io/en/}. Among them, 379 CCs have 
already been collected and discussed in \citet{da2023oxygen} (hereinafter \citetalias{da2023oxygen}). We added 66 
High-Resolution (HR) spectra for 22 CCs (proprietary plus public archives): 10 with $\mathrm{R_{GC}\lesssim 6\, kpc }$ 
and 4 with $\mathrm{R_{GC}\gtrsim 10\,kpc}$.

   \subsection{Spectral sample} 
   \label{spectral sample}
    In the spectroscopic sample presented in \citetalias{da2023oxygen}, individual heliocentric distances were estimated using the following methods: 
    Gaia DR3 parallaxes prior-weighed according to \citet{bailerjones21}, which also includes the \citet{lindegren21} bias (72\%), 
    W1-band Period-Luminosity (PL) relations (25\%), 
    K-band PL (2\%), and J-band PL (1\%). This ranking was adopted in \citetalias{da2023oxygen} to favor purely geometrical distances over distances from PL relations. On the other hand, the ranking of PLs is W1-K-J to minimize the impact of extinction, which is less severe at redder passbands.     
    The \citetalias{da2023oxygen} sample includes 1,285 HR spectra and more than 80\% have a signal-to-noise ratio per pixel greater than 100. For a comprehensive description of 
    the \citetalias{da2023oxygen} spectroscopic sample the reader is referred to \citet{da2022new} and \citetalias{da2023oxygen}.
    
    The \citetalias{da2023oxygen} dataset was complemented with 22 CCs for which 40 HARPS-N, 10 UVES and 16 ESPaDOnS spectra were available from public archives: 35 HARPS-N are proprietary spectra collected in the 
    Stellar Population Astrophysics\footnote{The Stellar Population Astrophysics is an ongoing project 
    based on a Large Programme conducted at the Telescopio Nazionale Galileo (TNG) using the HARPS-N 
    and GIANO B echelle spectrographs for about 74 nights from 2018 to 2021} (\citealt{origlia19}) 
    programme, while the other 5 were available in the TNG database and analysed in \citet{ripepi21}. 
    The additional UVES and ESPaDOnS spectra have been analysed in \citet{trentin23, Trentin24B}, \citet{martin15}, 
    \citet{andrievsky16} and \citet{kovtyukh22}. Details of the spectroscopic sample, such as the spectrographs, the resolution, 
    and the spectral coverage, are provided in Table~\ref{spec_sample}. We estimated the individual distances by using the same 
    approach adopted by \citetalias{da2023oxygen}: Gaia EDR3 parallaxes for 20 CCs and W1-band PL for 2 CCs. For two CCs (ASAS 181024-2049.6 and ASAS SN J065046.50-085808.7) 
    we estimated a different distance from \citet{andrievsky16} and \citet{trentin23}. 
    The new distance estimates are discussed in more detail in the Appendix~\ref{distances}.
  
   \subsection{Spectral synthesis method}
   \label{spectral_synthesis_method}
   We utilized a Python version of Spectroscopy Made Easy (SME, \citealt{piskunov17}) for spectral synthesis, i.e. PySME\footnote{\texttt{pysme-astro}=0.6.20, https://github.com/MingjieJian/SME} (\citealt{wehrhahn21}). PySME generates synthetic spectra based on a given set of atmospheric parameters, specified spectral intervals, and spectral resolution. It determines the optimal values for the selected atmospheric parameters by fitting the synthetic spectra to the observed data, taking into account the data uncertainties. The free parameters in the fitting process can include one or more stellar parameters, specific elemental abundances, and parameters related to atomic transitions in the line list. PySME provides two uncertainties according to two different methods. The first is based on SME statistics, i.e. uses a metric based on the distribution of the derivatives for each free parameter, estimating the cumulative distribution function of the generalized normal distribution (\citealt{piskunov17}). The second is based solely on the least-squares fit. Further details on PySME are presented in Appendix~\ref{details_synthesis}. 

   We adopted the line list presented in Table~\ref{linelist} to estimate the atmospheric parameters, including 
   Fe~I, Fe~II, Ti~I and Ti~II lines. Since CCs are intrinsically variable stars, they do not have fixed benchmark atmospheric parameters, but instead span a range of values throughout their pulsation cycle. For this reason, the validation of our strategy and line list includes non-variable stars with well-established atmospheric parameters. We first tested this line list in a NARVAL 
   solar spectrum with resolution R$\sim$68,000 and in HARPS and NARVAL spectra of 
   16 Gaia benchmark stars, both dwarfs and giants 
   (\citealt{blancocuaresma14}, \citealt{casamiquela25}). Spectral lines of CCs are broader with respect to dwarfs and to red giants because of extended convective envelopes and pulsation, 
   which are parameterised by higher microturbulence ($\mathrm{v_{mic}}$) and 
   macroturbulence ($\mathrm{v_{mac}}$) values (\citealt{da2023oxygen}; \citealt{luck81}, \citeyear{luck85}; \citealt{bersier96}). 
   Consequently, spectral lines that show minimal blending in the Sun may experience more 
   significant blending effects in CCs. Fig. \ref{benchmark} shows the difference between the atmospheric parameters derived in this work and the reference values for the Sun and the benchmark stars provided by \citet{blancocuaresma14}, which take into account any NLTE corrections (\citealt{jofre14}).\\
   The atmospheric parameters are estimated according to the following steps:
   \begin{itemize}
       \item Definition of the line masks, with information on the spectral coverage (segments), lines and continuum regions, and lines used for the RV adjustment;
       \item Definition of the continuum and RV optimisation, we optimised the RV and the continuum level in each segment ($\mathrm{sme.vrad\_flag="each"}$, $\mathrm{sme.cscale\_flag = "linear"}$, $\mathrm{sme.cscale\_type = "match+mask"}$);
       \item Definition of the atmosphere model, the abundance scale and the NLTE grids. We used the grid of MARCS model atmospheres (\citealt{gustafsson2008grid}), that PySME-package defines as 'marcs2012'. We used solar abundances from \citet{asplund09} and the most updated versions of the NLTE grids \footnote{https://zenodo.org/records/3888394} (\citealt{amarsi2020galah}). In order to have supersolar [N/Fe] and subsolar [C/Fe], typical of stars that have experienced mixing episodes during their evolution such as CCs, we used A(N)=8.23 and A(C)=8.2 for [Fe/H]=0.0;
       \item Spectral fitting in two different steps. In the first step, the free parameters are $\mathrm{T_{eff}}$, $\mathrm{\log(g)}$, $\mathrm{[Fe/H]}$, $\mathrm{v_{mac}}$ and A(Ti), while in the second step the free parameters are $\mathrm{[Fe/H]}$, $\mathrm{v_{mic}}$, $\mathrm{v_{mac}}$ and A(Ti). 
   \end{itemize}

A two-steps fitting procedure based on our selected line list is adopted to disentangle the contribution of each fitting parameters. In the first step, a simultaneous optimisation is performed to account for line broadening due to $\mathrm{v_{mac}}$, which in CCs can reach substantial values (5-30 km $\mathrm{s^{-1}}$ or higher). This approach provides reliable estimates of $\mathrm{T_{eff}}$ and $\mathrm{\log g}$ for the second step, while also yielding informed priors on $\mathrm{v_{mac}}$, $\mathrm{[Fe/H]}$, and $\mathrm{[Ti/H]}$. By splitting the fitting into two lower-dimensional procedures, we reduced the computational cost while ensuring that the derived parameters were consistent with theoretical expectations for pulsating stars (see Appendix \ref{details_synthesis}).

We focus on the elements for which NLTE grids are available in PySME, that are C, N, O, Na, Mg, Al, Si, S, K, Ca, Ti, Mn, Fe, Cu, Ba (\citealt{amarsi2020galah}, \citeyear{amarsi22}, \citeyear{amarsi25}; \citealt{mallinson24}; \citealt{caliskan25}). The line list highlighted in Table \ref{linelist_abu} is built from Gaia-ESO database according to the two quality parameters $\mathit{gfflag}$ and $\mathit{synflag}$ (further details in Appendix \ref{details_synthesis}). We used a NARVAL solar spectrum and three UVES spectra of red clump stars of M67, in order to validate our line list on giant stars for which the chemical abundances are well established (\citealt{occaso24}). In PySME, we use the abundances of the specific element and the macroturbulence as free parameters. The final [X/Fe] value is the median value of the [X/Fe] values of each line. Table~\ref{x_fe_reference_1} includes the [X/Fe] values of the calibrating stars.

We estimated the sensitivity of each spectral line to the respective chemical abundance, computed as the mean abundance variation over 10 CCs spectra, in response to changes in the atmospheric parameters. We varied $\mathrm{T_{eff}}$ of $\pm$50 K and $\pm$100 K, $\mathrm{\log(g)}$ of $\pm$0.1 and $\pm$0.2 dex, $\mathrm{[Fe/H]}$ of $\pm$0.05 and $\pm$0.1 dex, and $\mathrm{v_{mic}}$ of $\pm$0.2 and $\pm$0.4 km~s$^{-1}$. The results are listed in Table~\ref{sensitivitytable}

\section{Abundance Analysis}  
\label{abundance_analysis}
This section deals with the estimate of atmospheric parameters and the comparison with similar estimates available in the literature. Furthermore, we also discuss elemental abundances and their radial gradients with a comparison of the NLTE and LTE results. Finally, we present the kinematic properties of the sample.

\subsection{Atmospheric parameters}
\begin{figure}[!ht]
    \centering
    \includegraphics[width=0.98\linewidth]{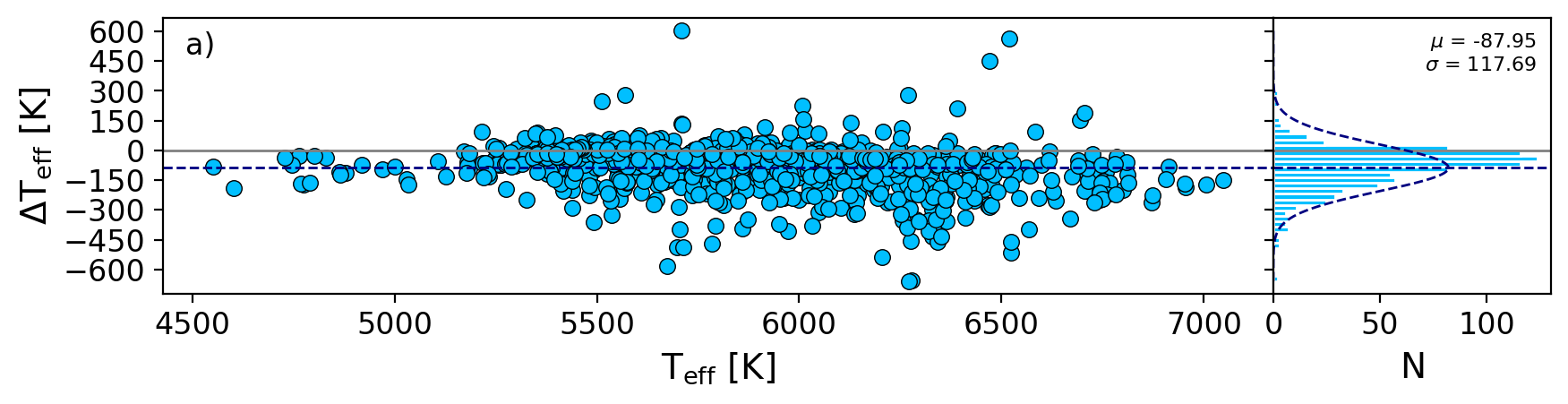}
    \includegraphics[width=0.98\linewidth]{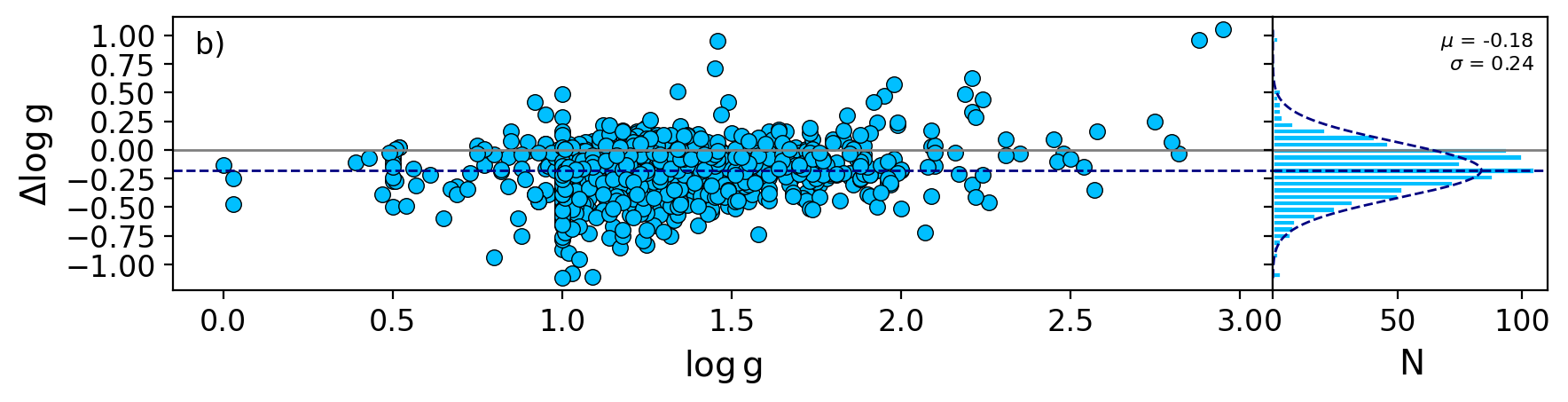}
    \includegraphics[width=0.98\linewidth]{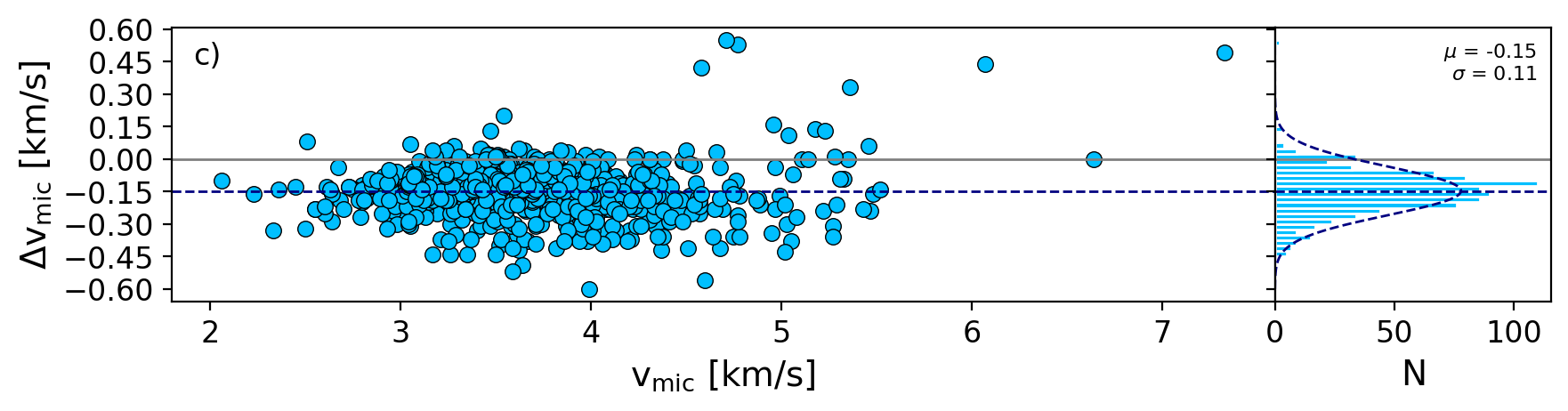}
    \includegraphics[width=0.98\linewidth]{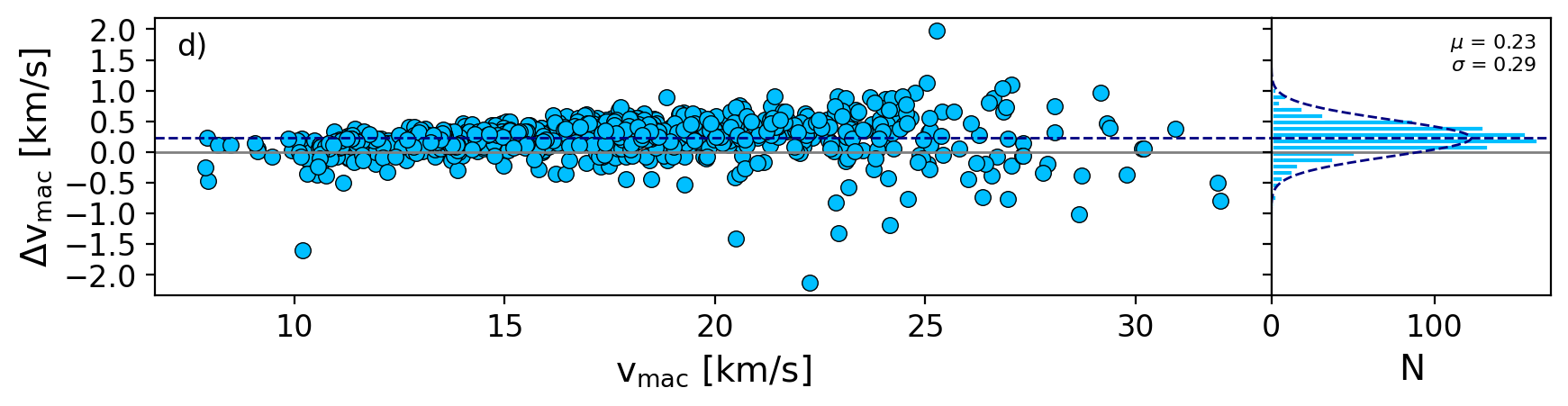}
    \includegraphics[width=0.98\linewidth]{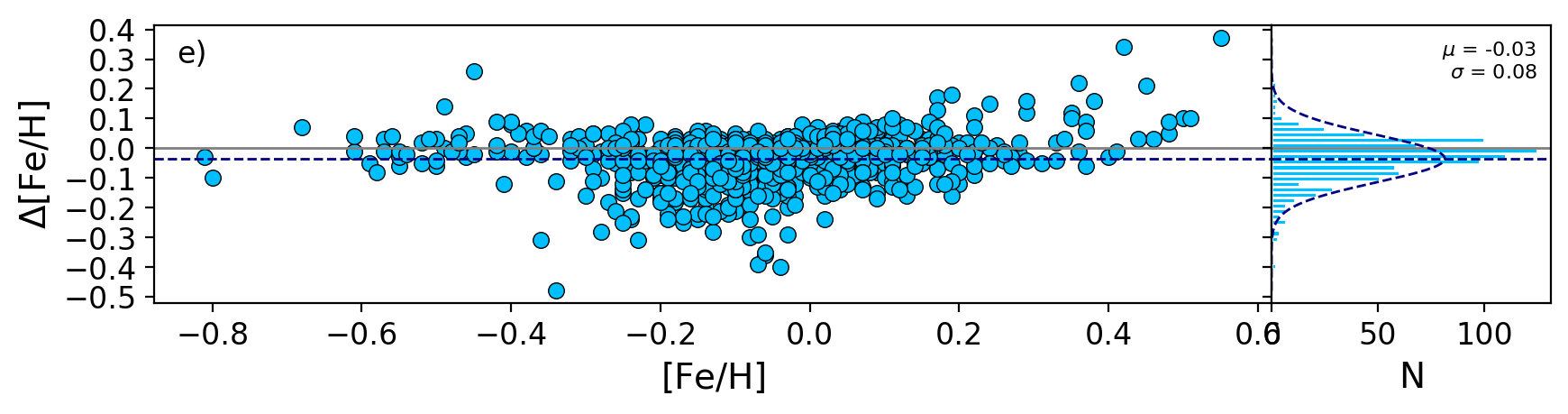}
    \includegraphics[width=0.98\linewidth]{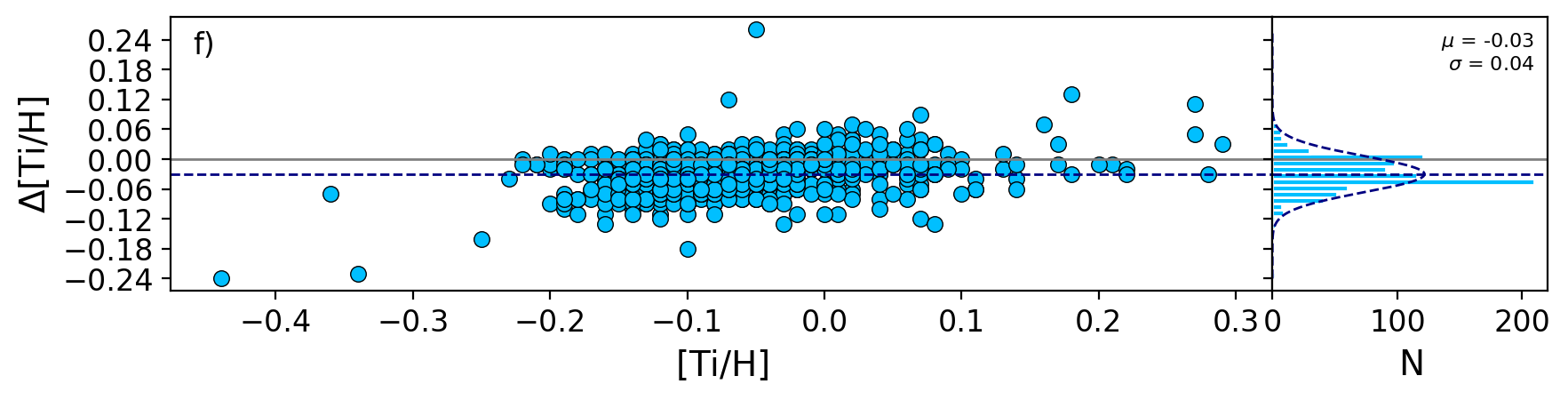}
    \caption{Comparison of atmospheric parameters derived under NLTE and LTE assumptions. Each panel displays the difference (NLTE – LTE) as a function of the corresponding parameter on the left, and the distribution of the difference on the right. A blue-dotted Gaussian fit is overplotted on each histogram, the mean and standard deviation are labelled in the top-right corner. The blue-dashed horizontal line shows the mean difference, while the grey solid line the null difference.}
    \label{lte_plots}
\end{figure}

We estimated the atmospheric parameters in LTE once the NLTE grids were turned off, leaving the rest of the code unchanged. Fig.~\ref{lte_plots} shows the difference of the atmospheric parameters and of iron and titanium abundances derived under NLTE and LTE assumption, with histograms. Effective temperature, surface gravity and microturbulence velocity are the most affected by the departures, 
while Fe and Ti abundances display minimal variations, indeed the differences are, on average smaller than $-0.03$~dex. Although, the difference between NLTE and LTE abundances is modest and the standard deviation is, on average, smaller than 0.08~dex, the individual differences can be of the order of $\pm0.4/0.5$~dex. Individual differences are in several cases larger than random errors, thus suggesting that the NLTE approach gives narrower metallicity distribution functions (smaller standard deviations).

Previous investigations addressing NLTE effects on the atmospheric parameters of FGK-type stars (see \citealt{ruchti13}; \citealt{kovalev19}) show that NLTE analyses generally yield higher $\mathrm{T_{eff}}$, $\mathrm{\log g}$ and $\mathrm{[Fe/H]}$ values, particularly at low metallicities, as well as systematically higher microturbulent velocities of about $\mathrm{0.1}$-$\mathrm{0.2\, km\, s^{-1}}$ for $\mathrm{[Fe/H]\approx -1}$. Cepheids are supergiants and the $\mathrm{v_{mic}}$ is intrinsically much higher. The increase is supported by data plotted in Fig.~\ref{lte_plots}, where $\mathrm{v_{mic}}$ is the only parameter that shows a statistically significant offset ($\mathrm{-0.15\pm 0.11\, km \, s^{-1}}$), suggesting that microturbulence has a substantial impact on the determination of the other atmospheric parameters, which explains why we also find negative NLTE-LTE differences in $\mathrm{T_{eff}}$, $\mathrm{\log g}$ and $\mathrm{[Fe/H]}$.
Further discussion and a comparison with previous studies is presented in Appendix \ref{test_nlte}.

\begin{figure}[!ht]
   \centering
   \includegraphics[width=8cm]{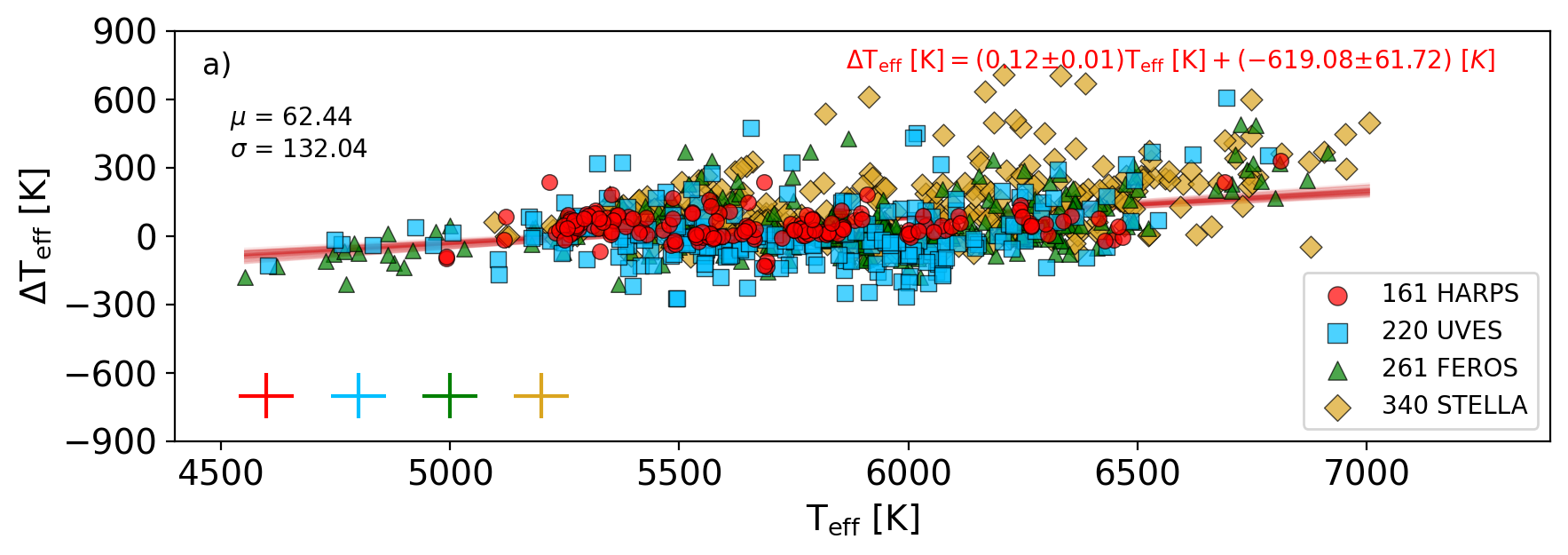}
   \includegraphics[width=8cm]{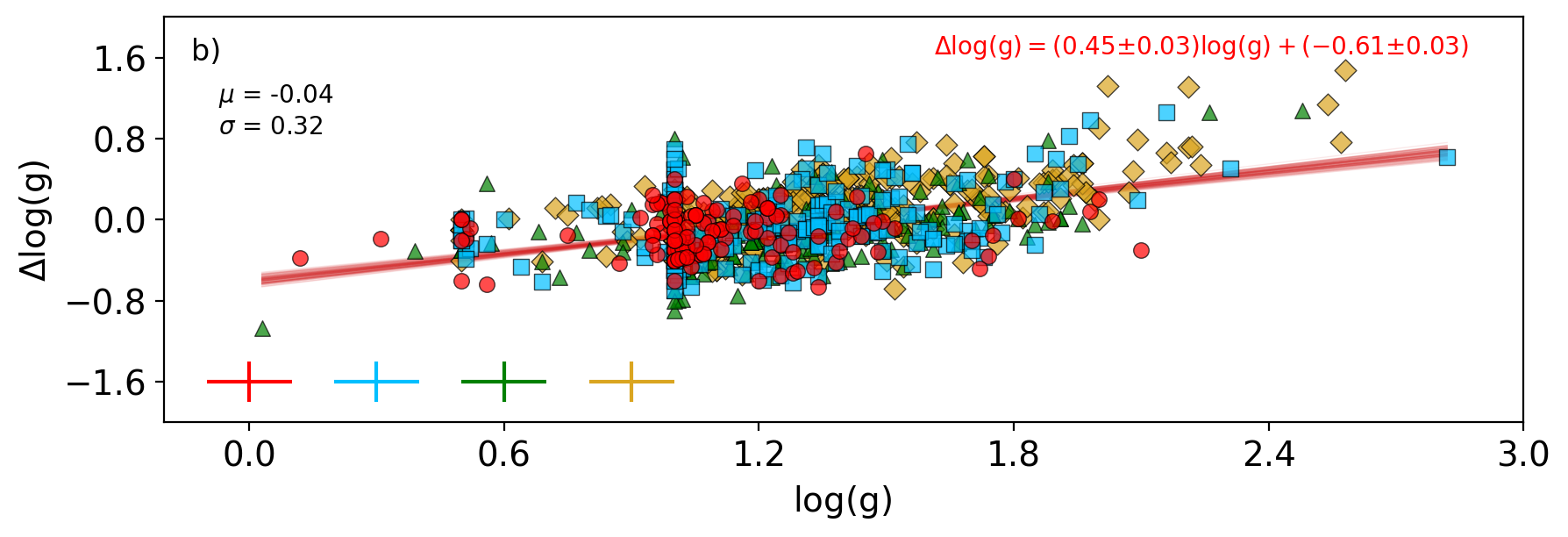}
   \includegraphics[width=8cm]{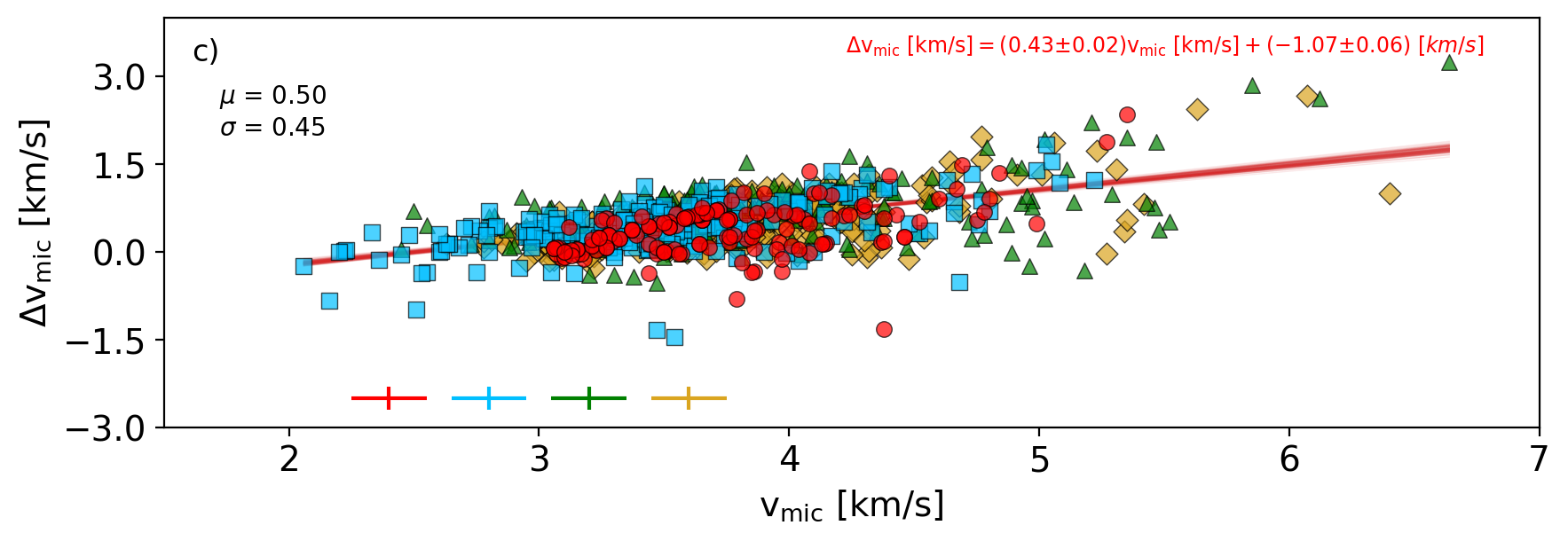}
   \includegraphics[width=8cm]{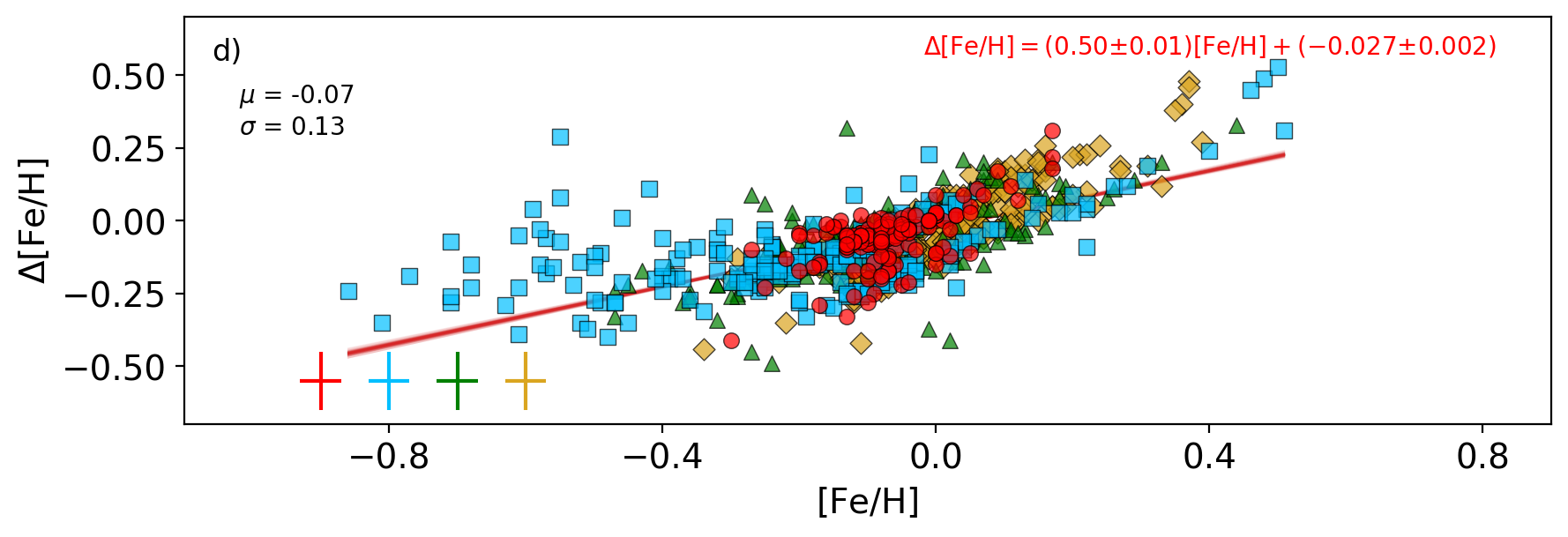}

      \caption{Panel a) -- Difference between the $\mathrm{T_{eff}}$
      from current estimates and \citetalias{da2023oxygen}, plotted as a function of the current $\mathrm{T_{eff}}$. Spectra from different 
      datasets are marked with different colours. The error bars are plotted in 
      the bottom left corner. The red line shows the linear fit and the coefficients 
      are displayed on top of each panel. The mean and the standard deviation are 
      also labelled.
      Panel b) -- Same as Panel a), but for the surface gravity. 
      Panel c) -- Same as Panel a), but for the microturbulent velocity (km~s$^{-1}$).
      Panel d) -- Same as Panel a), but for the iron abundance (dex).
    }
         \label{comp_ronaldo}
   \end{figure}

To constrain possible differences with similar investigations we also compared our NLTE 
atmospheric parameters and abundances with those provided by \citetalias{da2023oxygen} by using the same 
spectra  (Fig.~\ref{comp_ronaldo}).
The LTE atmospheric parameters provided by \citetalias{da2023oxygen} were estimated by using pyMOOGi (\citealt{adamow17}), a Python 
version of the MOOG code (\citealt{sneden73}, \citeyear{sneden2002}), atmosphere models from ATLAS (\citealt{castelli2004new}), and the ARES code (\citealt{sousa07}, \citeyear{sousa15}) to measure the equivalent widths (EWs). In \citetalias{da2023oxygen}, the authors used a large line list composed of 521 Fe~I and Fe~II lines based on different datasets and validated on calibrating Cepheids \citep{proxauf2018new, da2022new}. In contrast, our line list consists of 36 Fe~I, Fe~II, Ti~I and Ti~II, and was built in such a way as to be able to estimate atmospheric parameters via synthesis (see Sec \ref{spectral_synthesis_method} and Appendix \ref{details_synthesis}).
Moreover, they derived $\mathrm{T_{eff}}$ by using the line depth ratio method (\citealt{kovtyukh07, proxauf2018new}), and $\mathrm{\log(g)}$, $\mathrm{v_{mic}}$, and $\mathrm{[Fe/H]}$ following the classical approach, i.e. consecutive iterations to reach ionization equilibrium for Fe~I and Fe~II lines and no trends in the Fe~I abundances versus the EWs. 
We found positive trends in the difference of the atmospheric parameters, and indeed the slopes range from $+0.12\pm 0.01$ for the effective temperature (panel a) to $0.45\pm 0.03$ for the  surface gravity (panel b) and to  $0.43\pm 0.02$ for the microturbolence (panel c), but the mean differences are well within 1$\sigma$. The difference in iron abundance also shows a positive slope, which can be attributed both to differences in spectral analysis and to the fact that, as expected, the more metal-poor Cepheids appear even more metal-poor when analysed using a NLTE approach compared to LTE (\citealt{bergemann2012non}; \citealt{hansen2013lte}). However, the mean difference is quite small ($-0.07\pm0.13$). 

\begin{figure}[!ht]
   \centering
   \includegraphics[width=9cm]{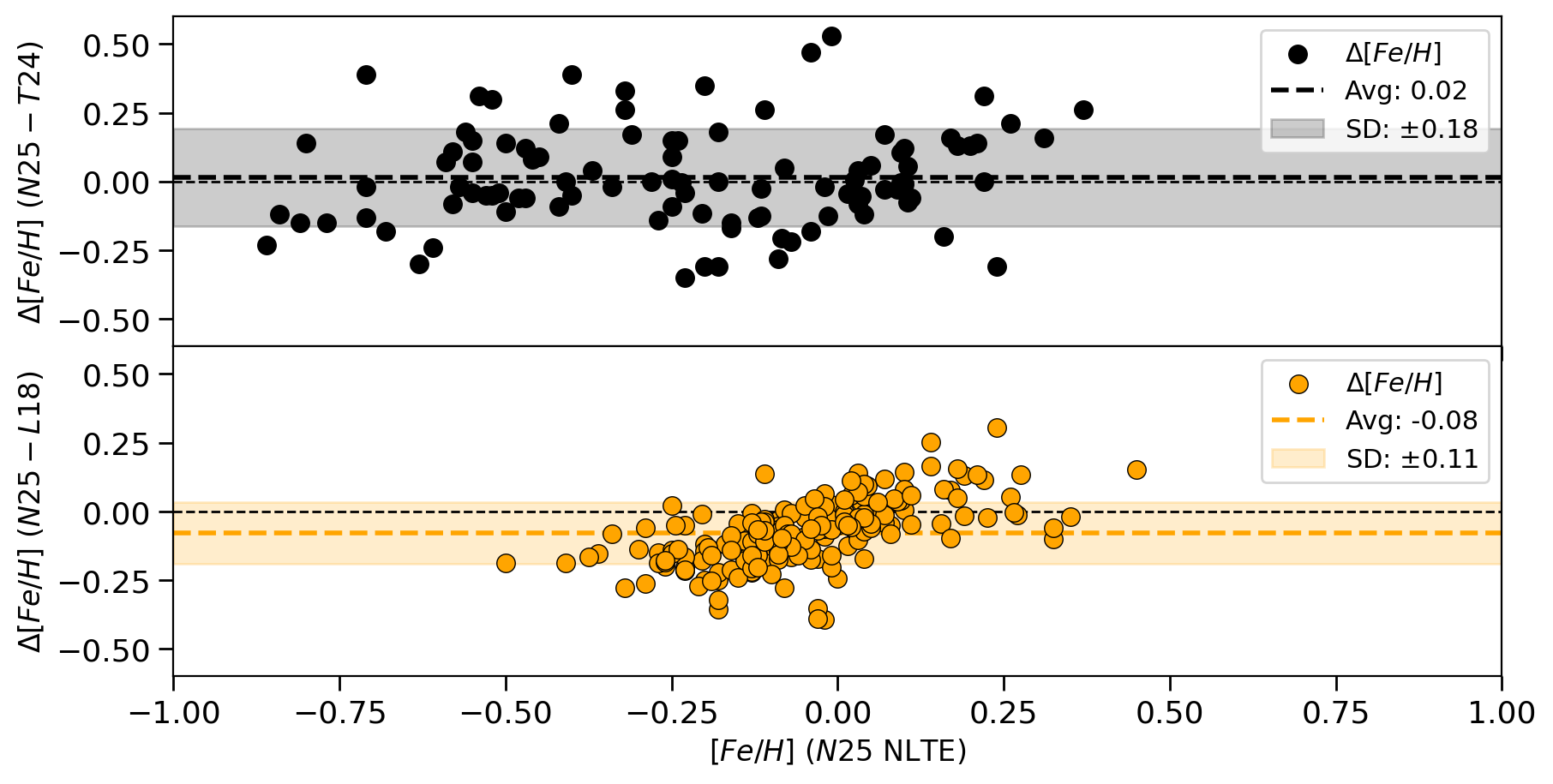}
      \caption{Comparison of our metallicity with \citet[top panel]{Trentin24B} and with \citet[bottom panel]{luck18}.}
         \label{comp_luck_trentin}
   \end{figure}

Furthermore, we compared our iron abundance with those from previous studies, specifically 
we do have 101 Cepheids in common with \citet{Trentin24B} and 209 Cepheids in common with 
\citet{luck18}. The top panel of Fig.~\ref{comp_luck_trentin} shows that the comparison with 
\citet{Trentin24B} covers a broad range in iron abundance ($-0.85\leq\mathrm{[Fe/H]}\leq 0.40$ dex)
and the difference does not show any trend, while the comparison with \citet{luck18} 
(bottom panel of the same figure) covers a narrower range in iron abundance ($-0.50\leq\mathrm{[Fe/H]}\leq0.50$ dex), 
but the difference displays a clear positive trend. The mean differences are quite small 
$0.02\pm0.18$~dex \citep{Trentin24B} and $-0.08\pm0.11$~dex \citep{luck18}, but the individual 
differences, once again, reach $\pm$0.4 dex.

The trends and offsets observed in comparison with the literature (Figs.~\ref{comp_ronaldo}~and~\ref{comp_luck_trentin}) reflect the methodological nature of the discrepancies, which arise primarily from the adoption of different fitting strategies and line lists, and secondarily differences in the underlying physical assumption. This conclusion is supported by the tests presented in Appendix~\ref{details_synthesis} and Appendix~\ref{test_nlte}, where a more detailed discussion of the atmospheric parameters is provided.

\subsection{Iron radial gradient}
\begin{figure}[!ht]
   \centering
   \includegraphics[width=9.cm]{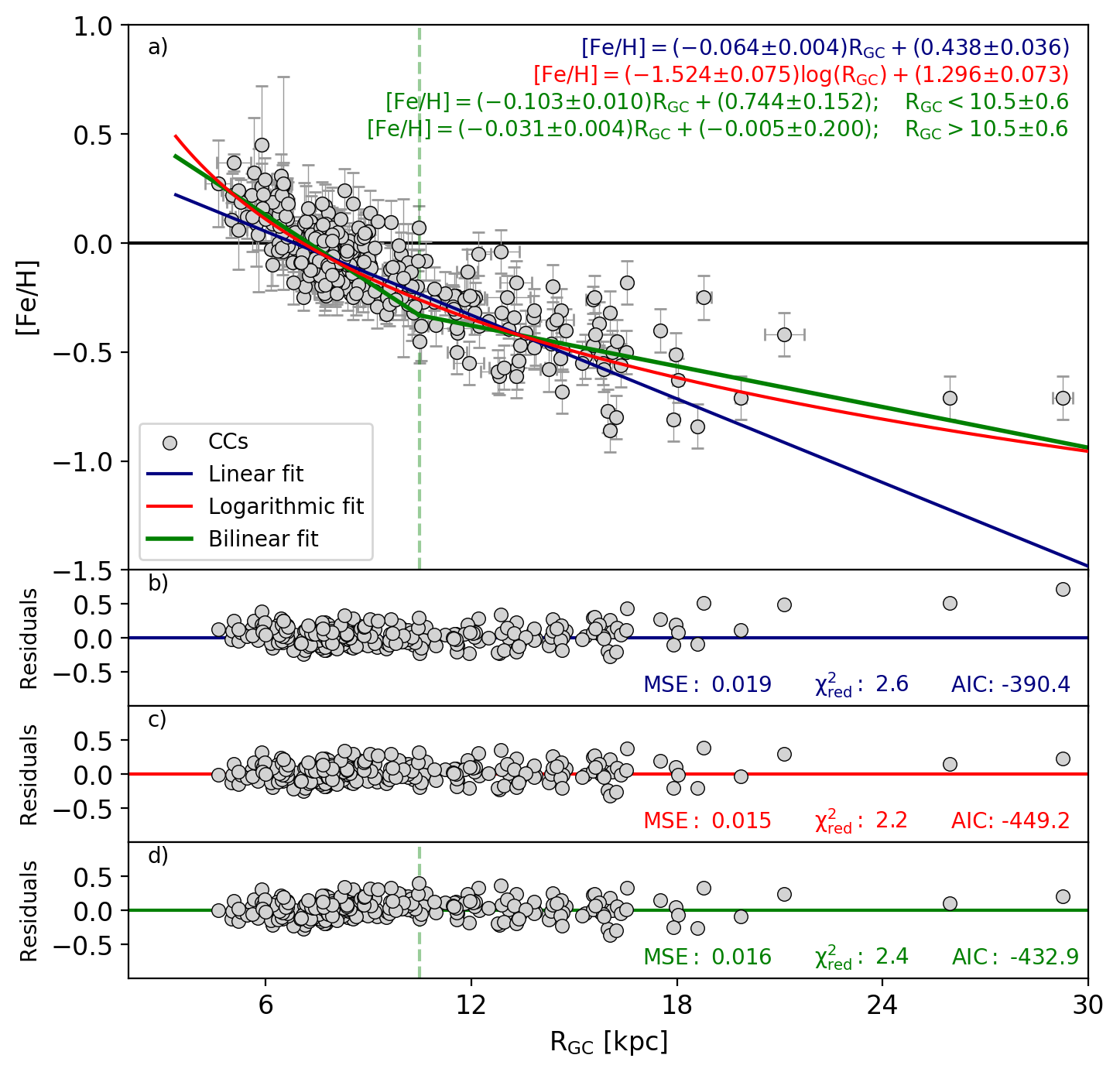}
      \caption{Panel a) -- Iron radial gradient for classical Cepheids as a function of the 
      Galactocentric distance. The linear, the logarithmic and the bilinear fits are shown respectively 
      in blue, red and green. The vertical green dashed line shows the 'knee' of the bilinear fit. 
      The coefficients of the fits are labelled. 
      Panel b) -- residuals of the linear fit. The values of the mean square error, the reduced chi-squared 
      and the AIC score are labelled.
      Panel c) -- Same as panel b), but for the logarithmic fit. 
      Panel d) -- Same as panel b), but for the bilinear fit.}
         \label{iron}
   \end{figure}

   The Galactocentric distances covered by our sample range from 4.6 kpc to 29.3 kpc, with four CCs located in the outskirt of the thin disk ($\mathrm{R_{GC}>20}$~kpc). To investigate the iron radial gradient we adopted three different analytical fitting functions 
   (Fig.~\ref{iron}) by taking into account errors on both axes: $\mathrm{[Fe/H]\propto R_{GC}}$, $\mathrm{[Fe/H]\propto \log R_{GC}}$, and a bilinear function.

   Moreover, we adopted a Bayesian inference framework to 
   estimate model parameters and their uncertainties. Specifically, given data $D$ and parameters $\theta$, 
   the posterior probability is:
   \begin{equation*}
   p(\theta|D)\propto \mathcal{L}(D|\theta)\pi(\theta)  ,  
   \end{equation*}
   where $\pi(\theta)$ is the prior and $\mathcal{L}$ the likelihood. We work with the log-posterior $\log p(\theta|D)=\log\mathcal{L}(D|\theta)+\log\pi(\theta)+\mathrm{const}$. We sampled the posterior with the affine-invariant Markov Chain Monte Carlo (MCMC) sampler implemented in the \texttt{emcee} library to obtain parameter point estimates (MAP) and credible intervals (the 68\% credible region from the marginal posterior). In particular, we adopted mixture models (\citealt{foremanmackey14}), for which the per-data-point likelihood is given by a weighted sum of a foreground (signal) and a background (outlier) component. Writing $Q$ for the mixing fraction (the prior probability that a datum belongs to the background component), the contribution of datum $x_i$ is:
   \begin{equation*}
       \mathcal{L}_i=(1-Q)\mathcal{L}_{fg}(x_i,y_i,\sigma_{x,i},\sigma_{y,i}|\theta_{fg})+Q\mathcal{L}_{bg}(x_i,y_i,\sigma_{x,i},\sigma_{y,i}|\theta_{bg}),
   \end{equation*}
   and the total log-likelihood is $\log\mathcal{L}=\sum_i\log\mathcal{L}_i$. The parameter $Q$ therefore controls the relative weight of the background/outlier model and robustifies the inference against points that are inconsistent with the foreground model. In Gaussian implementations the components are typically normal distributions with their own means and variances (optionally including an extra 'jitter' term added in quadrature to account for underestimated measurement errors). Numerical evaluation of $\log\mathcal{L}$ should use stable log-sum-exp arithmetic to avoid underflow when one component is much smaller than the other. Convergence of the MCMC chains was verified via autocorrelation-time estimates and visual inspection of trace plots.
   
   We provide the mean square error, the reduced chi-squared and the Akaike's information criterion (AIC) scores for each fit. The AIC is a method for evaluating and comparing statistical models, which provides a measure of the quality of a statistical model's estimate, taking into account both the goodness of the fit and the complexity of the model. It is defined as $\mathrm{AIC}=2k-2\log L$, where $k$ is the number of parameters of the model and $L$ is the maximized value of the total likelihood. Models with lower AIC should be preferred. 

Data plotted in the top panel of Fig.~\ref{iron} display quite clearly that the bilinear and the logarithmic fit 
reproduce quite well the radial gradient of CCs when moving from the innermost to the outermost disk regions. Note that the agreement 
for the bilinear fit is expected, since this fit has more degrees of freedom when compared with the linear and the 
logarithmic fit. However, the bilinear fit is also prone to possible systematics, since the edge of the two different intervals in 
Galactocentric distances is arbitrary. The maximized likelihood is obtained at a value of the 'knee' of $\mathrm{10.5\, kpc}$ (green dashed vertical line). In any case, the logarithmic fit is the most accurate analytical representation of the radial gradient, since the statistical parameters (MSE, reduced chi-squared, AIC score) attain their smallest values when compared with the other fits. 

\begin{figure}[!ht]
    \centering
   \includegraphics[width=9cm]{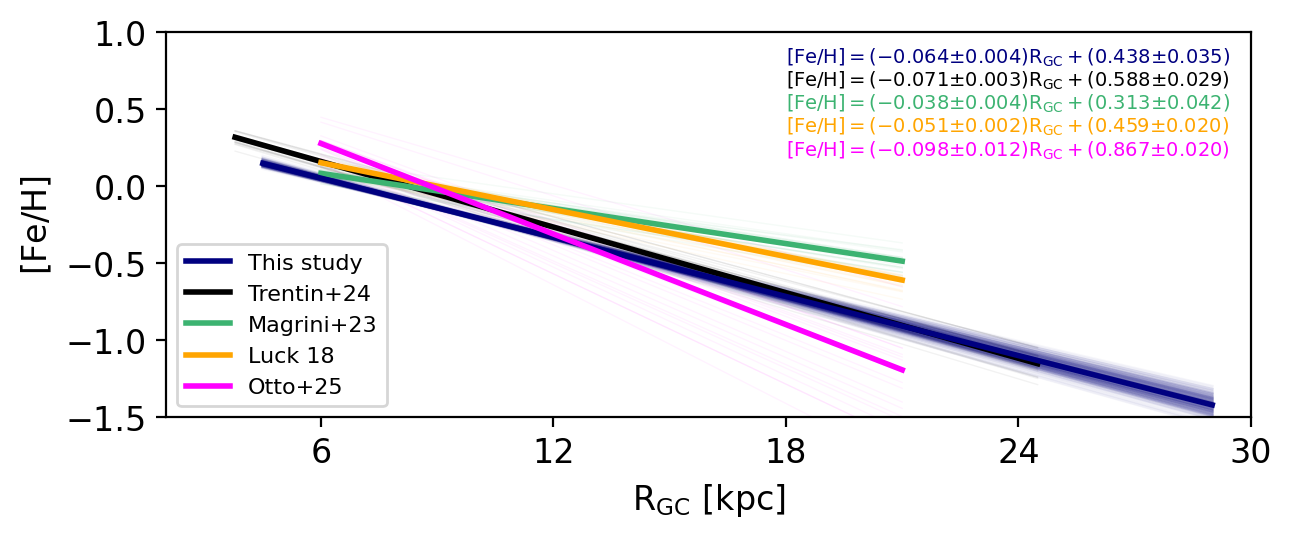}
      \caption{Comparison between this study, \citet{Trentin24B}, \citet{magrini23}, \citet{luck18} and \citet{otto25} in the [Fe/H] vs $\mathrm{R_{GC}}$ plane.}
    \label{log_lin}
\end{figure}

   Fig.~\ref{log_lin} shows the comparison between our linear iron radial gradient with those available in the literature, revealing a slope of $-0.064$~dex/kpc, which is steeper than the slopes provided by \citet[$-0.038$]{magrini23} by using open clusters and by \citet[$-0.051$]{luck18} by using CCs, and shallower than the slopes provided by \citet[$-0.071$]{Trentin24B} based on CCs and by \citet[$-0.098$]{otto25} by using open clusters. 
   
\begin{figure}[!ht]
    \centering
   \includegraphics[width=9cm]{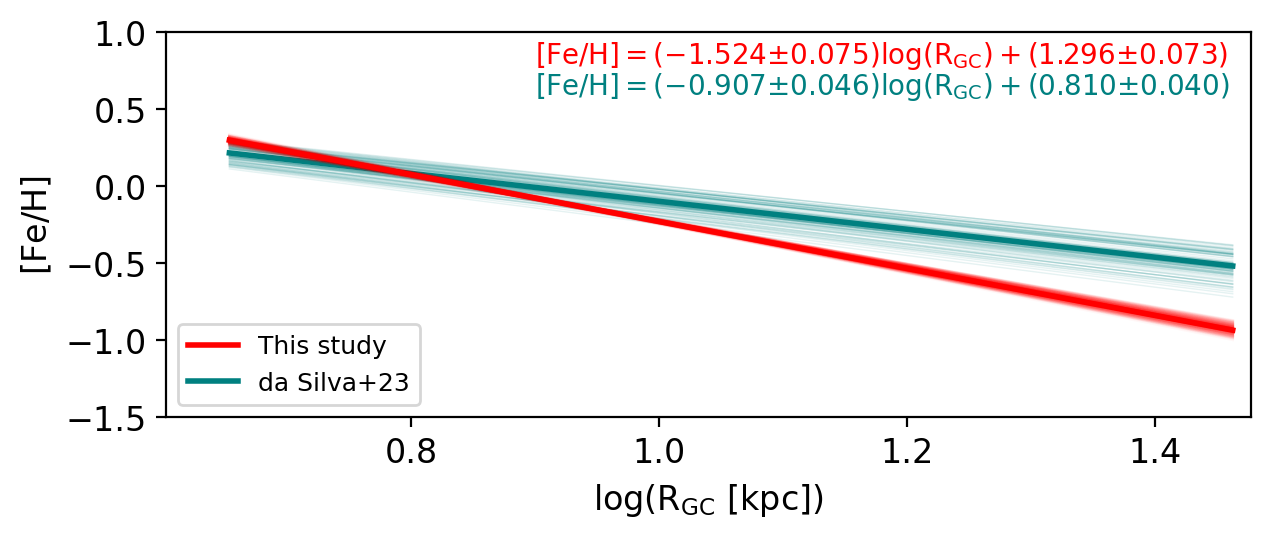}
      \caption{Comparison between the current iron radial gradient and that of \citetalias{da2023oxygen} in a log-log plane.}
    \label{log_log}
\end{figure}

Fig.~\ref{log_log} displays the comparison of the current iron radial gradient with that of \citetalias{da2023oxygen} in a  $\log-\log$ plane. The slope of the current gradient is steeper ($-1.53$ vs $-0.91$). The two radial profiles nearly overlap for $\mathrm{R_{GC}}\lesssim10$ kpc, but the current one becomes steeper in the outer disk. This difference is due to several reasons: the different spectral analysis method (including a different line list), the increased number of CCs at $\mathrm{R_{GC}\approx15-20\,kpc}$, which were not included in the \citetalias{da2023oxygen} sample, and NLTE effects, since the outer disk Cepheids are the most metal-poor stars in the current sample.


To overcome possible problems in the choice of the adopted analytical function to describe the iron 
radial gradient, we decided to use the Gaussian Process Regression (GPR) to model the data without 
assuming a predefined functional form. 
The GPR is a Bayesian non-parametric method for modelling an unknown function $f(x)$ in a regression 
setting, starting from a Gaussian process prior:   
  \begin{equation*}
        f(x) \sim \mathit{GP}(m(x), k(x,x'))
  \end{equation*}
  defined by a mean function $m(x)$ and a covariance (kernel) function $k(x,x')$. Observational data $y=f(x)+\epsilon$ (with noise $\epsilon\sim\mathit{N}(0,\sigma^2)$) induce a predictive posterior distribution for $f$ at new input points. This posterior yields both a predictive mean (the best estimate of the function) and a predictive variance (quantifying uncertainty in the estimate). The kernel function encodes assumptions about smoothness, length scales and correlation structure of the function. Hyper-parameters of the kernel are typically inferred by maximizing the marginal likelihood (or by Bayesian integration). Therefore, GPR adaptively fits complex, non-linear relationships while providing principled uncertainty quantification. In short, GPR offers a flexible and rigorous framework for regression when uncertainty estimates are required and when one prefers to avoid strong parametric assumptions on the functional form.
  
  We employed GPR by using Gpy\footnote{https://gpy.readthedocs.io/en/deploy/}. 
  After testing different kernel functions, we adopted GPy.kern.Matern32 according to the distribution dispersion and different sampling across the x-axis. Furthermore, we optimised the variance and the length-scale of the kernel function.
  
  \begin{figure}[!ht]
   \includegraphics[width=1\linewidth]{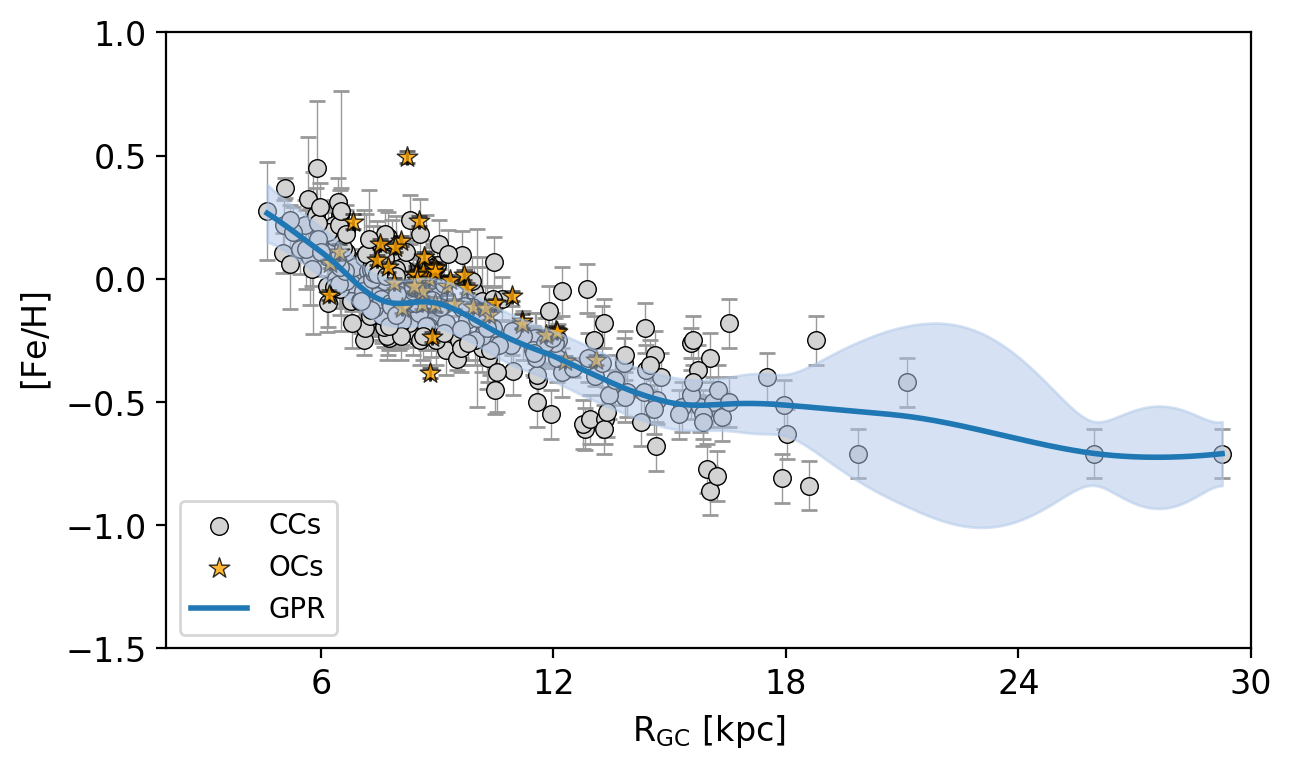}
      \caption{Iron abundances of CCs as a function of $\mathrm{R_{GC}}$, shown in grey. Orange symbols display open clusters younger than 400 Myr from \citet{otto25}. GPR applied to the CCs distribution is highlighted in blue, while the shaded region highlights the uncertainties on the model.}
    \label{gpr}
\end{figure}

Data plotted in Fig.~\ref{gpr} display several interesting features worth being discussed in more detail. The GPR fits quite well the distribution of CCs as a function of the Galactocentric distance and this outcome applies to both the inner and the outer disk Cepheids. The GPR shows a particular trend in the data across the Solar circle, a 'shoulder' at $\mathrm{R_{GC}}\sim$8~kpc. Moreover, GPR also shows a steady change in the slope for $\mathrm{R_{GC}}\sim$ 14-16 kpc. This evidence indicates that the linear fit fails to take account the radial gradient variations of the CCs in the outer disk, thus suggesting that the change is an intrinsic feature when moving into the outer disk. The radial distribution of open clusters younger than 400 Myr collected by \citet{otto25} agrees quite well with the radial distribution of Cepheids. This agreement is expected, since CCs are stellar tracers younger than $\sim$300~Myr. Unfortunately, the range in Galactocentric distances covered by this sample is too small when compared with CCs, but the agreement is very promising. 

\subsection{Radial gradients of chemical abundances}
\begin{figure}[!ht]
   \centering
   \includegraphics[width=9cm]{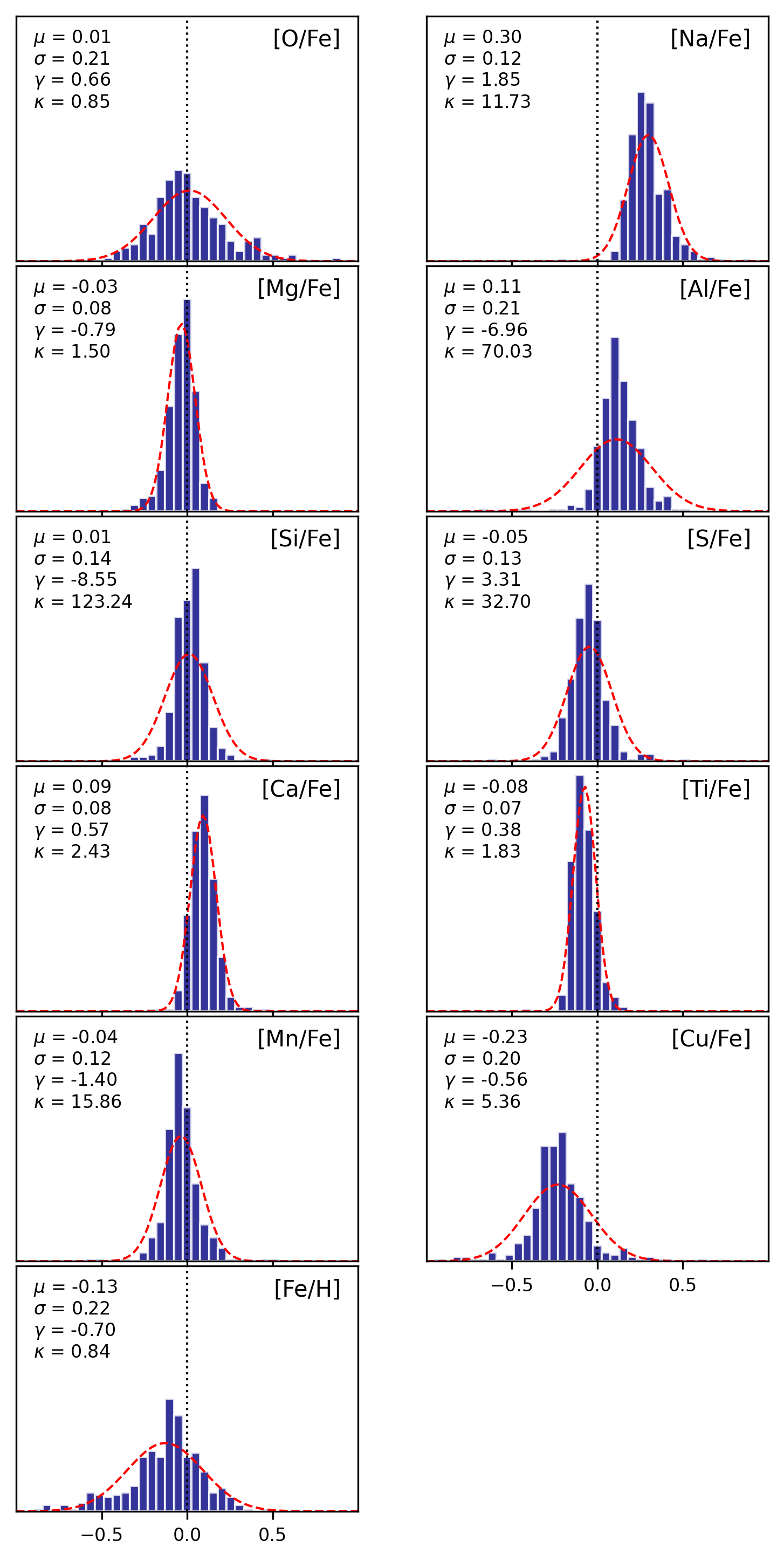}
      \caption{Distribution of chemical abundance ratios ([X/Fe]). The red dashed line shows the 
      Gaussian fit. Mean value ($\mu$), standard deviation ($\sigma$), skewness ($\gamma$) and kurtosis ($\kappa$) are labelled on the top 
      left corner of each panel. The bottom left panel shows the iron distribution function.}
         \label{hist}
   \end{figure}

Distributions of the chemical abundance ratios are shown in Fig.~\ref{hist}, together with 
overlaid Gaussian fits (dashed red lines). The present results indicate Solar abundance 
ratios, within 1$\sigma$, for four $\alpha$-elements (O, Mg, Si, S) and for one 
light metal (Al). In contrast, [Na/Fe] ratio shows a clear overabundance of +0.30$\pm$0.12 dex, while [Cu/Fe] appears underabundant, with a mean value of $\mathrm{[Cu/Fe]=-0.23\pm 0.20}$ dex. The Na overabundance is expected, since Cepheids are intermediate-mass stars and the low--mass tail experiences the first dredge-up along the red giant branch (see Fig.~C13 in \citealt{bono2024cepheids}). 
A similar Na overabundance was also reported by \citet{Trentin24B}, who found a mean value of $\mathrm{[Na/Fe]=+0.39\pm0.16}$ dex, whereas their [Cu/Fe] measurements yielded a mean value of $\mathrm{[Cu/Fe]=+0.10\pm0.27}$. \citet{otto25}, analyzing open clusters who derived abundances for both main-sequence and giant stars, obtained mean values of $\mathrm{[Na/Fe]=-0.16\pm0.72}$ dex and $\mathrm{[Cu/Fe]=-0.04\pm0.45}$ dex. In passing, an overabundance in [Na/Fe] was also found by \citet{genovali15}.

\begin{figure*}[!ht]
   \centering
   \includegraphics[width=9cm]{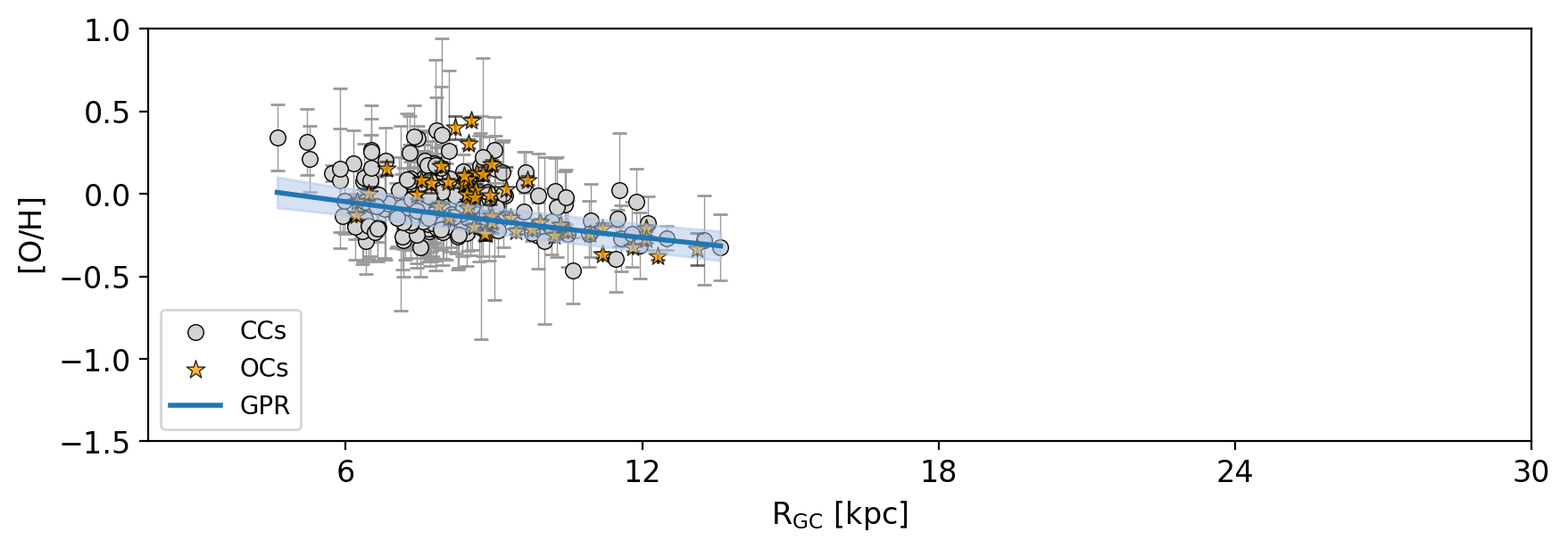}
   \includegraphics[width=9cm]{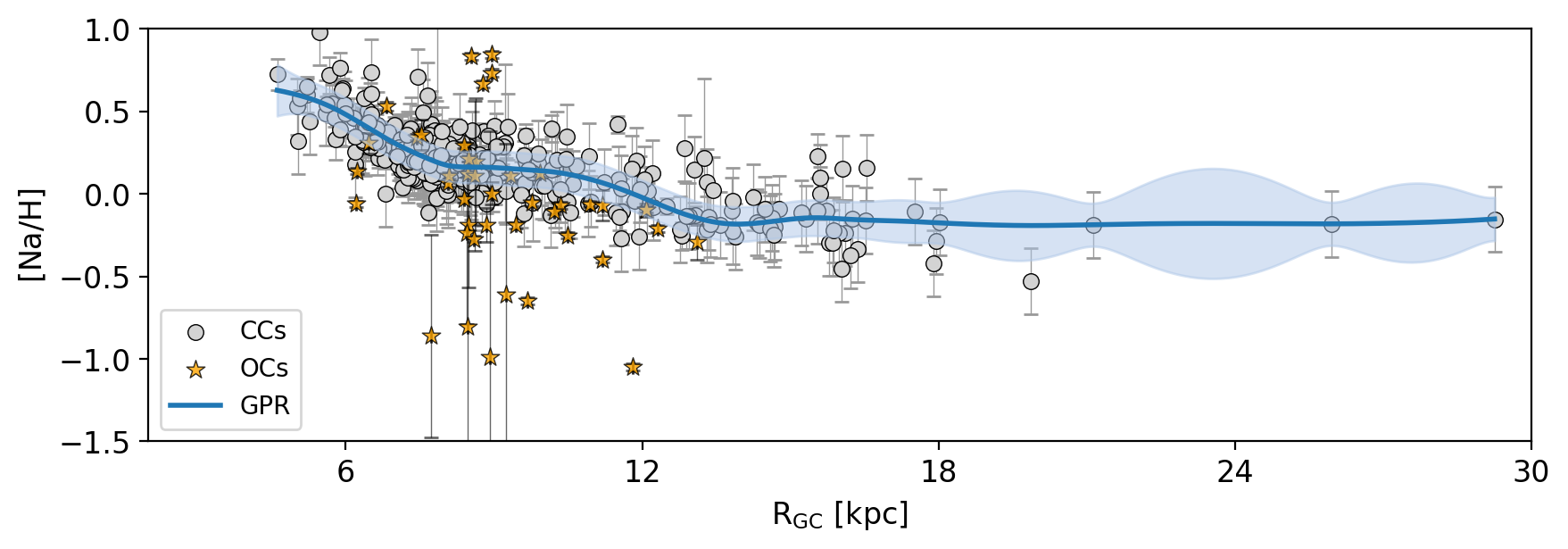}
   \includegraphics[width=9cm]{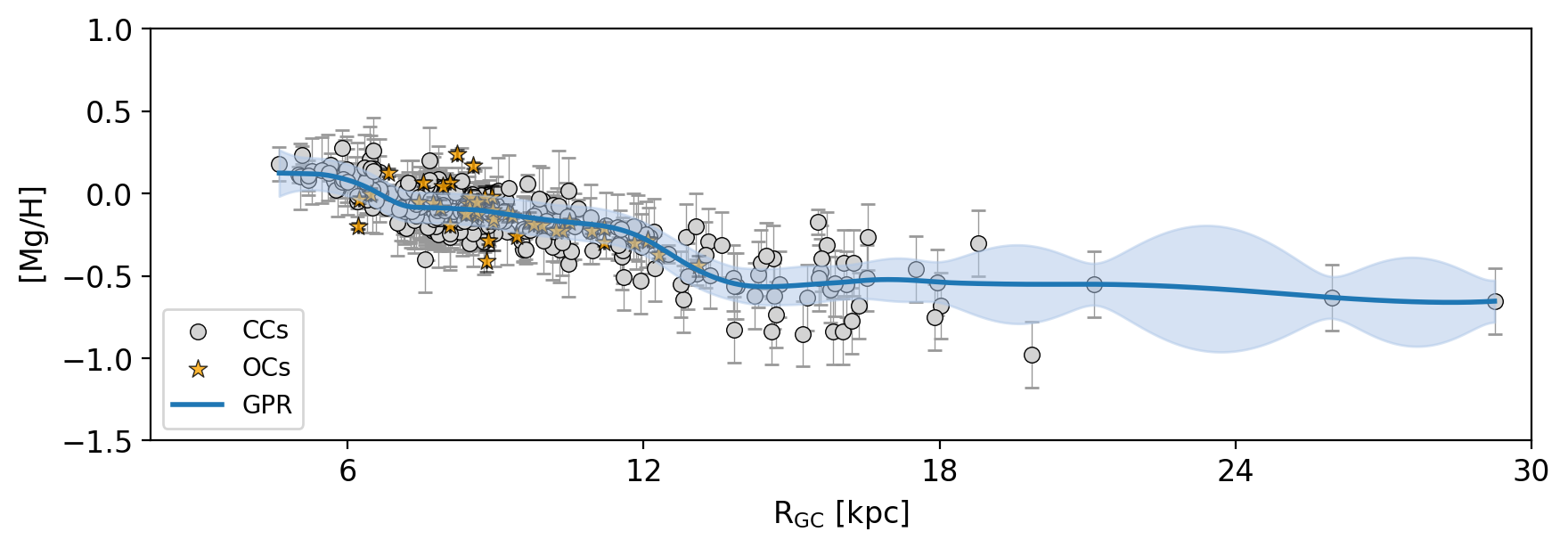}
   \includegraphics[width=9cm]{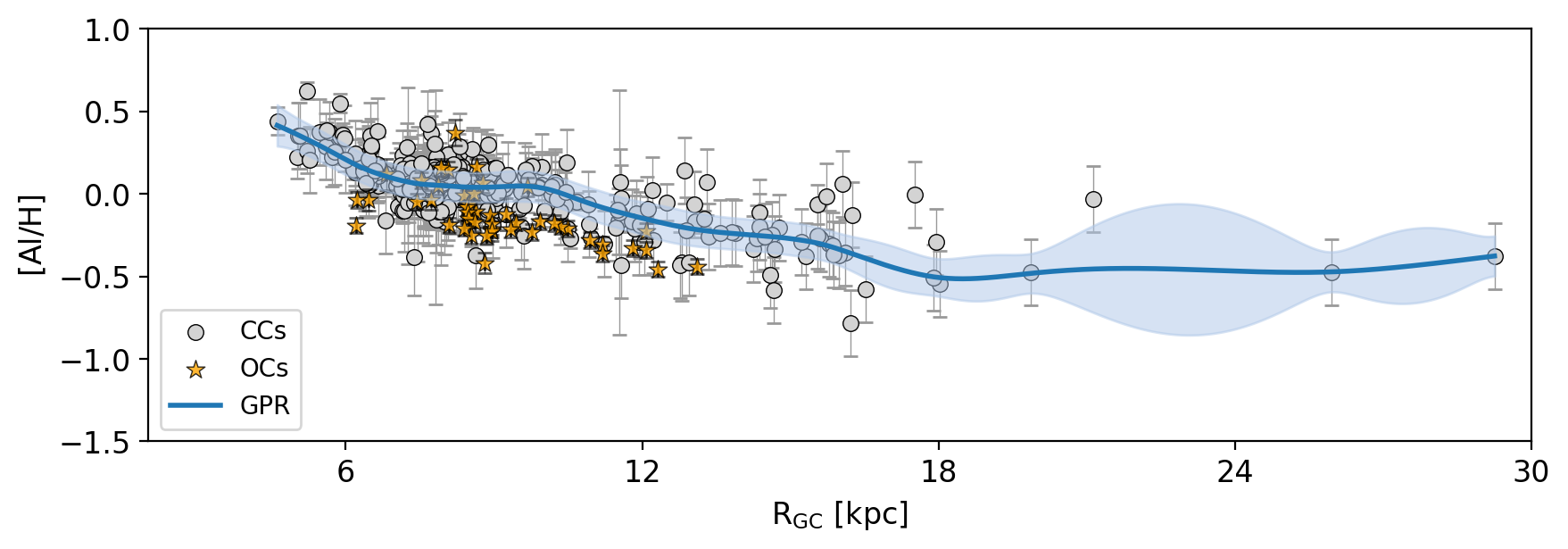}
   \includegraphics[width=9cm]{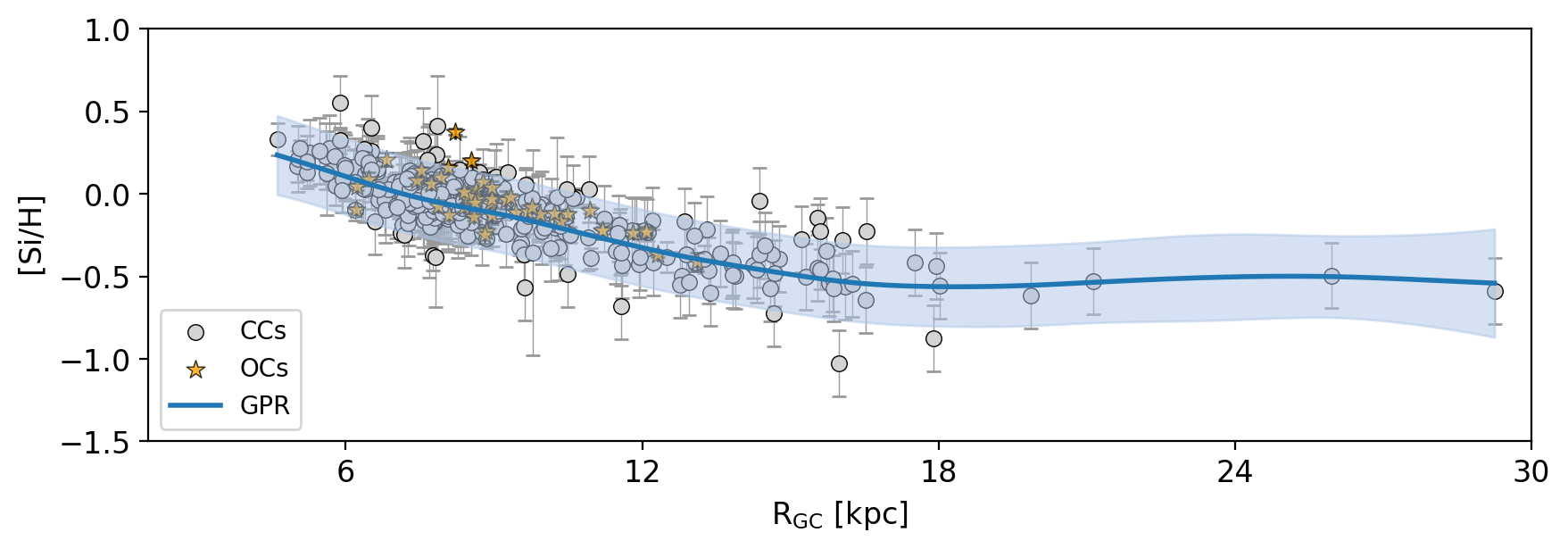}
   \includegraphics[width=9cm]{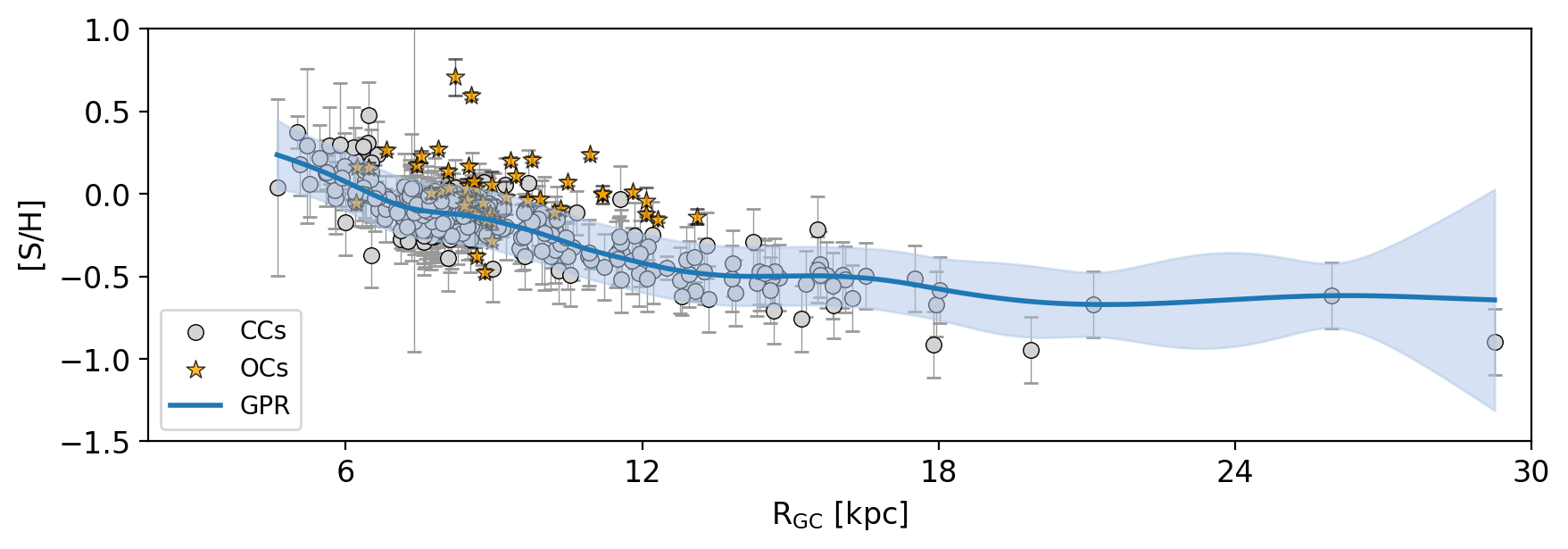}
   \includegraphics[width=9cm]{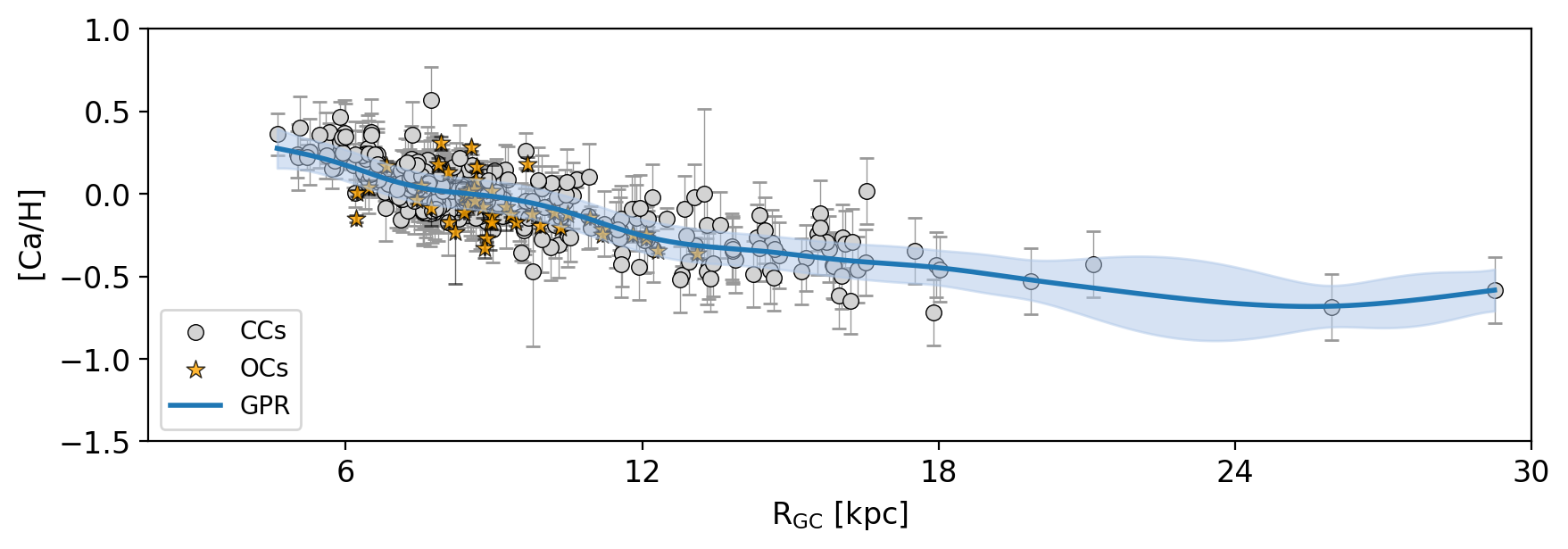}
   \includegraphics[width=9cm]{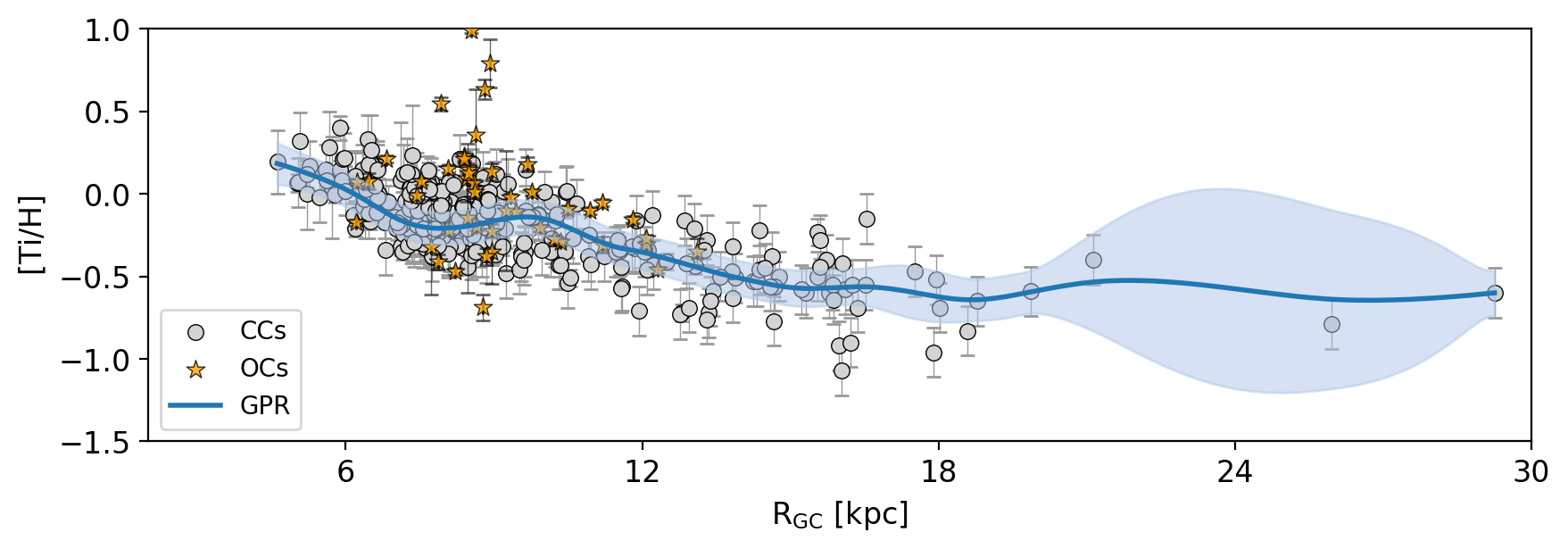}
   \includegraphics[width=9cm]{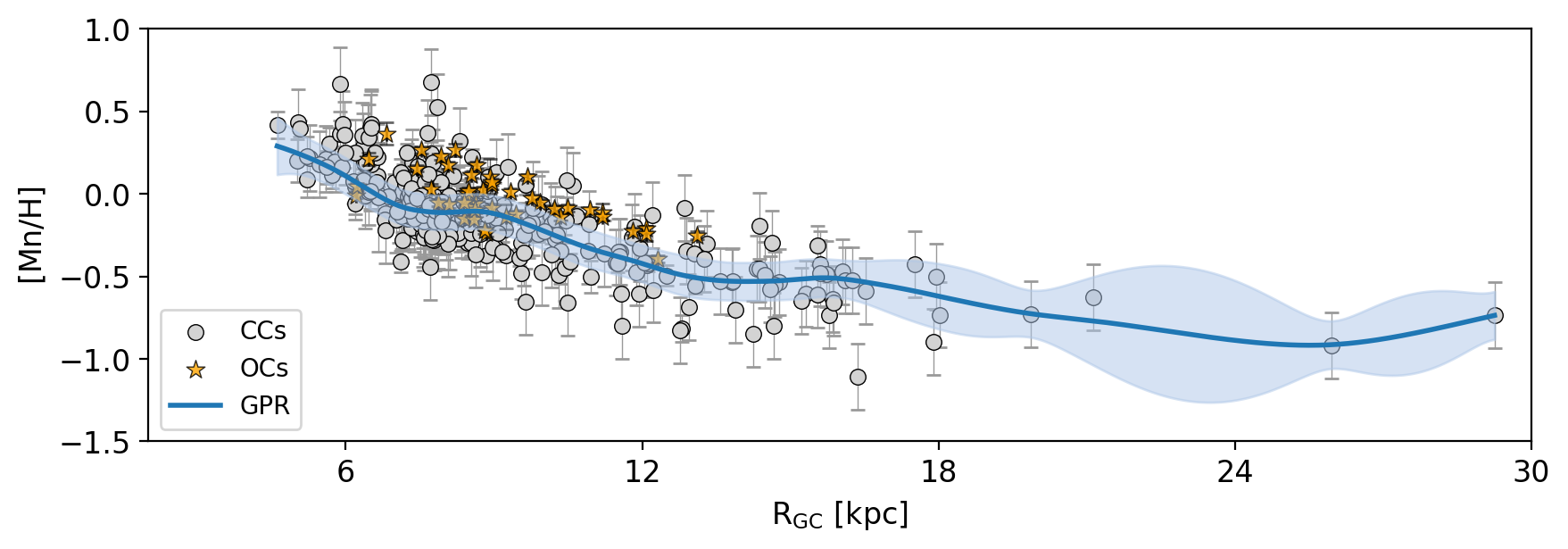}
   \includegraphics[width=9cm]{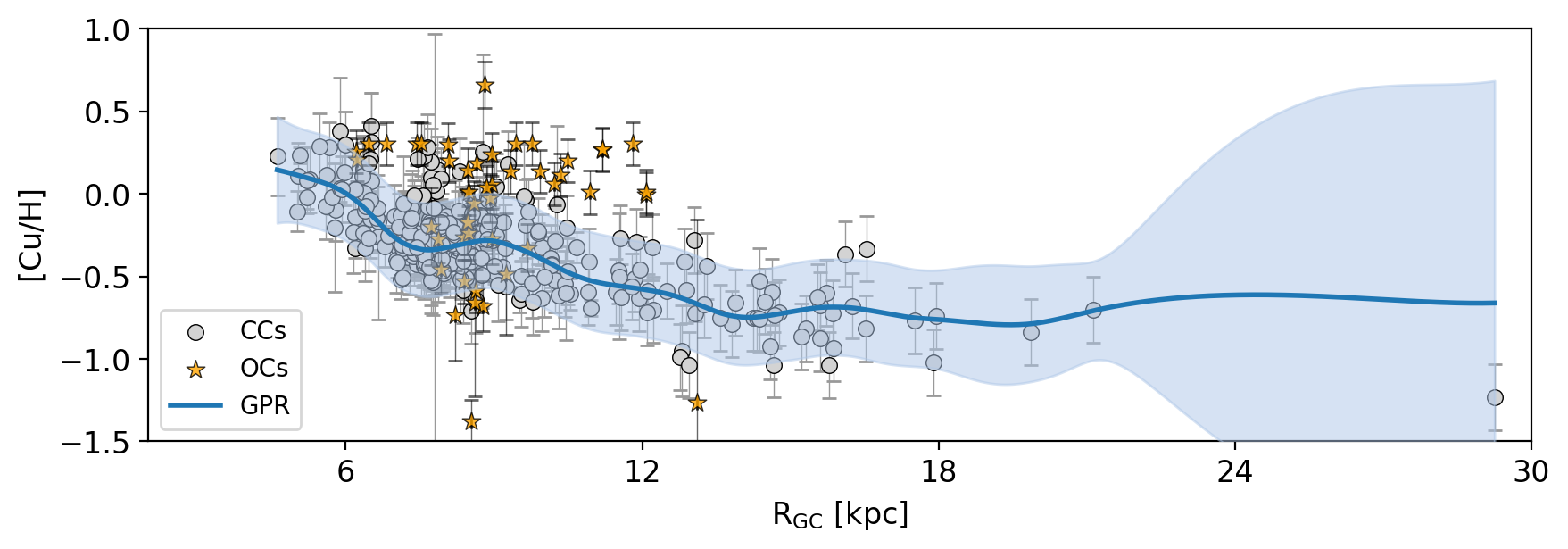}
      \caption{Chemical gradients with GPR modelling for O, Na, Mg, Al, Si, S, Ca, Ti, Mn and Cu. Colours and symbols are the same of Fig. \ref{gpr}.}
         \label{gradients_gpr}
   \end{figure*}

 Figure~\ref{gradients_gpr} shows the radial abundance gradient of ten out of the eleven investigated 
 chemical elements as a function of $\mathrm{R_{GC}}$, with blue solid lines representing the GPR models. Similar to iron, a clear flattening of the gradients at $\mathrm{R_{GC}}$ $\approx$ 14-16 kpc is obtained for all the elements, but for oxygen. Unfortunately,  O abundances are only limited to CCs located within 14~kpc, since our oxygen-determination is based on the triplet at 7770$\AA$ which is only covered by FEROS, STELLA and ESPaDOnS spectra in our spectroscopic dataset.
 
\begin{figure}[!ht]
   \includegraphics[width=4.4cm]{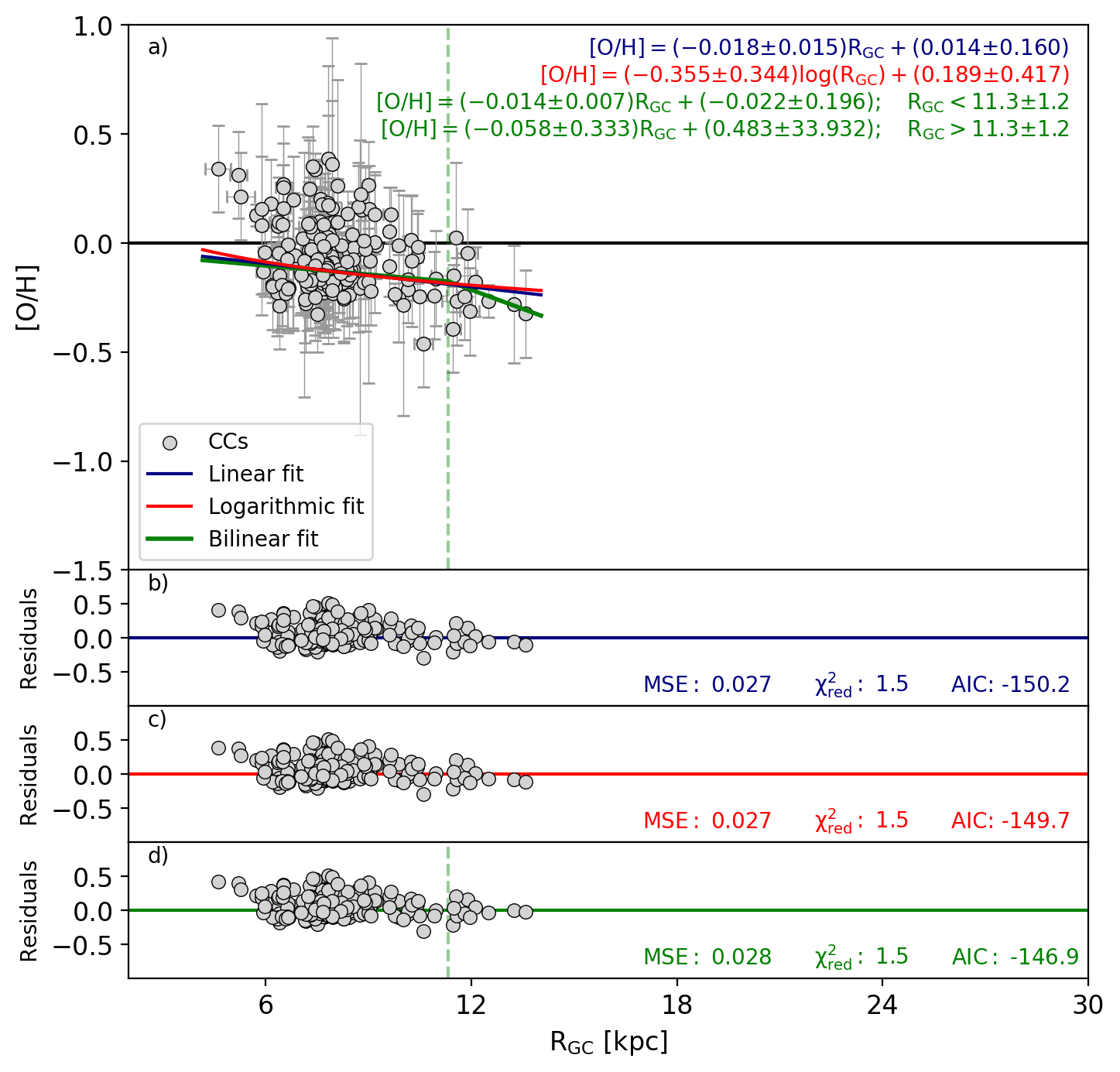}
   \includegraphics[width=4.4cm]{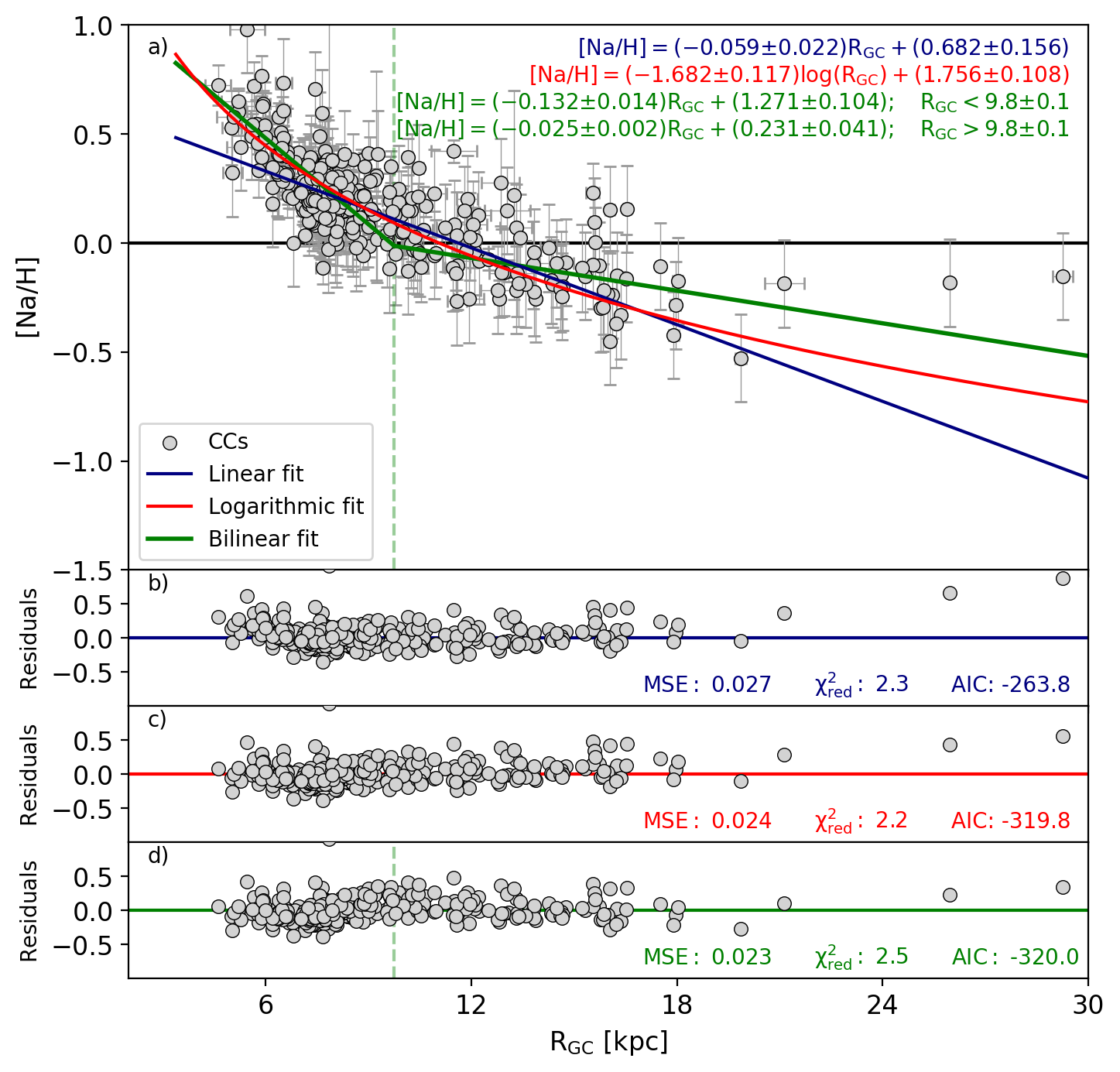}
   \includegraphics[width=4.4cm]{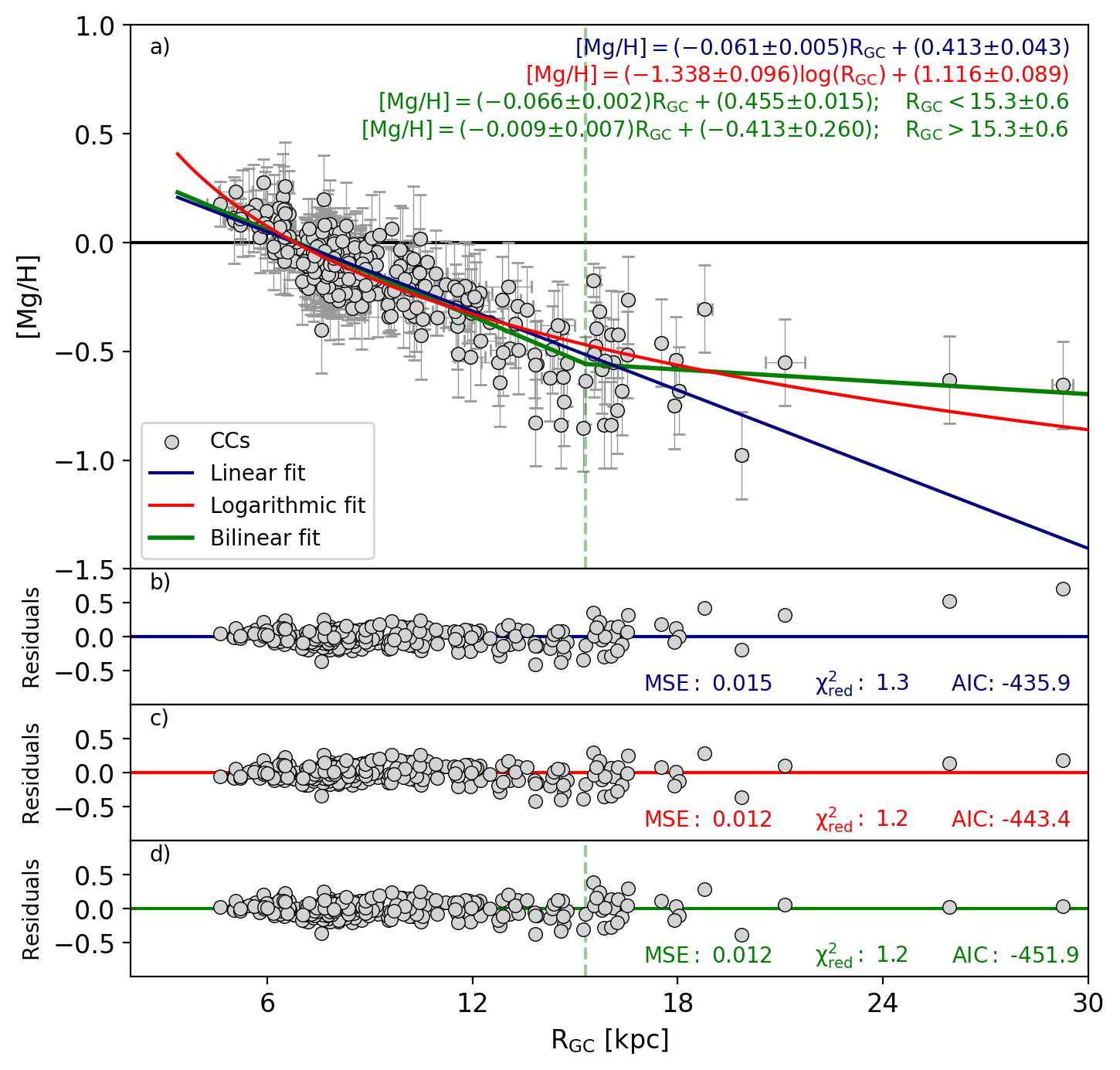}
   \includegraphics[width=4.4cm]{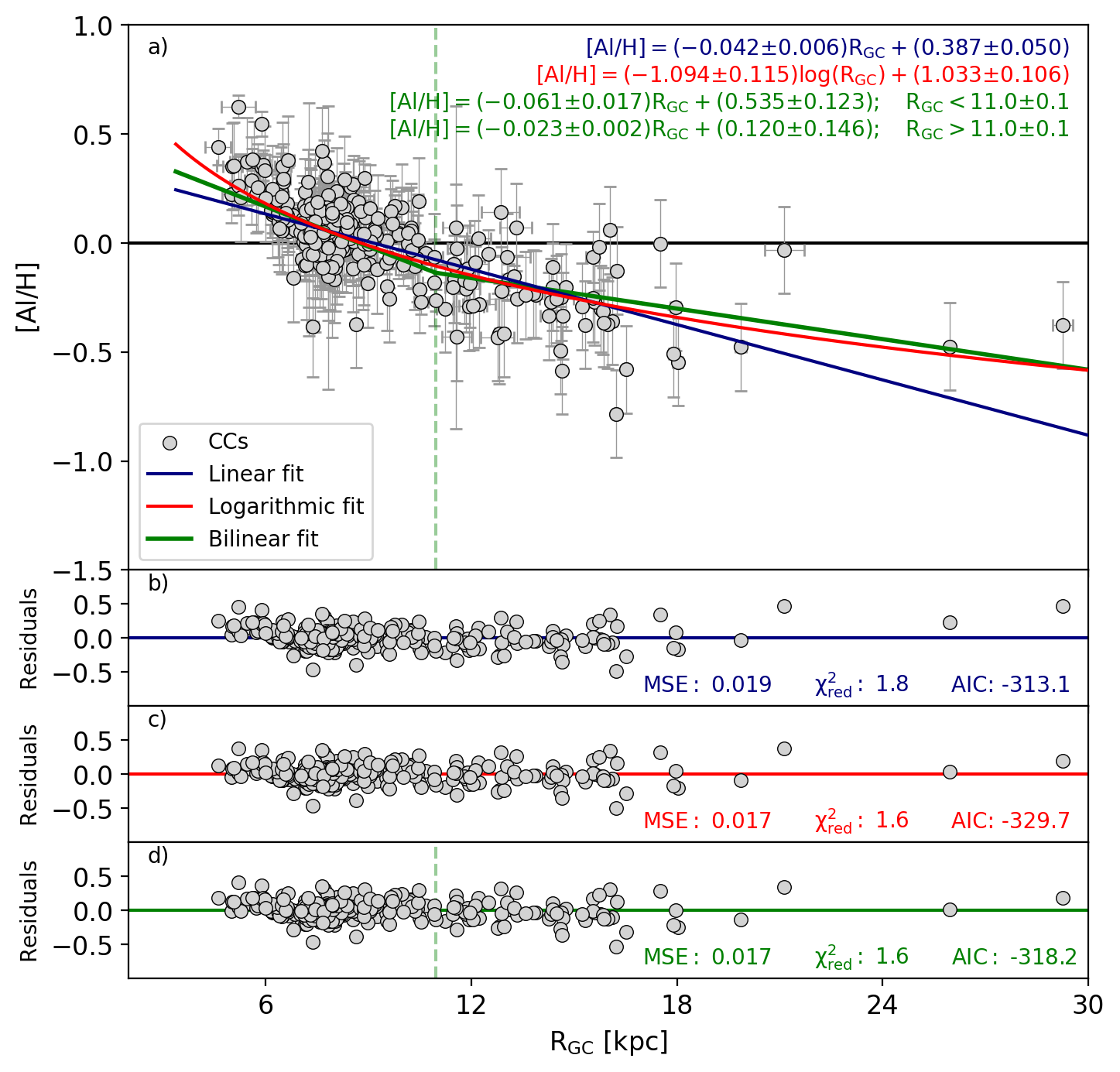}
   \includegraphics[width=4.4cm]{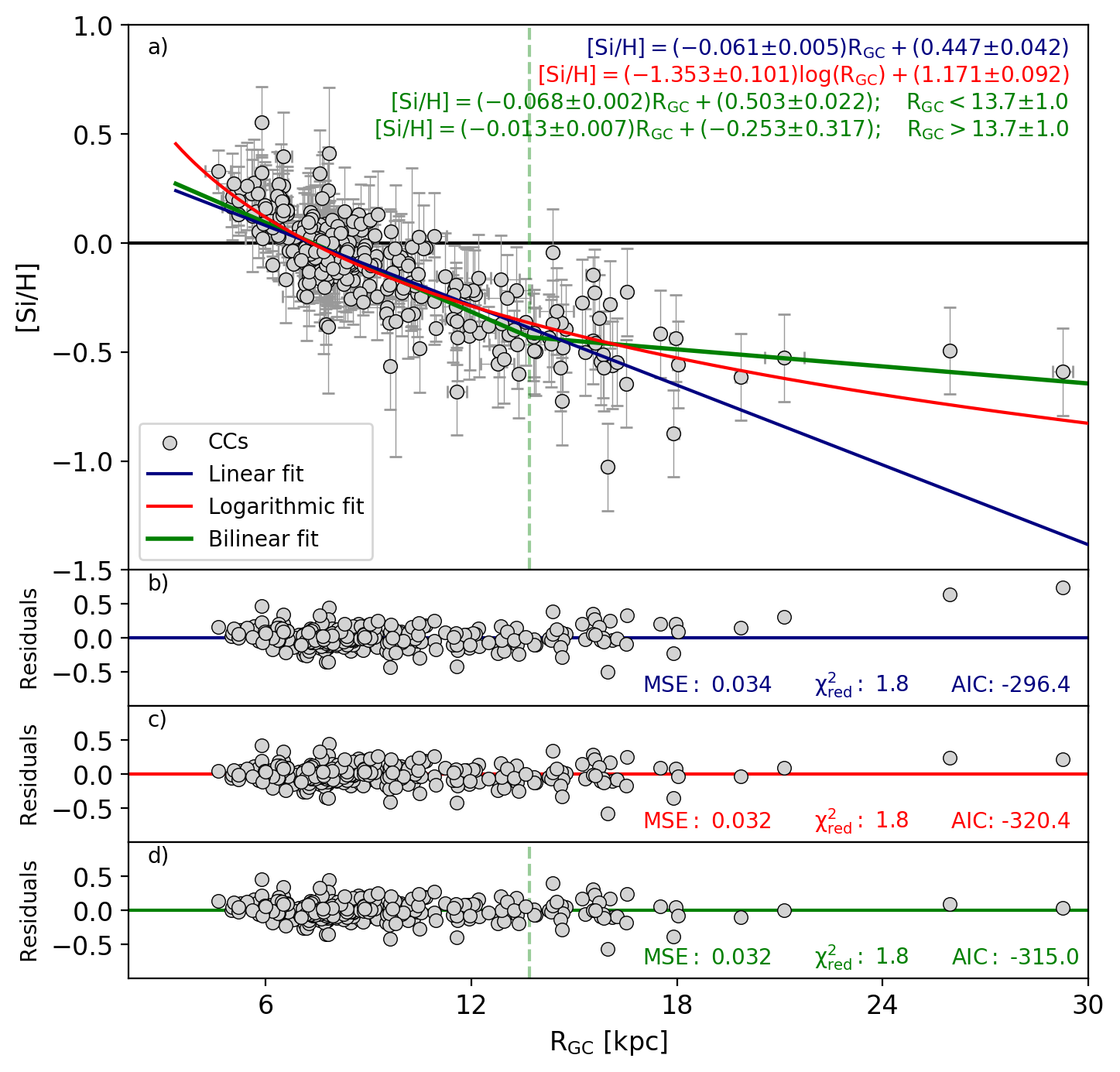}
   \includegraphics[width=4.4cm]{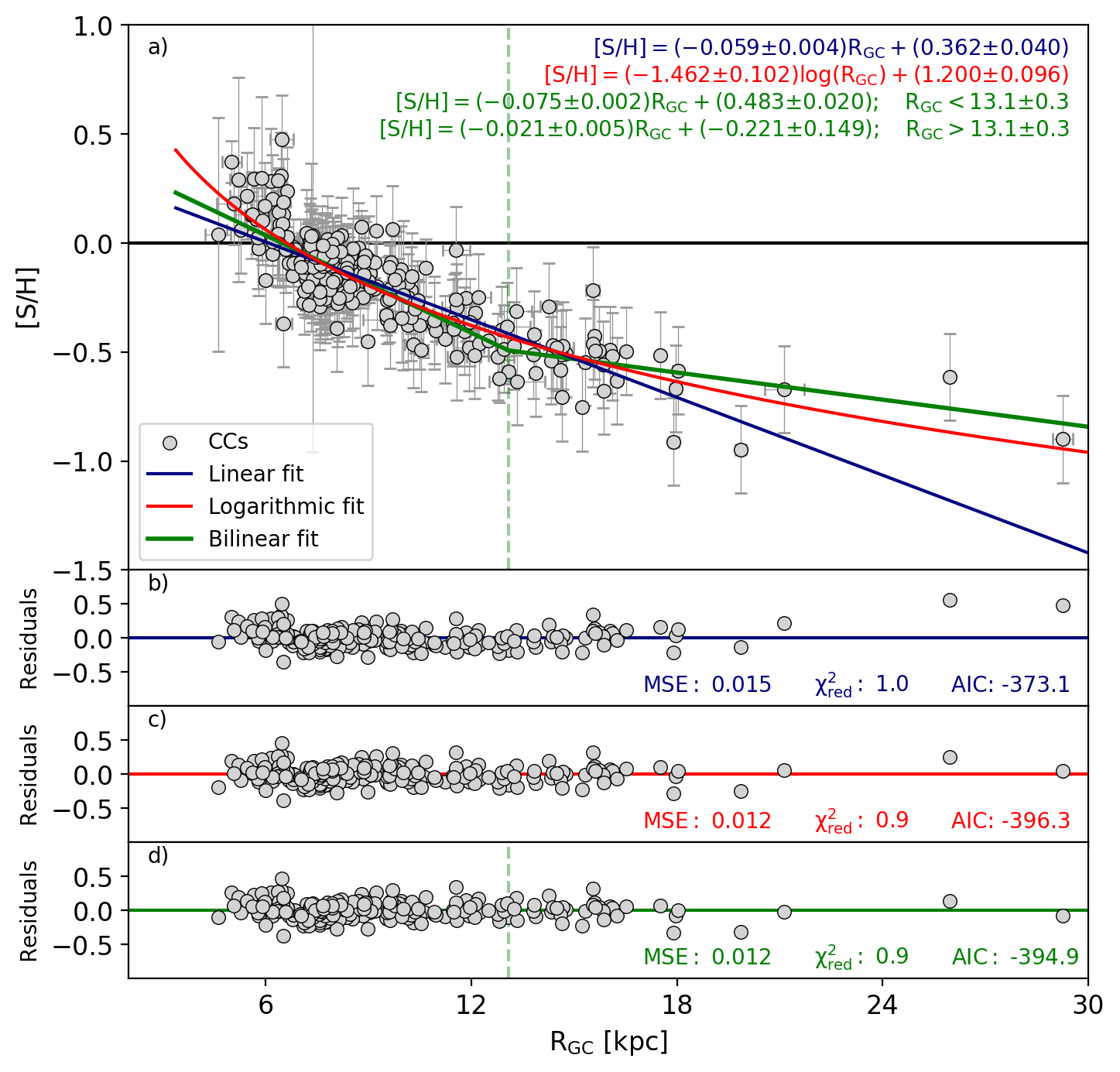}
   \includegraphics[width=4.4cm]{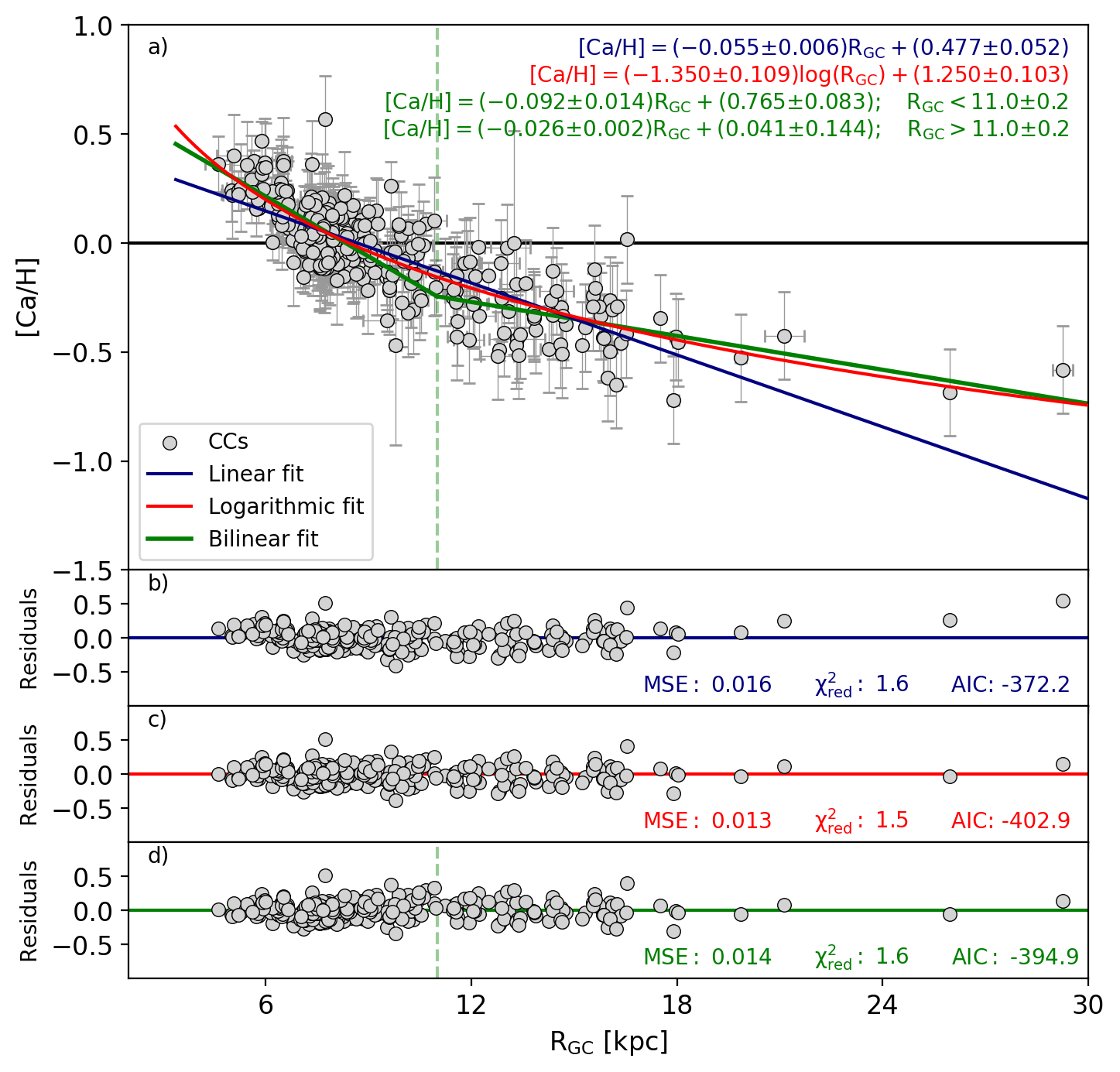}
   \includegraphics[width=4.4cm]{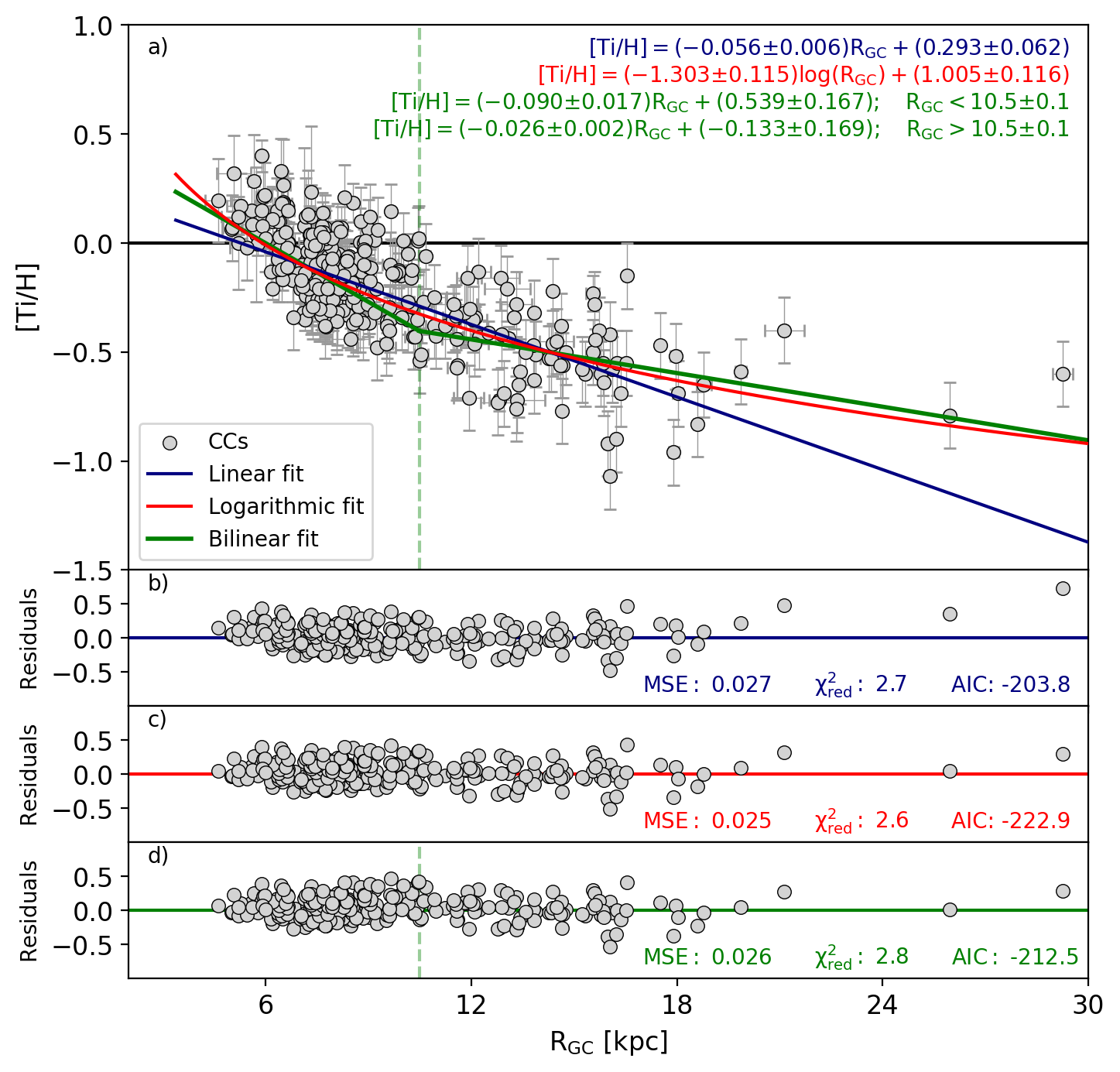}
   \includegraphics[width=4.4cm]{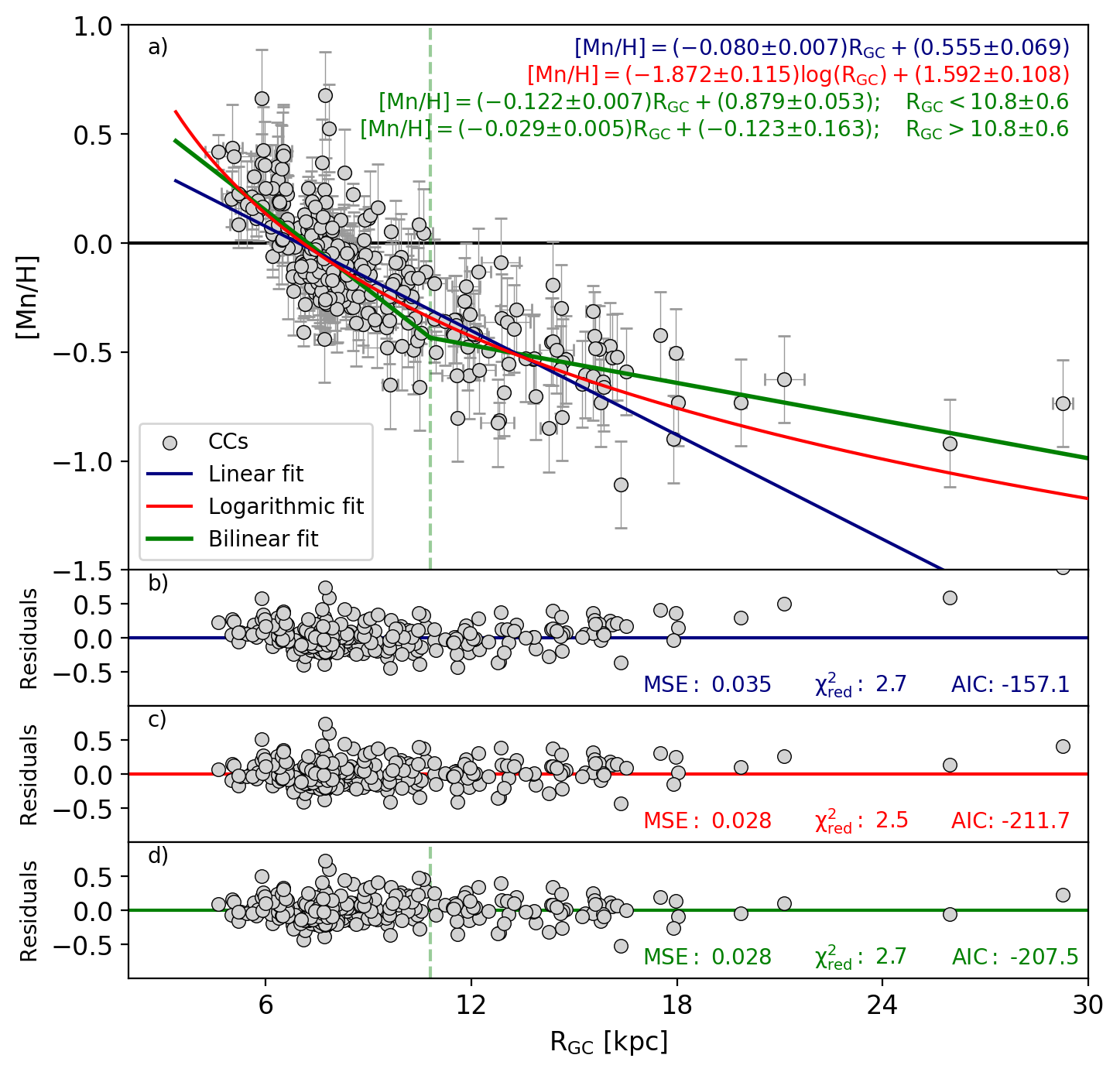}
   \includegraphics[width=4.4cm]{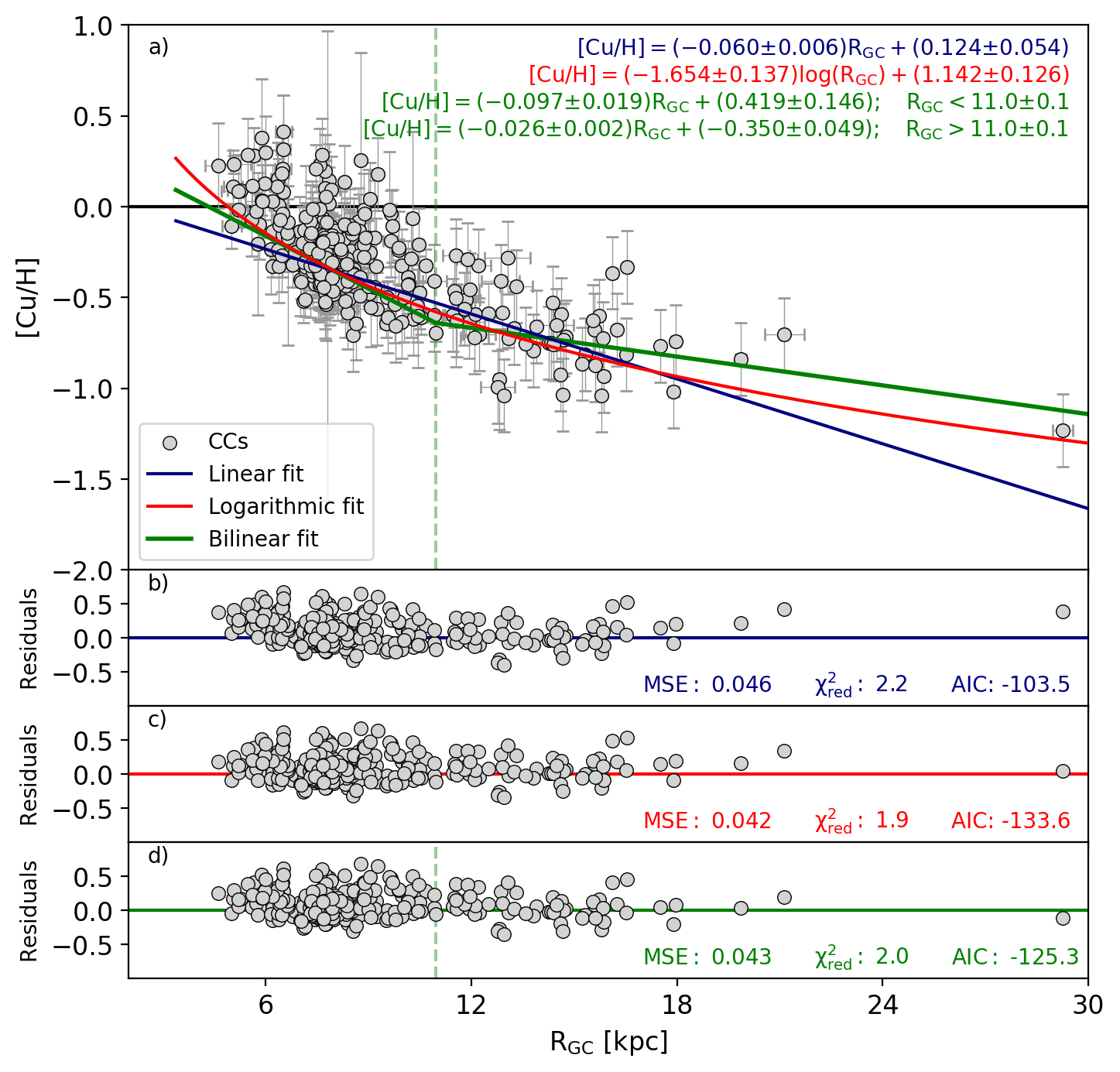}
      \caption{Chemical gradients with linear, logarithmic and bilinear fits for O, Na, Mg, Al, Si, S, Ca, Ti, Mn and Cu. Colours and symbols are the same of Fig. \ref{iron}. 
      }
    \label{gradients_threemodels}
   \end{figure}

 Flattening of the gradients is also supported by data plotted in Fig.~\ref{gradients_threemodels}, where the linear, logarithmic and bilinear fits are shown. The AIC scores support the logarithmic fit for Al, Si, S, Ca, Ti, Mn and Cu among the three models. Moreover, the linear fit is always the least suitable model of the three, while Na and Mg favor the bilinear model with respect to the logarithmic one, but with a different knee position (9.8 vs 15.3 kpc).
 
\begin{figure*}[!ht]
   \includegraphics[width=1.\linewidth]{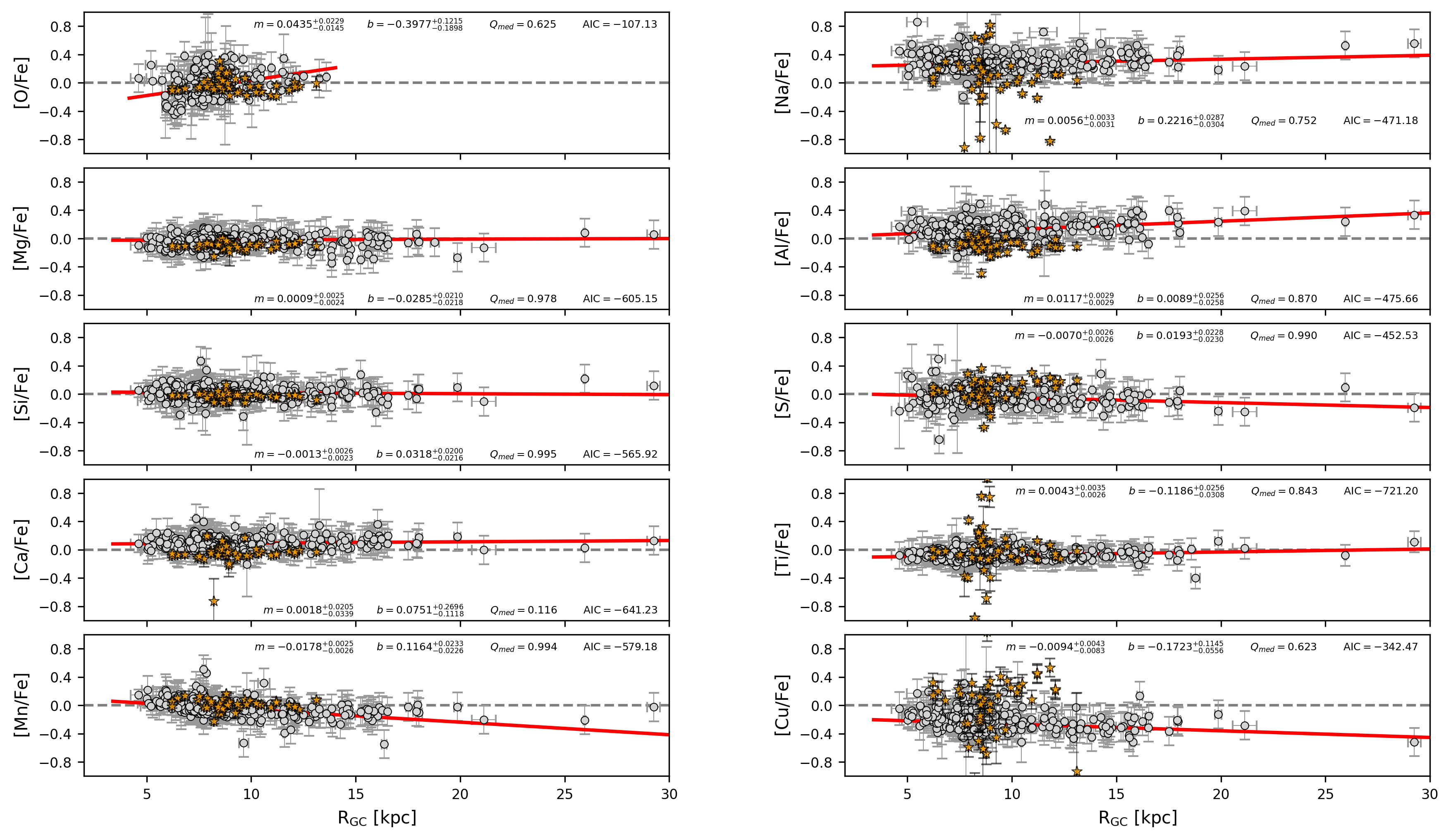}
      \caption{Element over iron gradients for O, Na, Mg, Al, Si, S, Ca, Ti, Mn and Cu. Iron abundances as a function of RGC for CCs are shown in grey, while orange symbols depict open clusters younger than 400 Myr from \citet{otto25}. The red profile represents the linear fit with parameters given in the top-right box for each panel.}
         \label{el_fe_gradient}
   \end{figure*}

To further investigate the radial variation of the abundance gradients, Fig.~\ref{el_fe_gradient} presents 
the different [X/Fe] ratios as a function of Galactocentric distance. The radial trends remain approximately constant 
for most of the investigated elements across the entire disk. The abundance ratio slopes are vanishing, suggesting that these elements follow the same radial behaviour as iron.
Three elements, however, deviate from this pattern: Al, Mn, and Cu. Mn and Cu display well-defined negative 
gradients ($-0.018\pm0.003$, $-0.009\pm0.005$~dex $\mathrm{kpc^{-1}}$, respectively), while Al shows a positive gradient 
($+0.012\pm0.003$~dex~$\mathrm{kpc^{-1}}$). The negative trend observed for Mn supports the behaviour found for the iron-peak elements (Mn, Co, Ni) in OCs by \citet{otto25}, whereas the nearly flat [X/Fe] ratios for O, Mg, Si, S, and Ca (all with slopes <0.01~dex~$\mathrm{kpc^{-1}}$) further confirm the overall homogeneity of $\alpha$-element enrichment across the thin disk. \citet{genovali15} also reported nearly constant trends for Na, Al and Si. In contrast, they found mild evidence of positive slopes for Mg ($\mathrm{0.015\pm0.006\, dex\, kpc^{-1}}$) and Ca ($\mathrm{0.028 \pm 0.004\, dex\, kpc^{-1}}$). However, their results were based on a smaller Cepheid sample, characterized by higher dispersion and a more limited spatial coverage. 

\begin{figure*}[!ht]
\includegraphics[width=1.\linewidth]{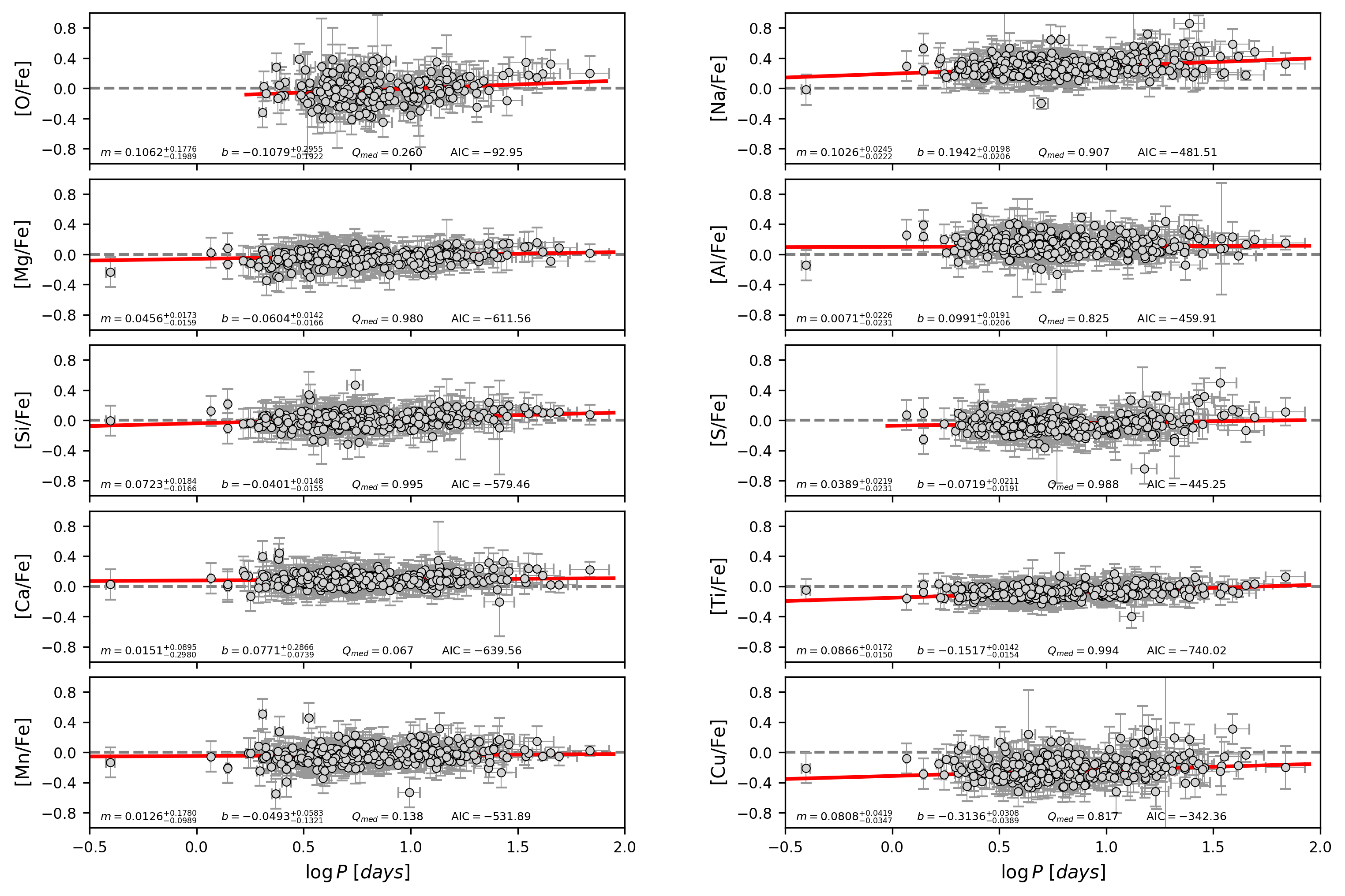}
 \caption{Abundance ratios ([X/Fe]) as a function of the logarithmic period for O, Na, Mg, Al, Si, S, Ca, Ti, Mn, and Cu. The red  lines display the linear fit and their coefficient are plotted in the bottom left box of each panel.}
         \label{el_logP}
\end{figure*}

Figure~\ref{el_logP} shows for the investigated elements the abundance ratios as a function of $\log$(P) along with a linear fit. CCs obey to a well defined Period-Age relation \citep[][and references therein]{bono05}. The individual age steadily increases when moving from 
long- to short-period Cepheids. The abundance ratio of all the elements remain approximately constant across the full age range of the Cepheids in the sample. Only O, Na, Si, Ti, and Cu show a weak positive trend, 0.11$\pm$0.20, 0.10$\pm$0.03, 0.07$\pm$0.02, 0.09$\pm$0.02, 0.08$\pm$0.05, respectively. For oxygen, the higher dispersion and the shorter age coverage may affect the result. In other words, age-related effects are weak or negligible in the radial gradients and abundance ratios presented above. \citet{genovali15} found a slight negative trend for Ca (-0.15), in contrast with the null slope obtained in our study. In \citetalias{da2023oxygen}, they found a slope of 0.08$\pm$0.02 in the [S/Fe] versus $\mathrm{\log P}$ plane, while we obtained a weaker positive trend of 0.04$\pm$0.02. This suggests that the overall consistency of the abundances in our sample, which reduces the dispersion, allows for a more accurate assessment of whether a trend is present. 

\begin{figure}[!ht]
   \includegraphics[width=1.\linewidth]{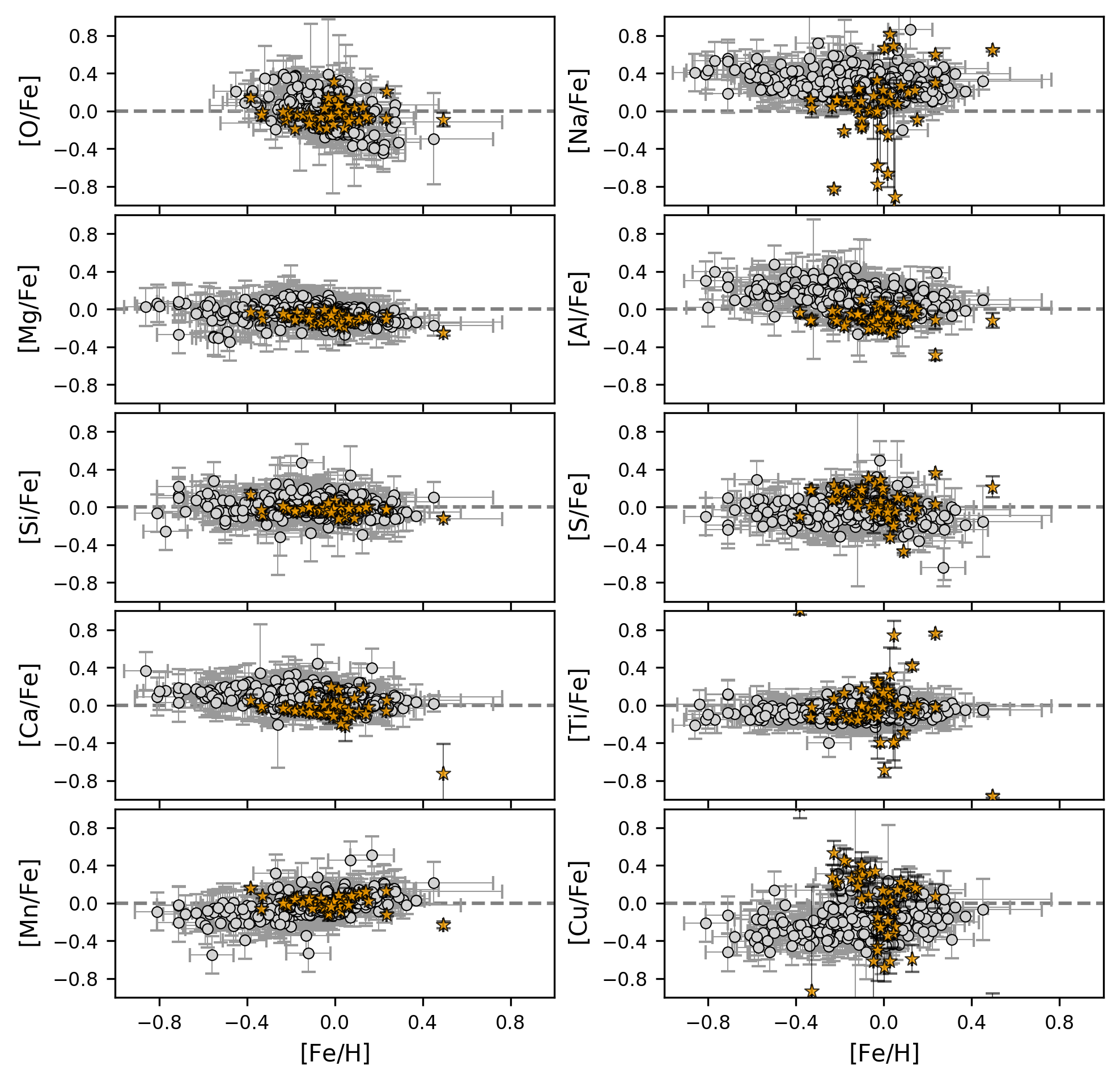}
      \caption{Chemical planes for O, Na, Mg, Al, Si, S, Ca, Ti, Mn and Cu. Colours and symbol are the same as in Fig. \ref{gpr}.}
         \label{el_fe_fe}
   \end{figure}

In Fig.~\ref{el_fe_fe} the [X/Fe] vs [Fe/H] planes are shown. We emphasize the positive trend observed for Mn. Manganese is mainly produced in thermonuclear supernovae (Th-SNe), with yields that increase with metallicity (\citealt{badenes08}; \citealt{kobayashi09}), which explains the positive trend. Sodium, copper and aluminum are predominantly produced in core-collapse supernovae (CC-SNe); however, we observe a slight positive trend for Cu and a negative trend for Na and Al, reflecting their different metallicity dependences. \citet{kobayashi20} predict a constant or slightly negative trend for Na and Al, and a slight positive trend for Cu over the metallicity range we cover ($\mathrm{-1<[Fe/H]<0.5\, dex}$), in agreement with our findings. Moreover, for Na and Al, the slight positive trend in [Na/Fe] and [Al/Fe] versus $\mathrm{R_{GC}}$ corresponds to a negative trend in the [Na/Fe] and [Al/Fe] versus [Fe/H] plane. Similarly, the negative trends of [Cu/Fe] and [Mn/Fe] with Galactocentric distance correspond to positive trends in the [Cu/Fe] and [Mn/Fe] versus [Fe/H] planes. For oxygen, magnesium, silicon, sulfur, calcium, and titanium, we expect a roughly constant trend in the metal-poor region of the sample and a decreasing trend at higher metallicities. This behaviour arises because these elements are predominantly produced in CC-SNe, whereas metal-rich stars are increasingly affected by Th-SN enrichment, which reduces their [X/Fe] ratios. Since Si, S, Ca and Ti also receive contributions from Th-SNe, the negative trend in the metal-rich region is weak. In contrast, O and Mg are more affected and show a higher decreasing trend.

\subsection{Radial gradients: LTE versus NLTE}
\begin{figure}[!ht]
    \centering
    \includegraphics[width=8cm]{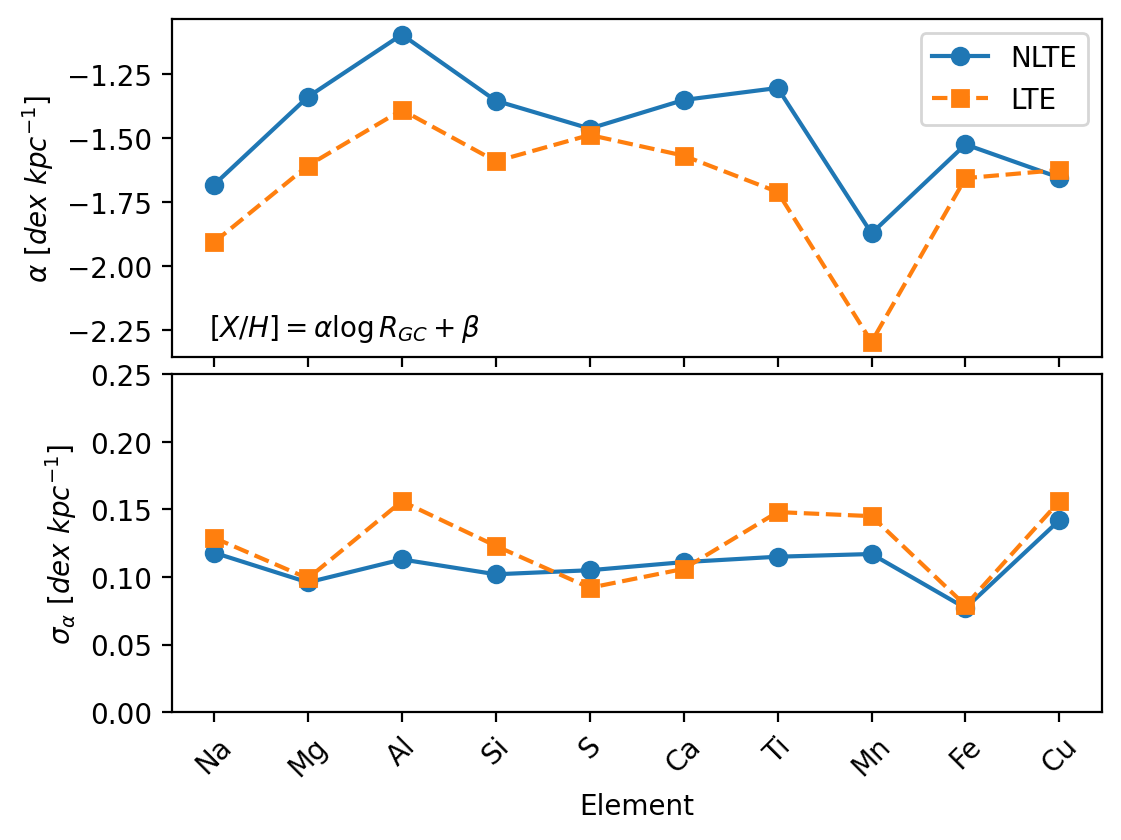}
    \caption{NLTE (blue line) and LTE (orange line) chemical gradients slopes ($\alpha$) of the logarithmic fit for each element (top panel) and their uncertainties (bottom panel).}
    \label{slopes_lte}
\end{figure}

We measured the chemical abundances under the LTE assumption and performed the logarithmic fit of the LTE abundance gradients. Figure~\ref{slopes_lte} reveals that for almost all the elements the NLTE slopes of the logarithmic fit are shallower. Only S and Cu show a similar slope, which suggests negligible NLTE effects for the lines employed for these two elements. This is expected for S, since we used an S I triplet that is weakly prone to NLTE effects (\citealt{duffau2017gaia}). NLTE have a minor impact on Cu I lines in stars with $\mathrm{[Fe/H]}\geq -1$ dex. \citet{shi18} found a mean abundance increase of about +0.07 dex when applying NLTE corrections, a result previously suggested by \citet{yan16}.   

\begin{figure}[!ht]
    \includegraphics[width=4.4cm]{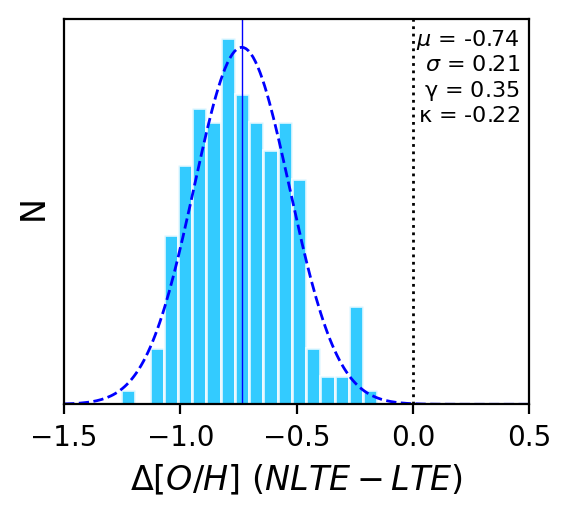}
    \includegraphics[width=4.4cm]{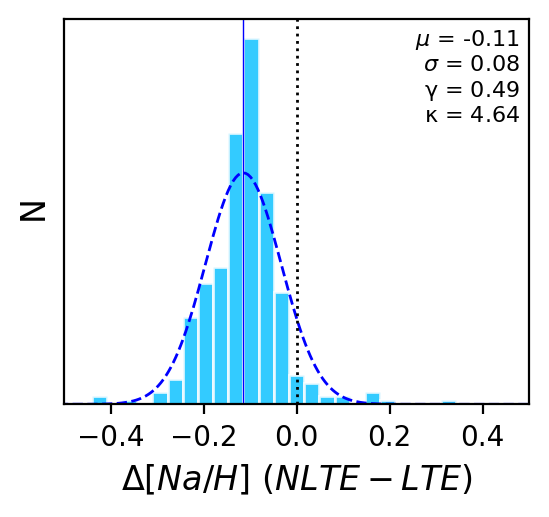}
    \includegraphics[width=4.4cm]{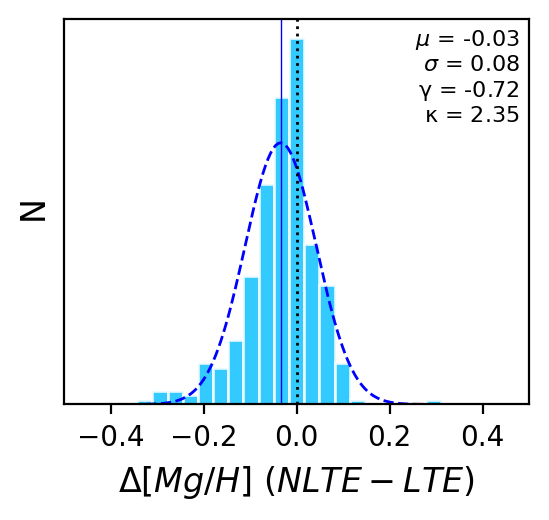}
    \includegraphics[width=4.4cm]{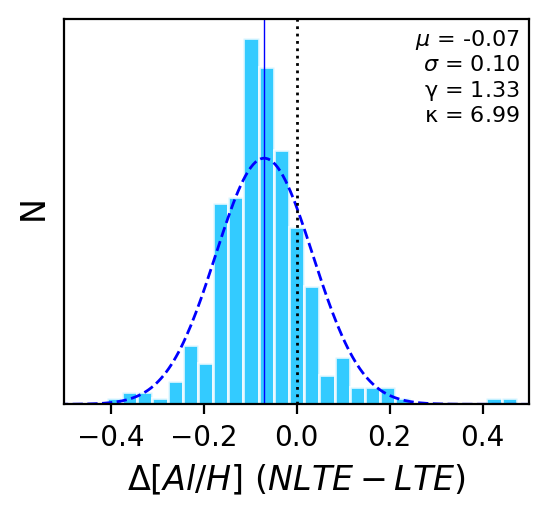}
    \includegraphics[width=4.4cm]{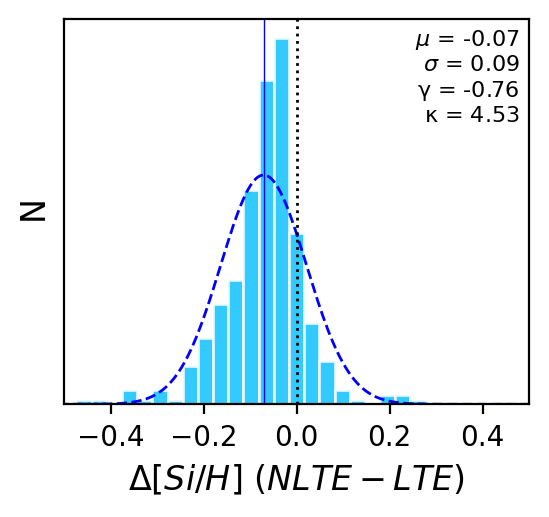}
    \includegraphics[width=4.4cm]{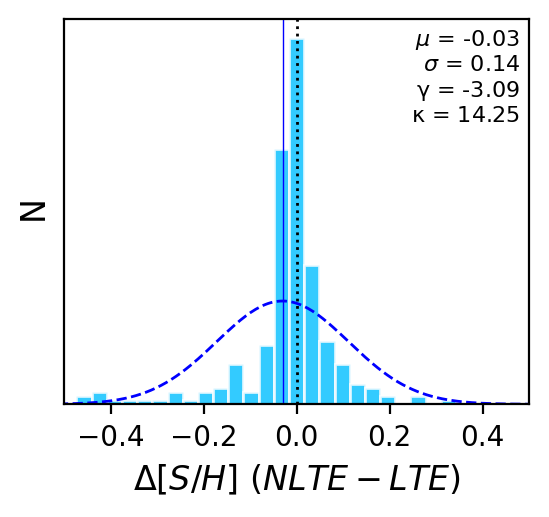}
    \includegraphics[width=4.4cm]{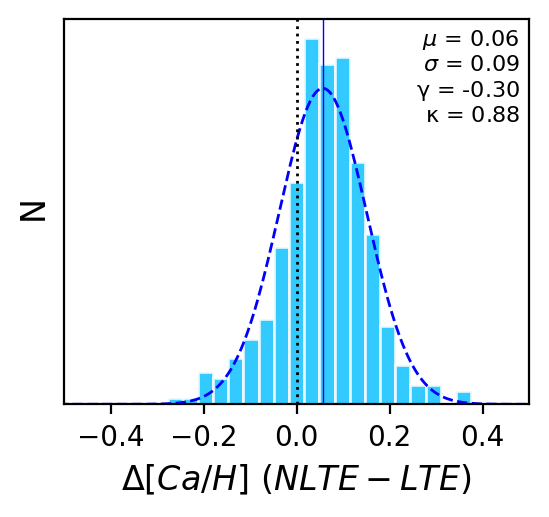}
    \includegraphics[width=4.4cm]{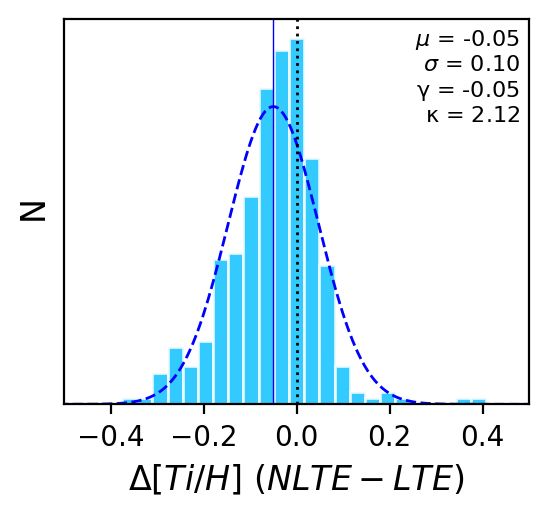}
    \includegraphics[width=4.4cm]{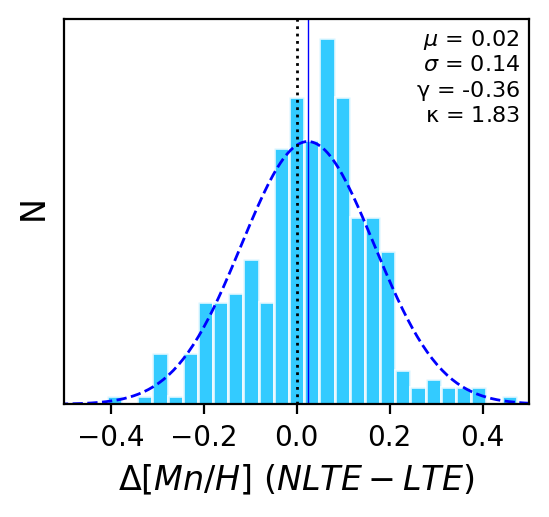}
    \includegraphics[width=4.4cm]{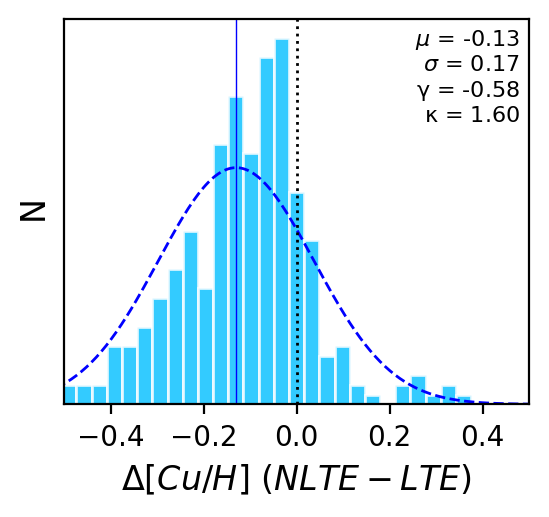}
    \caption{Histograms of the difference between NLTE and LTE abundances. The x-axis spans the range $-0.5$ to $+0.5$ dex for all elements, except for oxygen, for which the range is $-1.5$ to $+0.5$ dex. Mean value ($\mu$), standard deviation ($\sigma$), skewness ($\gamma$) and kurtosis ($\kappa$) of the overlaid gaussian fit (blue dashed line) are indicated in the top-right corner of each panel.}
    \label{nlte_vs_lte_abu}
\end{figure}

The histograms in Fig.~\ref{nlte_vs_lte_abu} illustrate the comparison of the NLTE and LTE abundances where, on average, lower NLTE abundances for O ($-0.74\pm0.21$~dex), Na ($-0.11\pm0.08$~dex), Al ($-0.07\pm0.10$~dex), Si ($-0.08\pm0.15$~dex) and Cu ($-0.13\pm0.17$~dex), and more centred distributions for Mg ($-0.03\pm0.08$~dex), S ($-0.02\pm0.10$~dex), Ca ($0.06\pm0.09$~dex) and Mn ($0.02\pm0.14$~dex). Our results indicate that the oxygen lines used in this study (OI 777 nm triplet) are strongly affected by NLTE, consistent with \citet{amarsi16}, who reported corrections of about 0.6~dex or higher at solar metallicities. For the other elements, the NLTE corrections are more moderate and agree with the findings of \citet{amarsi2020galah}. 

We emphasise that in our analysis, the derivation of abundances is not performed at fixed atmospheric parameters: each run (LTE or NLTE) adopts a consistent set of atmospheric parameters optimised for that assumption. As a consequence, the comparison of abundances between NLTE and LTE cannot be attributed solely to the direct NLTE line corrections for a given element, since the underlying atmospheric parameters themselves also differ. In appendix~\ref{test_nlte} we further discuss the NLTE abundances for Mn and O.

\begin{figure*}[!ht]
   \includegraphics[width=9cm]{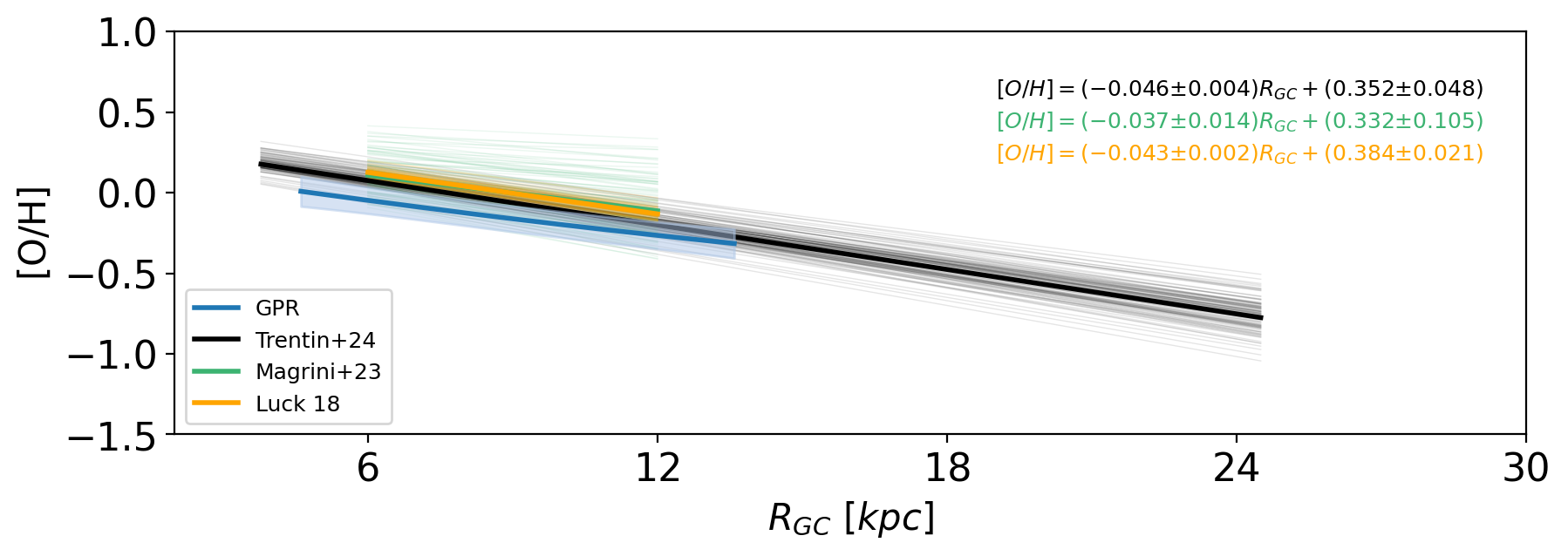}
   \includegraphics[width=9cm]{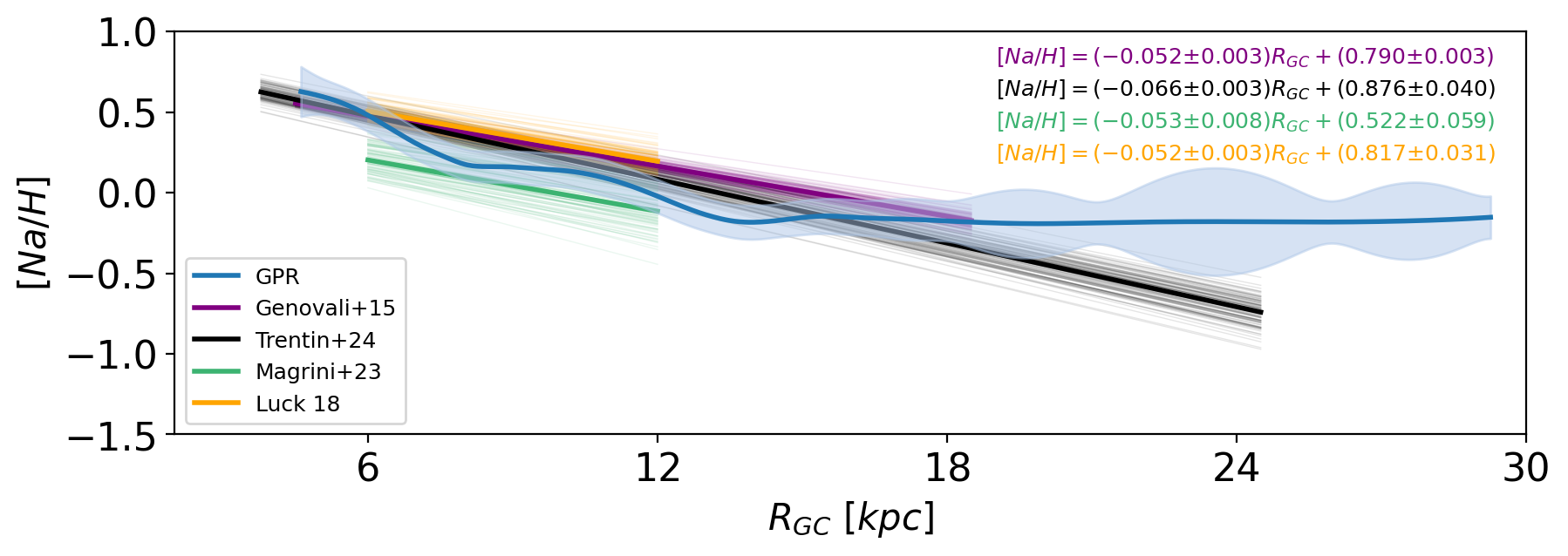}
   \includegraphics[width=9cm]{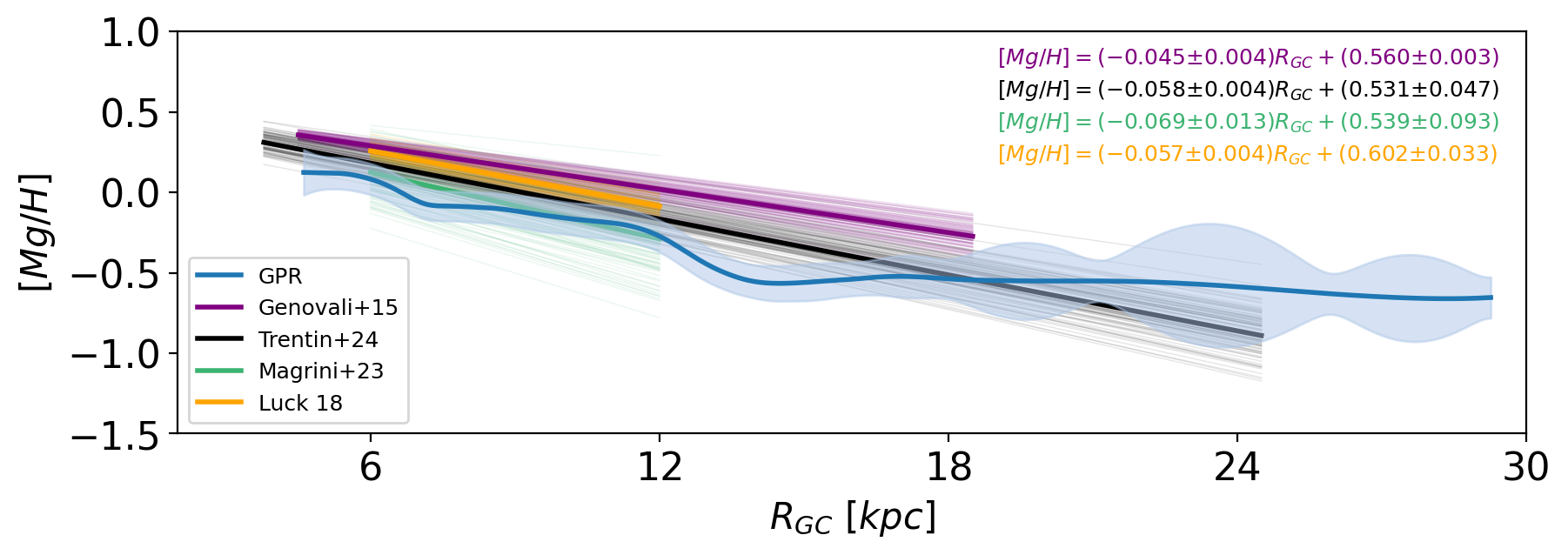}
   \includegraphics[width=9cm]{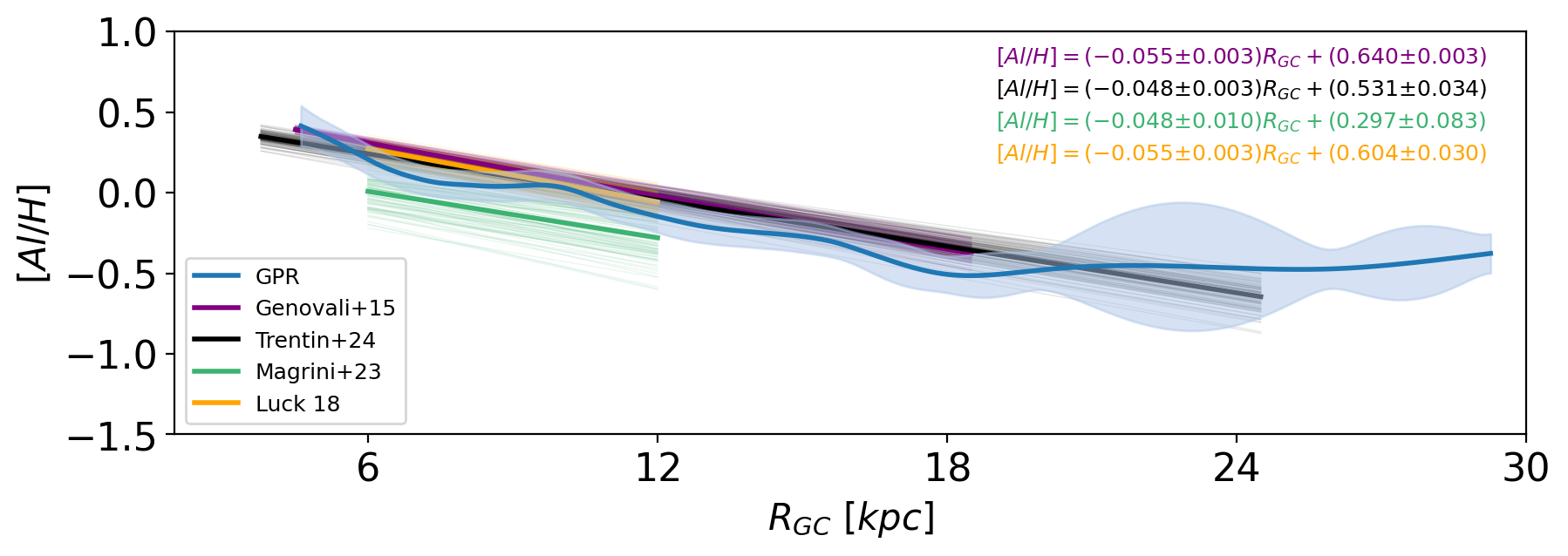}
   \includegraphics[width=9cm]{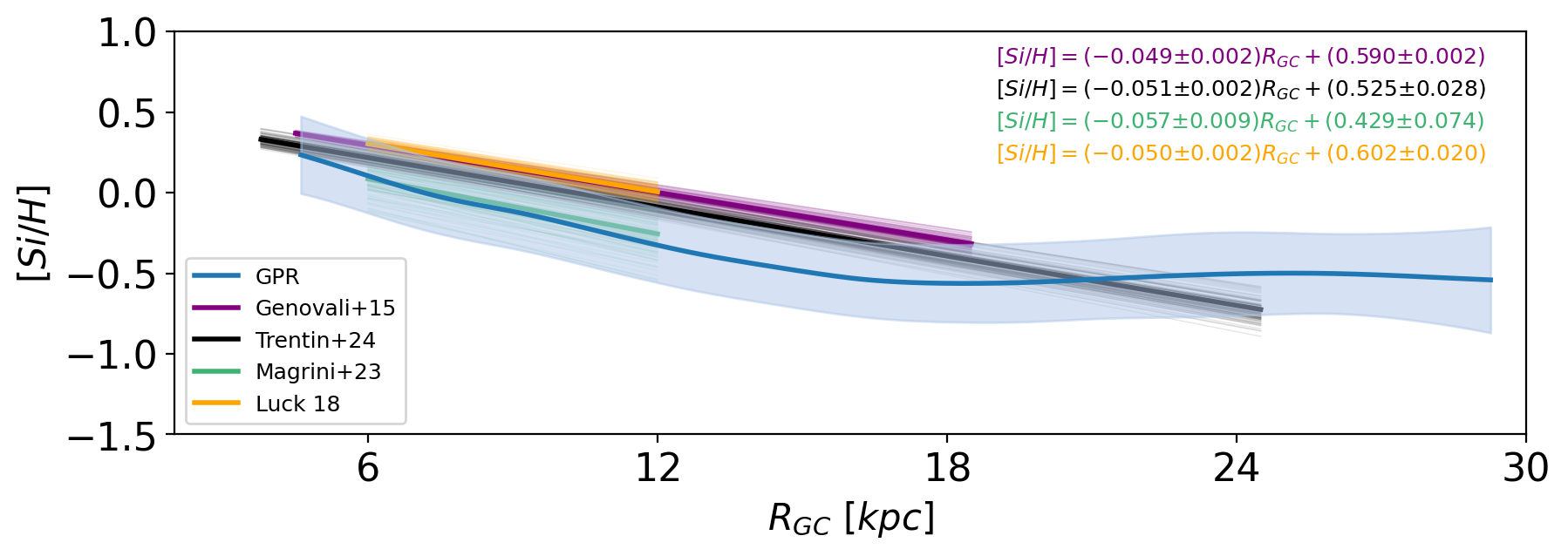}
   \includegraphics[width=9cm]{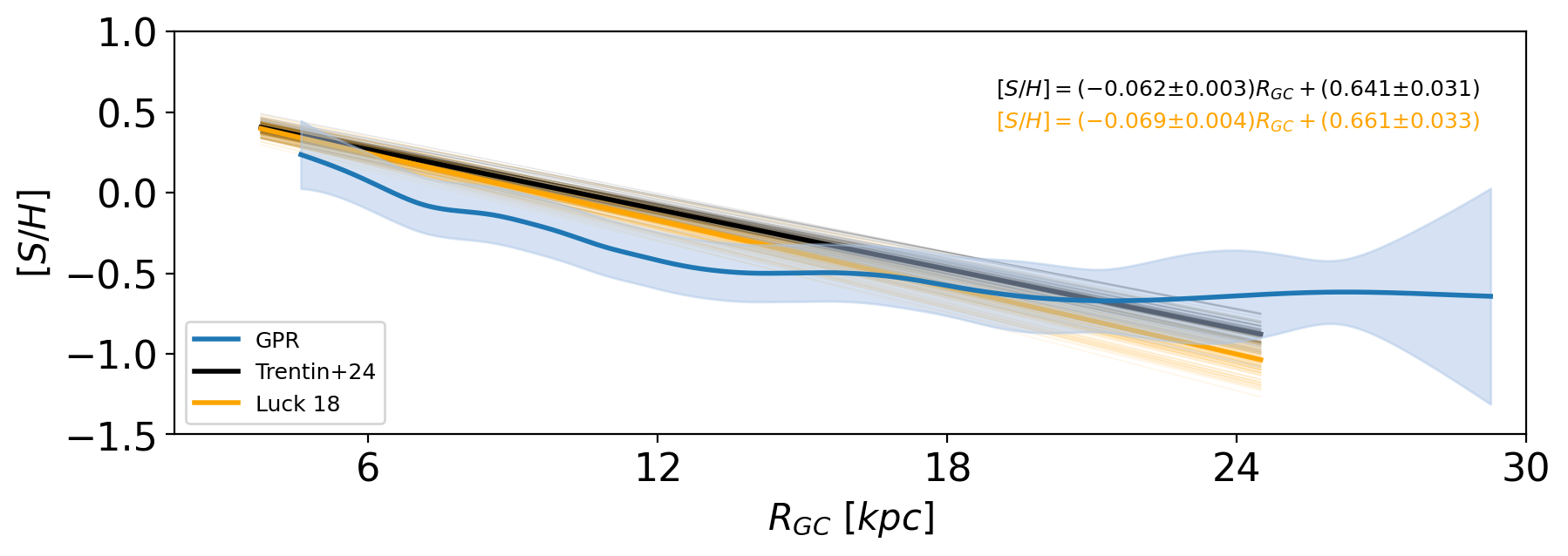}
   \includegraphics[width=9cm]{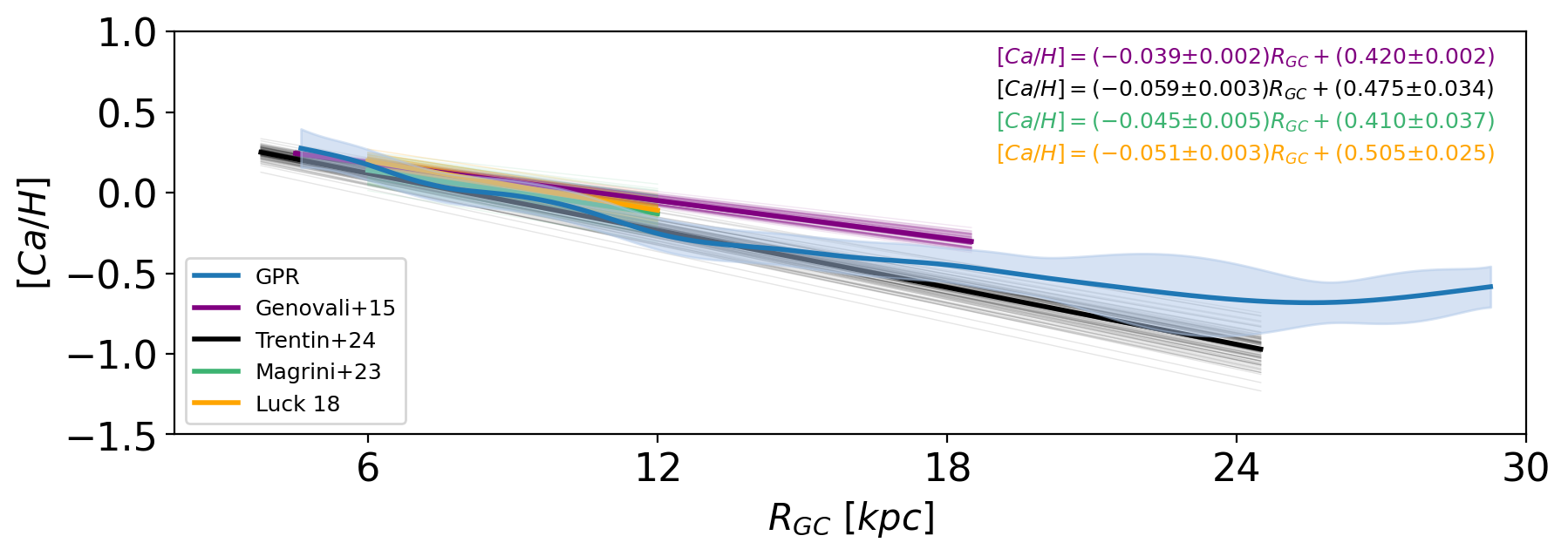}
   \includegraphics[width=9cm]{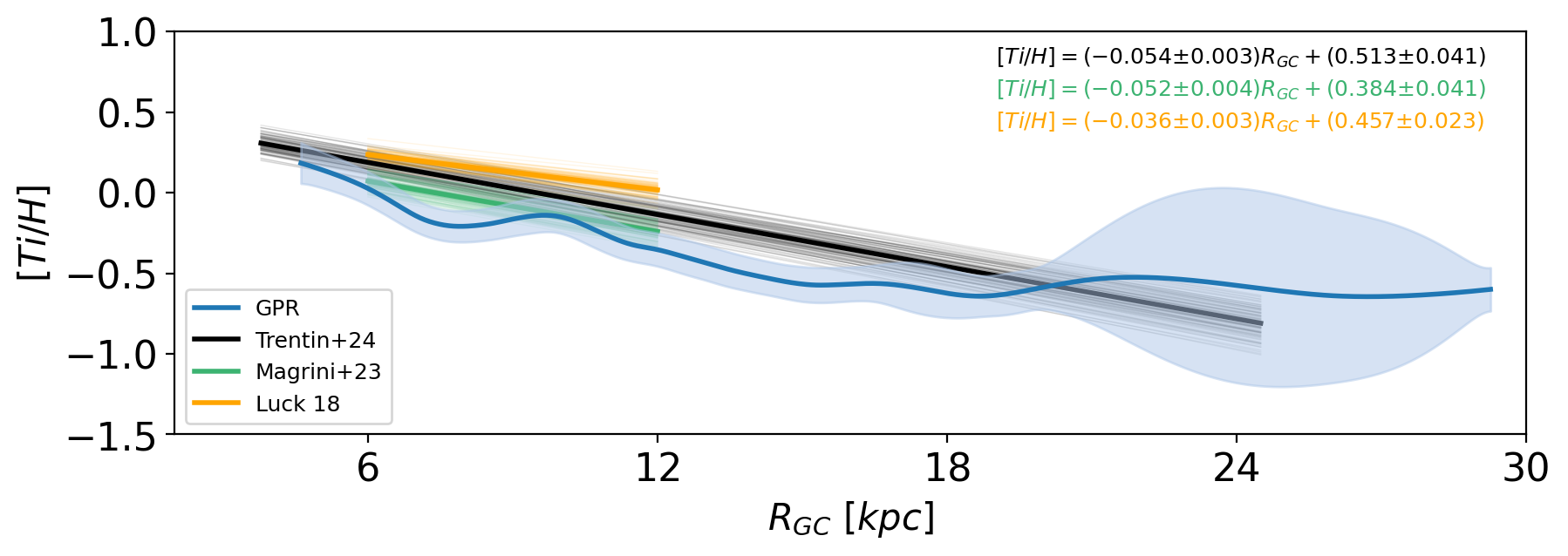}
   \includegraphics[width=9cm]{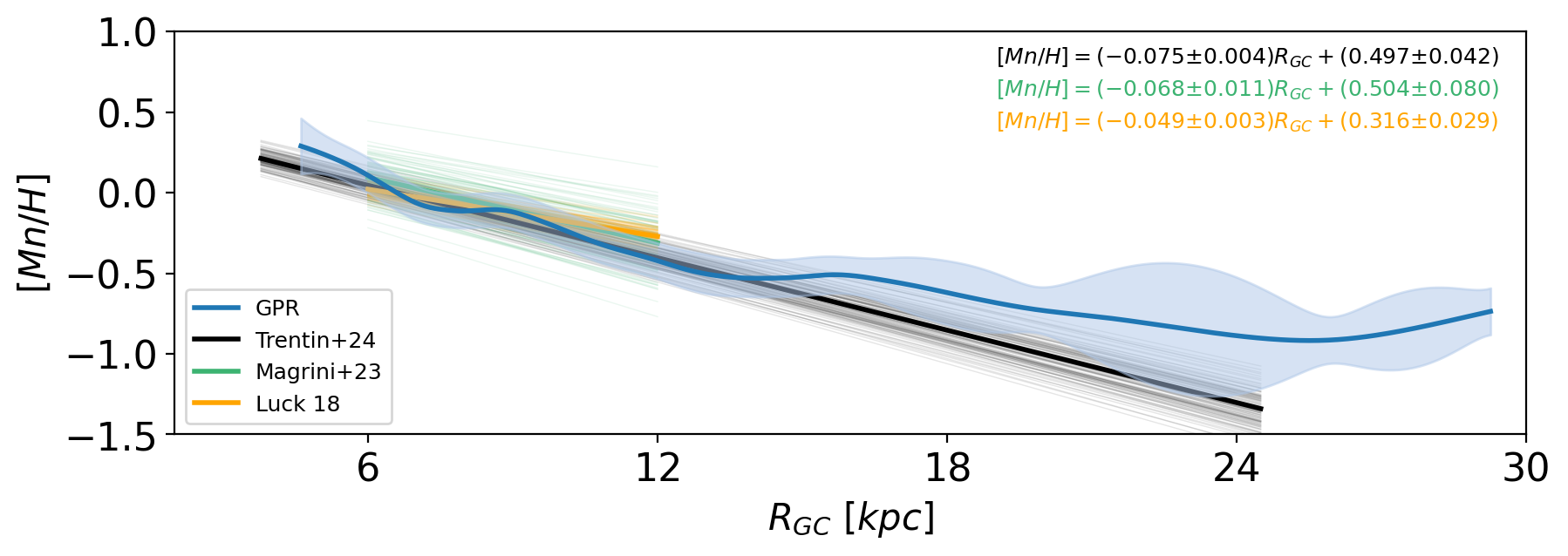}
   \includegraphics[width=9cm]{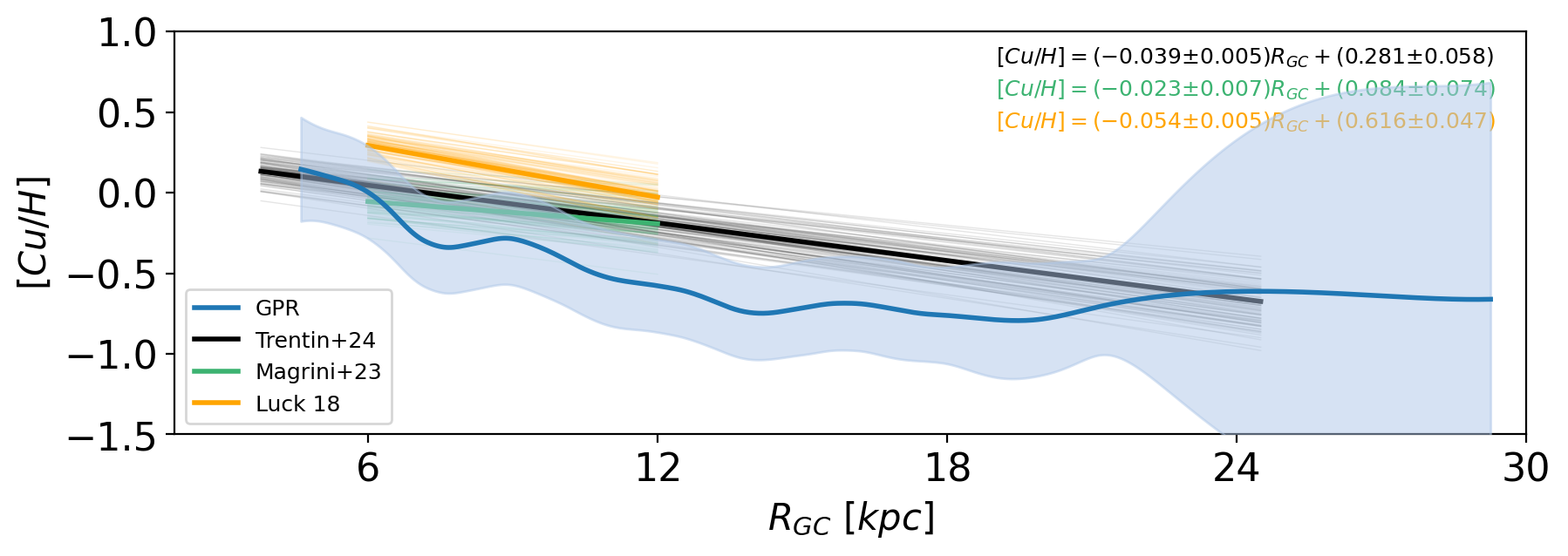}
      \caption{Comparison of the current GPR profiles with similar findings available 
      in the literature based either on CCs by \citet{genovali15}, \citet{luck18} and 
      \citet{trentin24A} or on open clusters by \citet{magrini23}.}
         \label{comparisons}
   \end{figure*}

In Fig.~\ref{comparisons}, we compare the slopes of our chemical gradients with those reported in previous studies. A general agreement is observed in the inner disk for most elements, although a slight underabundance is found for sulfur and copper when comparing the GPR profiles to the linear trends from other works. Moreover, the GPR profiles provide additional information on the behaviour of Na, Mg, Al, S, Ca, Ti, Mn and Cu, revealing a shoulder at the solar circle similar to that obtained in the iron GPR profile. The GPR profiles also suggest a change of slope around 14-16 kpc, resulting in a flatter gradient in the outer disk compared to the linear trends of previous studies, particularly for Na, Mg, Al, Si, S, Ca, Ti, Mn and Cu.  

\begin{figure*}[!ht]
    \includegraphics[width=9cm]{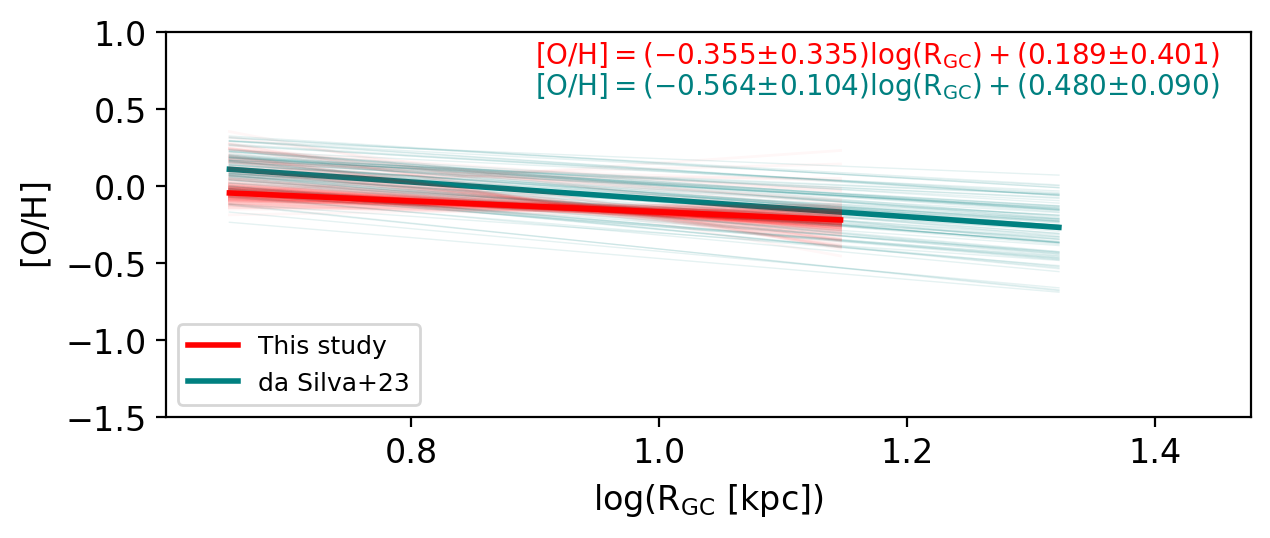}
    \includegraphics[width=9cm]{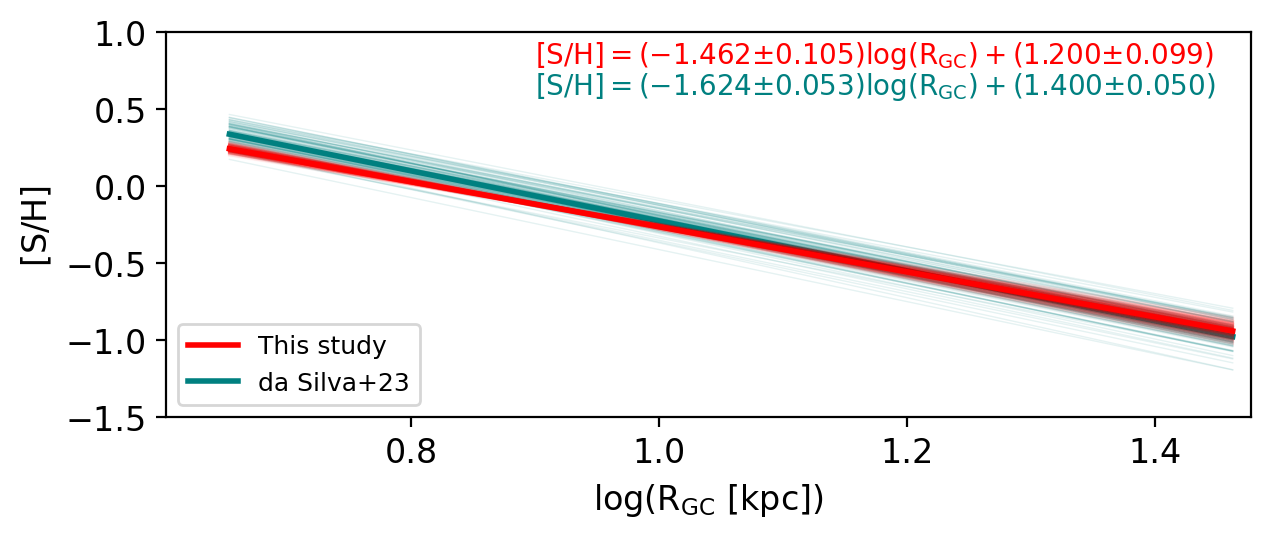}
    \caption{Comparison of the current oxygen and sulphur radial gradients 
    (logarithmic scale) and \citetalias{da2023oxygen} investigation.}
    \label{S_O_ronaldo}
\end{figure*}

In Fig.~\ref{S_O_ronaldo}, we emphasize the agreement with the \citetalias{da2023oxygen} logarithmic profiles for oxygen and sulfur. 
According to \citetalias{da2023oxygen}, the radial gradient of sulfur is steeper than that of iron, and sulfur is on average under-abundant compared to oxygen. This discrepancy points to the need for a revision of chemical evolution models, particularly regarding sulfur yields from massive stars, in order to reproduce observed trends in the outer Galaxy. However, our results confirm the general behaviour of sulfur but show a profile that is more consistent with iron and the other elements, rather than reproducing the steeper trend reported by \citetalias{da2023oxygen}. 

\subsection{Kinematic properties}
The kinematic properties  of CCs provide fundamental insights to constrain their true nature 
of thin disk stellar population. Their orbits were integrated by using the \texttt{galpy} code, 
a Python library dedicated to Galactic dynamics (\citealt{bovy15}). 
Positional parameters (RA, DEC, distance) and proper motions are mainly based on Gaia DR3. Individual distances have been discussed in more detail in Sect.~\ref{spectral sample}, while 
radial velocities are mainly based on current spectroscopic measurements.
The galactic potential employed is MWPotential2014.

To improve the sampling across the four quadrants, the current sample was complemented with 
CCs analysed by \citet{luck18} and by \citet{Trentin24B} for which homogeneous elemental 
abundances are available. We did not include the 16 CCs located at 3-5.6 kpc and analysed by using high resolution NIR spectra by \citet{matsunaga23}, because of possible systematics between NIR and optical spectroscopy. Moreover, the quoted samples have a sizable number of objects 
in common with the current sample and they have been adopted to move their chemical 
abundances into our metallicity scale. 

\begin{figure*}[!ht]
  \centering
   \includegraphics[width=0.9\linewidth]{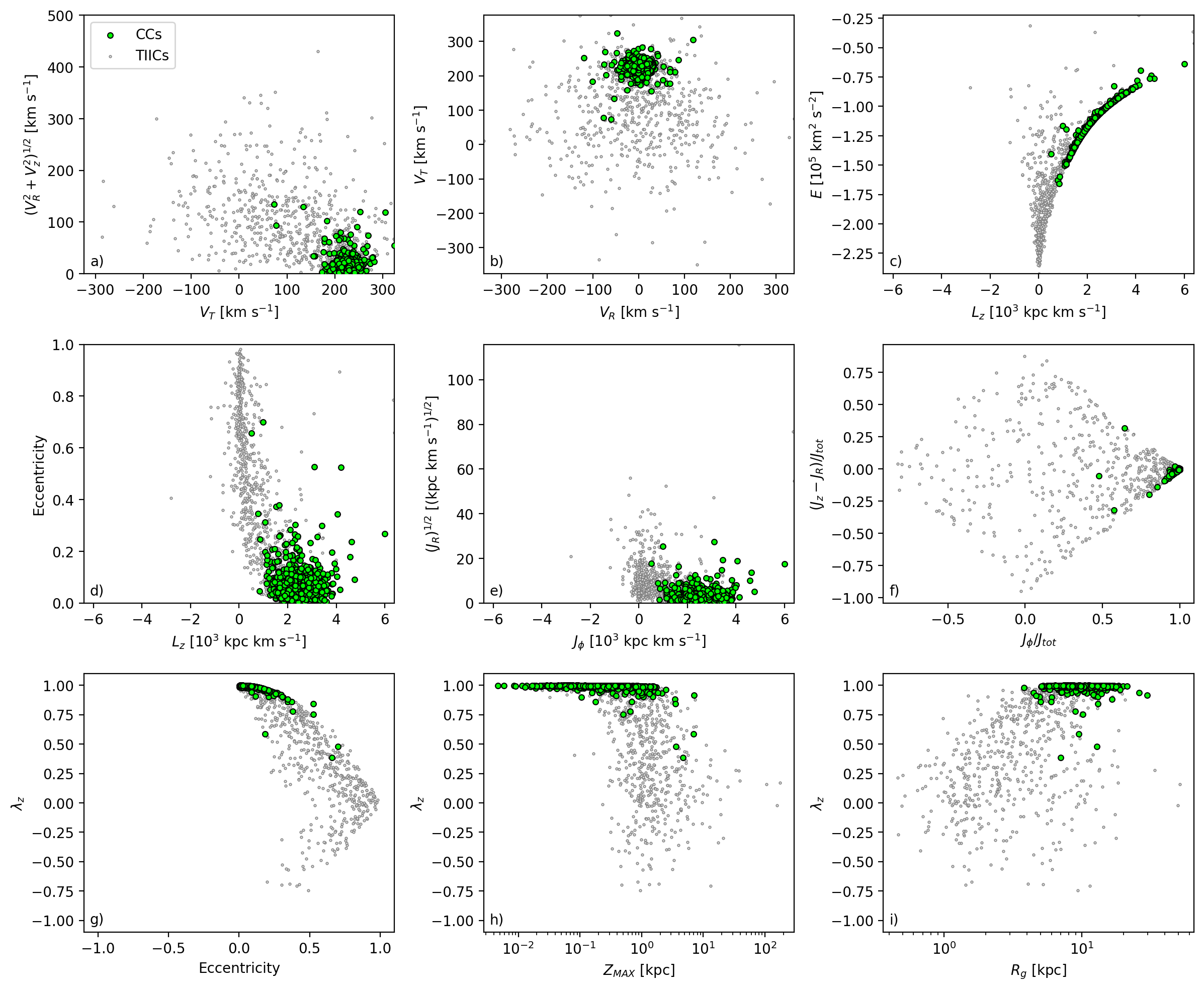}
      \caption{Kinematic planes for CCs (green circles) and a sample of Galactic Type II Cepheids (TIICs, grey dots). Panel a)-[Toomgre diagram]: $\mathrm{V_T}$ as a function of the sum in quadrature of $\mathrm{V_R}$ and $\mathrm{V_z}$. Panel b): $\mathrm{V_T}$ as a function of $\mathrm{V_R}$. Panel c)-[Lindblad diagram]:  $\mathrm{E}$ as a function of $\mathrm{L_Z}$. Panel d): eccentricity as a function of $\mathrm{L_Z}$. Panel e): the square root of $\mathrm{J_R}$ as a function of $\mathrm{J_{\phi}}$. Panel f)-[Action diamond diagram]: $\mathrm{(J_z-J_R)/J_{tot}}$ versus $\mathrm{J_\phi/J_{tot}}$, where $\mathrm{J_{tot}=|J_\phi|+J_R+J_z}$. Panel g): circularity as a function of eccentricity. Panel h): circularity as a function of $\mathrm{Z_{max}}$. Panel i): circularity as a function of $\mathrm{R_{GC}}$.}
         \label{kinematics}
   \end{figure*}

Figure~\ref{kinematics} shows nine different 
kinematic planes with the orbital properties of CCs compared with a sample of Type II Cepheids (TIICs). These planes are based on kinematic diagnostics that are quite useful to characterize Galactic stellar populations, and in particular to identify stars that are associated with the different Galactic components  (\citealt{lane22}; \citealt{bonifacio24}; \citealt{bono25submitted}). According to a Galactocentric cylindrical frame, 
$\mathrm{V_R}$, $\mathrm{V_T}$ and $\mathrm{V_z}$ are the radial, the tangential and the vertical velocity, while 
$\mathrm{L_Z}$ and $\mathrm{E}$ are the angular momentum component along the Galactic Z-axis and the total orbital energy. 
$\mathrm{J_R}$, $\mathrm{J_{\phi}}$, and $\mathrm{J_z}$ are the radial, the azimuthal, and the vertical actions, while $\mathrm{J_{tot}}$ is the total action, defined as $\mathrm{J_{tot}=|J_{\phi}| + J_{R} + J_z}$. 
$\mathrm{\lambda_z}$ is the normalized angular momentum (circularity), defined as $\mathrm{\lambda_z=J_z/J_{max}(E)}$ where $\mathrm{J_{max}(E)}$ is the angular momentum of a circular orbit with the same maximum binding energy. 
Finally, $\mathrm{Z_{max}}$ denotes the maximum height along the orbit above the Galactic plane. 
The bulk of the kinematic sample has cold orbits typical of thin disk stars, as stated by the coherent distribution in the 9 planes. In particular, in the Toomgre diagram (panel a) our sample is grouped in the bottom-right corner, characterized by $\mathrm{V_{T}\simeq250\, km\, s^{-1}}$ and low perpendicular velocities, typical of the thin disk component. On the other hand, TIICs cover a broad range in transversal velocity and a fraction of them also attain negative transversal velocities suggesting that they are on retrograde orbits. The stark difference between the two samples is fully supported by the Lindblad plane (panel c) 
in which CCs are distributed along the tiny region typical of thin disk stars, while TIICs move from the region typical of hot stellar components with radial orbits and random motions. The same outcome applies to the action-diamond diagram (panel f), since CCs are grouped in the right corner, since the major contribution to the action is given by their $\mathrm{L_z}$. Furthermore, CCs show high circularities ($\mathrm{\lambda_z \gtrsim 0.8}$, 
which means almost circular orbits and strong rotation), low eccentricities and low $\mathrm{Z_{max}}$ values (panels g, h, i). 
The TIICs display a more complex behaviour, since they move from a warm stellar component more typical of thick disk stars to hot stellar components and to retrograde orbits. A more quantitative analysis will be provided in a forthcoming paper (Nunnari et al. in preparation).

\begin{figure*}[h]
\centering
   \includegraphics[width=1.\linewidth]{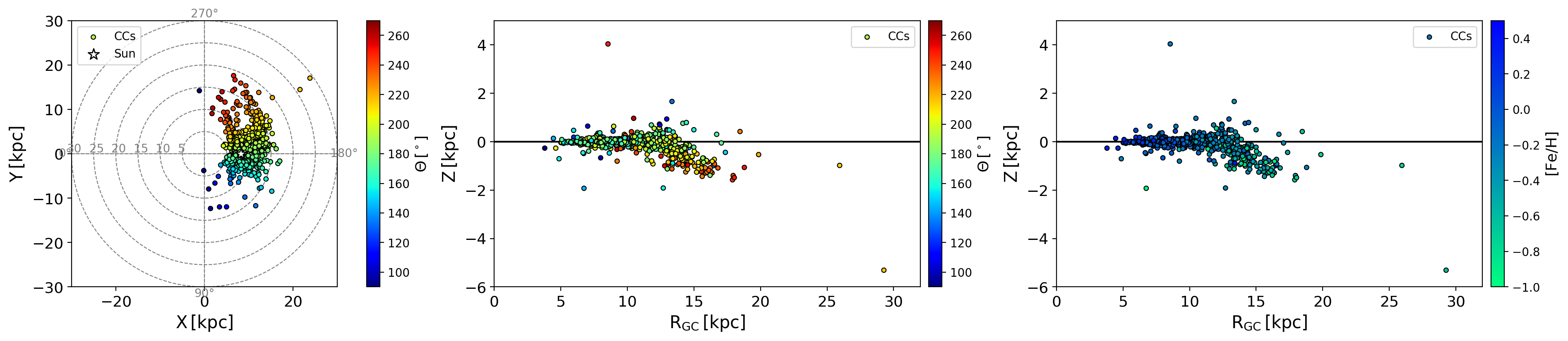}
      \caption{Left -- Distribution of the total sample of CCs onto the Galactic plane colour-coded according to the azimuth angle. 
      Central -- Distance perpendicular to the Galactic plane as a function of the 
      Galactocentric distance colour-coded according to the azimuth angle.
      Right -- Distance perpendicular to the Galactic plane as a function of the 
      Galactocentric distance colour-coded according to the metallicity.
      }
         \label{theta}
   \end{figure*}

The left panel of Fig.~\ref{theta} shows the radial distribution of CCs projected onto the Galactic plane. 
The symbols are colour-coded according to the azimuth angle $\Theta$. We note that stars with $\Theta\lesssim 180^{\circ}$ typically reach smaller Galactocentric radii than those at $\Theta\gtrsim 180^{\circ}$. This trend may arise from the intrinsic spatial distribution of stars in the sample (for example, due to the presence of spiral arms), from the selection function of our dataset, or from a combination of both effects. Data plotted in the central and right panel of the same figure display that for 
Galactocentric distances larger than 11-12~kpc the height above the Galactic plane 
steadily decreases. This finding is expected, since the variation is mainly caused 
by the Galactic warp. Specifically, it is known from previous observations that the Galactic midplane is bent downwards (i.e. Z<0) in the portion of the disk covered by our dataset. The reader interested in a more detailed discussion concerning the 
Galactic warp is referred to \citet{chen19}, \citet{skowron19}, \citet{lemasle22} and \citet{poggio25}. 

It is worth mentioning that the change in the slope of the radial metallicity gradient occurs at Galactocentric radii where the Galactic warp begins to become significant (i.e. approximately 11-12 kpc), tentatively suggesting that the two aspects might be related. It is important to note, however, that the spiral arms and bar-driven radial mixing can also potentially generate variations in the radial metallicity gradient (see e.g.,  \citealt{buder25}).

The orbital properties (radial velocity and circularity) of these objects fully support their 
association with the thin disk. However, more photometric and spectroscopic data are required 
to address this issue on a more quantitative basis. 

\section{Summary and conclusions}
\label{summary}
The present study represents one of the most comprehensive NLTE spectral analyses performed on CCs spanning an unprecedented range in Galactocentric distance (4.6–29.3 kpc) and elemental coverage including light, alpha, odd-Z, and iron peak elements. Such an approach is crucial because CCs, as young (ages younger than a few hundred Myrs), luminous super-giant stars, sample the current chemical enrichment of the Galactic thin disk with minimal contamination from older populations. We used the largest high-resolution optical spectral sample, for which we estimated the atmospheric parameters and chemical abundances. We constrained the iron and the chemical gradients for O, Na, Mg, Al, Si, S, Ca, Ti, Mn and Cu.

Previous studies available in the literature have modeled the chemical gradients of the thin disk, 
by assuming either a linear (\citealt{luck18}; \citealt{genovali2014fine}; \citealt{Trentin24B}) 
or a logarithmic (\citealt{da2023oxygen}) function. The fits based on a linear function 
overestimate the observed gradient in the inner disk ($\mathrm{R_{GC}}\lesssim 7$ kpc) and 
underestimate the observed gradient in the outer disk ($\mathrm{R_{GC}}\gtrsim 18$ kpc). 
The difference with previous investigations, based on a linear fit, is mainly due to the 
larger range in Galactocentric distances covered by the current sample. This means that we 
are dealing with 22 CCs with $\mathrm{R_{GC}}$ smaller than 6~kpc and 33 CCs with a $\mathrm{R_{GC}}$ larger 
than 15~kpc\footnote{The distance of the Cepheid ASAS J062939-1840.5 was estimated by using 
a mid-infrared (W1-band) PL relation and we found $\mathrm{R_{GC}}=29.3\pm0.3$~kpc. The distance 
of this object was also estimated by \citet{trentin23} using period-Wesenheit relations and they found 
$\mathrm{R_{GC}}\approx25$~kpc.}.

A flattening of the abundance gradient in the outer disk has been reported by several studies (e.g., \citealt{luck18}; \citealt{kovtyukh22}; \citealt{da2023oxygen}). A similar behaviour has also been observed in open clusters (\citealt{magrini23}; \citealt{otto25}), where a bilinear fitting approach has been adopted. Nevertheless, the AIC test performed by \citet{otto25} indicated that introducing two additional parameters in a bilinear model does not significantly improve the fit compared to a simple linear relation. In any case, such modelling inherently implies a break at a certain Galactocentric radius, raising a theoretical question about the physical origin of this discontinuity. For our sample, the best-fitting bilinear model applied to the iron gradient (Fig.~\ref{iron}) suggests a break, or "knee", at $\mathrm{R_{GC}}=10.5\pm0.4$ kpc. This model provides a better fit than the linear one, as indicated by smaller residuals and a lower AIC score. However, when the same bilinear approach is applied to the other elements, the inferred knee positions vary considerably, ranging from 8 to 15 kpc. The bilinear model yields the lowest AIC score only for Mg compared to alternative models (Fig.~\ref{gradients_threemodels}), although in this case the knee occurs at $\mathrm{R_{GC}}=15.3\pm0.7$ kpc, a value inconsistent with the other elements and with previous studies. 

In \citetalias{da2023oxygen}, a large sample of 356 CCs, including four at $\mathrm{R_{GC}}\gtrsim 20$ kpc, was analysed using a log–log relation for Fe, O, and S. In Fig.~\ref{iron}, we applied the same functional form to fit the iron gradient. As expected, this approach yields lower mean squared error, chi-squared, and AIC values compared to the linear and bilinear models, except in the case of Mg, as noted above. Furthermore, as shown in Fig.~\ref{log_log}, the slope we obtain differs from that reported in \citetalias{da2023oxygen}, likely due to differences in spectral analysis, such as the line list and the use of NLTE grids for estimating atmospheric parameters, and to the increased number of Cepheids located in the outskirt of the disk. Overall, the log-log fit appears to be the most reasonable among the three analytical models, although sodium and magnesium are better modeled by the bilinear fit (Fig.~\ref{gradients_threemodels}), suggesting that a single analytical function may not fully describe all gradients. For this reason, we also employed Gaussian Process Regression, which does not require any predetermined functional form. This modelling confirms that the radial abundance gradients in the thin disk generally flatten with increasing Galactocentric distance across all studied elements.

The flattening of the gradients is a feature that extends to extragalactic studies. Works on other disk galaxies have shown that, although the inner regions show a negative gradient, the radial distribution of metallicity flattens to a virtually constant value beyond the isophote radius. This behaviour has been found in the disk galaxy M83 (\citealt{bresolin09}, \citeyear{bresolin16}), and subsequently confirmed in NGC 1512 and NGC~3621 (\citealt{bresolin12}; \citealt{kudritki14}), where the oxygen abundance remains homogeneous and approaches an almost constant value in the outer regions. These observations suggest that flattening is a common feature of the metallicity gradient in spiral galaxy disks.

The current metallicity of CCs can be compared with that of B-stars located in the solar neighborhood. According to \citet{nieva12}, early B-type stars within 500 pc show a mean iron abundance of 7.52$\pm$0.03 dex, consistent with the solar value. This result is based on a NLTE spectral analysis of 29 early B-type stars. In the present work, the mean iron abundance of the 11 CCs located within 500 pc of the Sun is 7.49$\pm$0.13 dex, in perfect agreement with \citet{nieva12}. Differently, the 93 CCs located at the solar circle (i.e. all Cepheids with $7.7\lesssim \mathrm{R_{GC}} \lesssim 8.7$ kpc), exhibit a slightly iron underabundance of 7.42$\pm$0.10 dex,  but still consistent within 1$\sigma$.  

Our results for chemical abundance ratios as a function of $\mathrm{R_{GC}}$ reveal that [X/Fe] ratios remain approximately constant across the entire thin disk. The near-flat distributions, with average values within 0.10 dex and minimal slopes, indicate that most elemental abundances scale closely with iron throughout the Galactic disk. This consistency suggests that these elements trace the gas distribution in the thin disk in a coherent manner. Slight overabundances in Al and Na, underabundance in Cu, and the small but measurable slopes for Mn, and Cu point to subtle variations, yet they do not compromise the overall trend of roughly constant abundance ratios. The negative slopes observed for Mn and Cu are in agreement with previous findings for iron-peak elements (\citealt{otto25}).

Chemical gradients from previous studies, such as \citet{genovali2014fine} and \citetalias{da2023oxygen}, indicate that the outer disk exhibits a higher dispersion. This effect can be partially attributed to the Galactic warp (\citealt{kerr57}; \citealt{oort58}; \citealt{lemasle22}; \citealt{poggio25}), a large-scale distortion of the disk that affects the estimated Galactocentric distances. In regions where the warp is pronounced, Cepheids appear closer in Galactocentric distance than they would in a non-warped disk. However, \citet{lemasle22} noted that the dispersion introduced by the warp is too small to fully account for the larger spread around the mean metallicity gradient reported by \citet{genovali2014fine} and \citet{da2023oxygen}. Instead, this increased dispersion likely reflects larger uncertainties associated with distant stars, whose spectra typically have lower signal-to-noise ratio. In our study, the gradients do not show a significant increase in dispersion (Fig.~\ref{gradients_gpr}), in agreement with \citet{lemasle22}. It is worth noting that our sample contains fewer Cepheids in the outer disk ($\mathrm{R_{GC}}>12$ kpc), where the warp is stronger, than in the inner regions. Nevertheless, the trend visible in Fig.~\ref{theta} appears to be a direct consequence of the Galactic warp. Similar trends have also been observed using larger kinematic samples of Cepheids (\citealt{lemasle22}, Fig.~A.3; \citealt{poggio25}).

Results for the sample of the youngest open clusters (OCs) from \citet[age < 400~Myrs]{otto25} overlaps with that of Cepheids. Sodium is an exception, since Cepheids are on average overabundant in this element. It was suggested by \citet{takeda13} and confirmed by \citet{dasilva15} in a comparison between giant and dwarf field stars, that this overabundance is mainly due to stellar evolution effects: the first dredge-up brings some sodium from the interior to the atmosphere, altering its initial chemical composition (\citealt{ventura13, lagarde21}). In contrast, the chemical abundances of OCs have been measured in both giant and dwarf stars, with the latter still unaffected by significant mixing processes. Comparisons with other stellar populations, such as open clusters, generally reveal similar radial trends for iron-peak and alpha elements, further supporting the homogeneity of chemical enrichment in the thin disk, at least in the region between 5 and 14 kpc. These findings are consistent with previous studies (\citealt{magrini23}; \citealt{yong12}; \citealt{otto25}).

This study opens multiple possibilities for future investigation. In addition to the natural need of increasing the sample size, especially in the outer regions of the disk, we emphasize the following points:
\begin{itemize}
    \item  The chemical abundances of other elements, such as heavy elements produced by the s- and r-processes, have yet to be determined.  As discussed in \citet{Trentin24B}, some elements such as scandium, zinc, and zirconium show a bifurcation at the low-metallicity tail. This anomalous behaviour may arise from variations in spectral line analyses or from the presence of distinct stellar populations associated with different spiral arms. Since our sample includes Cepheids located in various spiral arms, a future study focusing on these elements could help clarify this issue. However, NLTE grids for these elements are not yet available in PySME.

    \item In the inner disk, for $\mathrm{R_{GC}}<5$ kpc, a double sequence can appear, one that follows the logarithmic profile and the other showing a change in the slope. Note that a significant decrease in the metallicity of the young stellar population, confined to the central and inner regions, has been found in barred spiral galaxies such as NGC~1365 and M83 (\citealt{sextl24}, \citeyear{sextl25}). Furthermore, \citet{andrievsky16} discussed a change in the slope at $\mathrm{R_{GC}}<5$ kpc, suggesting that it might be caused by a reduced star formation rate compared to the outer disk regions. To account for such a change of slope in the radial gradients, it has also been suggested that the inner disk regions have also been affected by dilution effects caused by the infall of more metal-poor gas
    \citep[][and references therein]{genovali15}. Further extensive sampling at small $\mathrm{R_{GC}}$ is crucial to investigate this feature of the gradient and its implications for the chemical enrichment of the Galactic bar.

    \item The [X/H] gradients exhibit similar slopes among the studied elements, and the [X/Fe] ratios remain approximately constant across the entire range of $\mathrm{R_{GC}}$ explored. This behaviour may indicate that these elements could serve as potential tracers of the underlying gas density profile. However, further validation is required to confirm these trends, and it remains to be established whether such a result can be extended to the disks of external galaxies. 

    \item The line list employed in this work serves as a tool for estimating atmospheric parameters and has been specifically adapted to optical spectra of Galactic Classical Cepheids. However, further investigation is required to extend the applicability of this method to the near-infrared (NIR) domain to fully exploit the NIR spectroscopic data and reach the more extinct regions. In addition, the metallicity range over which the selected lines have been tested should be broadened toward lower values, as the current analysis only covers the metallicity interval typical of the Galactic thin disk (-0.8 to +0.4 dex). This extension is essential to enable the application of the line list to more metal-poor Cepheids of Magellanic Clouds and the Local Group.
    
\end{itemize}

In conclusion, our results suggest that a single analytic function is not sufficient to model all the gradients. For this reason, here we provide the behaviour of each element gradient with a GPR modelling. However, when using an analytical function, a log-log profile should be preferred. Our results confirm that the radial abundance gradients in the thin disk flatten with radius across all studied elements. This is consistent with theoretical expectations of inside-out disk formation and continuous chemical enrichment, where inner regions exhibit higher metallicity and alpha-enhancements due to more rapid star formation. Nevertheless, the outer disk is still too poorly sampled to give a definitive conclusion.

The adoption of NLTE modelling throughout, from atmospheric parameters to elemental abundances, represents an important step forward relative to prior studies that often combined LTE atmospheric parameters with NLTE corrections. The method employed in this study could be extended to data that come from large spectroscopic surveys (4MOST\footnote{https://www.4most.eu/cms/home/}, \citealt{dejong16}) and from new instruments (MOONS\footnote{https://vltmoons.org/} at VLT, \citealt{cirasuolo12}; MAVIS\footnote{https://mavis-ao.org/mavis/} at VLT, \citealt{content22}). This will allow for an increase in sample sizes and extension of the spatial coverage, enabling finer chemical gradient mapping, especially in regions now poorly sampled such as the inner and outer disk, along with setting improved constraints on Galactic formation models.

\section{Data availability}
Tables \ref{linelist}, \ref{linelist_abu}, \ref{sensitivitytable}, \ref{ccs_info} and \ref{spectra_info}  are only available in electronic form at the CDS via anonymous ftp to \url{cdsarc.u-strasbg.fr} (130.79.128.5) or via \url{http://cdsweb.u-strasbg.fr/cgi-bin/qcat?J/A+A/}.

\begin{acknowledgements}
Part of the research activities described in this report were carried out with contribution of the Next Generation EU funds within the National Recovery and Resilience Plan (PNRR), Mission 4 - Education and Research, Component 2 - From Research to Business (M4C2), Investment Line 3.1 - Strengthening and creation of Research Infrastructures, Project IR0000034 – “STILES - Strengthening the Italian Leadership in ELT and SKA”, CUP C33C22000640006.
This research was supported by the Munich Institute for Astro-, Particle and BioPhysics (MIAPbP), which is funded by the Deutsche Forschungsgemeinschaft (DFG, German Research Foundation) under Germany´s Excellence Strategy – EXC-2094 – 390783311.
We thank the support of the project PRIN MUR 2022 (code 2022ARWP9C) ‘Early
 Formation and Evolution of Bulge and HalO (EFEBHO)’ (PI: M. Marconi), funded by the European Union—Next Generation EU, and from the Large grant INAF 2023 MOVIE (PI: M. Marconi).
This work has made use of the VALD database, operated at Uppsala University, the Institute of Astronomy RAS in Moscow, and the University of Vienna.
H.J. acknowledges support from the Swedish Research Council, VR (grant 2024-04989). E.P. is supported in part by the Italian Space Agency (ASI) through contract 2018-24-HH.0 and its addendum 2018-24-HH.1-2022 to the National Institute for Astrophysics (INAF). J.M.O. acknowledges support for this research from the National Science Foundation Collaborative Grants (AST-2206541, AST-2206542, and AST-2206543).
M.F. acknowledges financial support from the ASI-INAF agreement no. 2022-14-HH.0.
This research has also made use of the GaiaPortal catalogues access tool, ASI - Space Science Data Center, Rome, Italy (https://gaiaportal.ssdc.asi.it).

\end{acknowledgements}

\bibliographystyle{aa}
\bibliography{aanda}
\newpage
\clearpage
\begin{appendix}

\section{Tables}
\label{tables}
Table \ref{spec_sample} lists the properties of the spectroscopic sample. Lines used for atmospheric parameters, iron and titanium abundance estimation are listed in Table \ref{linelist}, while those used for chemical abundances are highlighted in Table \ref{linelist_abu}. Table \ref{x_fe_reference_1} shows the chemical abundances of the 3 giant stars of M67 and the Sun. Furthermore, Table \ref{sensitivitytable} lists the values of $\mathrm{\Delta[X/H]}$ for each spectral line computed as the mean abundance variation over 10 spectra, in response to changes in the atmospheric parameters. Finally, Table \ref{ccs_info} lists all the kinematic and chemical information for each Cepheid, with the flag "$\mathrm{newCC\_ flag}$" which determines whether the Cepheid has been added to the original \citetalias{da2023oxygen} sample, while the information and the results for each spectrum are highlighted in Table \ref{spectra_info}.
       \begin{table*}[h]
       \caption{Spectroscopic sample.}
       \centering
       \begin{tabular}{c c c c c }
       \hline
       \hline
    \hspace{0.5cm}Spectrograph\hspace{0.5cm}    & \hspace{0.5cm}N. of spectra\hspace{0.5cm}    &  \hspace{0.5cm}N. of stars\hspace{0.5cm}    &  \hspace{0.5cm}Resolution (R)\hspace{0.5cm}    &  \hspace{0.5cm}  Coverage [$\AA$]\hspace{0.5cm}    \\
            \hline
            HARPS\tablefootmark{a} & 231 & 29 & $\sim$ 115000 & [3790-6900] \\
            SES\tablefootmark{b} & 400 & 64 & $\sim$ 55000 & [3860-8800] \\
            FEROS\tablefootmark{c} & 339 & 161 & $\sim$ 48000 & [3500-9200] \\
            UVES\tablefootmark{d} & 373 & 224 & $\sim$ 40000 & [3800-9400] \\
            HARPS-N\tablefootmark{e} & 40 & 29 & $\sim$ 115000 & [3780-6910] \\
            ESPaDOnS\tablefootmark{f} & 16 & 13 & $\sim$ 81000 & [3700-10500] \\
            \hline
            \hline
            \end{tabular}
       \label{spec_sample}
       \tablefoot{
            \tablefoottext{a}{High Accuracy Radial velocity Planet Searcher (HARPS, \citealt{mayor03}).}
            \tablefoottext{b}{STELLA échelle Spectrograph  (SES, \citealt{strassmeier04, strassmeier10}).}
            \tablefoottext{c}{The Fiber-fed Extended Range Optical Spectrograph (FEROS, \citealt{kaufer99}).}
            \tablefoottext{d}{Ultraviolet and Visual Echelle Spectrograph (UVES, \citealt{dekker00}).}
            \tablefoottext{e}{High Accuracy Radial velocity Planet Searcher for the Northern hemisphere (HARPS-N, \citealt{cosentino12}).}
            \tablefoottext{f}{Echelle SpectroPolarimetric Device
for the Observation of Stars (ESPaDOnS, \citealt{donati06}).}
       }
   \end{table*}
    \begin{table}[h]
    \caption{Line list used to estimate the atmospheric parameters.}
        \centering
        \begin{tabular}{c c c c c c}
             \hline
             \hline
             El & Ion. state & $\lambda$ [$\AA$] & gfflag & synflag & $\mathrm{E_{low}\,[eV]}$ \\
             \hline
             Fe & 1 & 5217.389 & y & y & 3.211 \\
             ... & ... & ... & ... & ... & ... \\
             \hline
             Fe & 2 & 4993.350 & u & y & 2.807 \\
             ... & ... & ... & ... & ... & ... \\
             \hline
             Ti & 1 & 4913.613 & y & y & 1.873 \\
             ... & ... & ... & ... & ... & ... \\
             \hline
             Ti & 2 & 4874.009 & y & u & 3.095 \\
             ... & ... & ... & ... & ... & ... \\
             \hline
             \hline            
        \end{tabular}
        \tablefoot{The column labelled 'Ionization state' indicates whether the line corresponds to neutral atoms (value "1") or first-ionized species (value "2"). The full table is available at the CDS.}
        \label{linelist}
    \end{table}
    \begin{table}[h]
    \caption{Line list used to estimate the chemical abundances.}
        \centering
        \begin{tabular}{c c c c c c}
             \hline
             \hline
             El & Ion. state & $\lambda$ [$\AA$] & gfflag & synflag & $\mathrm{E_{low}\,[eV]}$ \\
             \hline
             O & 1 & 7771.944 &  & & 9.146 \\
             ... & ... & ... & ... & ... & ... \\
             \hline
             Na & 1 & 5688.205 & u & y & 2.104 \\
             ... & ... & ... & ... & ... & ... \\
             \hline
             Mg & 1 & 5711.088 & u & y & 4.346 \\
             ... & ... & ... & ... & ... & ... \\
             \hline
             Al & 1 & 5557.063 & u & u & 3.143 \\
             ... & ... & ... & ... & ... & ... \\
             \hline
             Si & 1 & 5645.613 & y & u & 4.930 \\
             ... & ... & ... & ... & ... & ... \\
             \hline
             S  & 1 & 6743.540 & u & u & 7.866 \\
             ... & ... & ... & ... & ... & ... \\
             \hline
             Ca & 1 & 5260.387 & y & y & 2.521 \\
             ... & ... & ... & ... & ... & ... \\
             \hline
             Mn & 1 & 5394.683 & y & y & 0.000 \\
             ... & ... & ... & ... & ... & ... \\
             \hline
             Cu & 1 & 5105.530 & y & u & 1.389 \\
             ... & ... & ... & ... & ... & ... \\
             \hline    
             \hline
        \end{tabular}
        \tablefoot{The column labelled 'Ionization state' indicates whether the line corresponds to neutral atoms (value "1") or first-ionized species (value "2"). For oxygen lines, the information on "gfflag" and "synflag" flags is not available. The full table is available at the CDS.}
        \label{linelist_abu}
    \end{table}
\begin{table*}[h]
    \caption{[X/Fe] values with errors of 3 red clump stars of M67 and the Sun.}
    \centering
    \begin{tabular}{lcccccccccccc}
        \hline
        \hline
        Target & \phantom{-0.01} & \phantom{0.01} & Mg & eMg & Si & eSi & Ca & eCa  & S & eS  \\
        \hline
        Gaia DR3 604918144751101440 &     &     & 0.01 & 0.08  & $-$0.02 & 0.14 & $-$0.08 & 0.05 & 0.04 & 0.10 \\
        Gaia DR3 604922164840316672 &     &     & $-$0.01 & 0.07  & $-$0.04 & 0.05 & $-$0.01 & 0.04 & 0.08 & 0.10 \\
        Gaia DR3 604917629355042176 &     &     & 0.01 & 0.08 & $-$0.01 & 0.14 & $-$0.03 & 0.06 & $-$0.01 & 0.10 \\
        Sun &     &     & 0.04 & 0.03 & 0.00 & 0.02 & 0.00 & 0.03 & $-$0.05 & 0.10 \\
        \hline
    \end{tabular}

    \begin{tabular}{lcccccccccccc}
        \hline
        \phantom{Target} & Ti & eTi & Mn & eMn & Al & eAl & Na & eNa & Cu & eCu  \\
        \hline
        Gaia DR3 604918144751101440 & $-$0.07 & 0.05 & $-$0.14 & 0.02 & $-$0.09 & 0.07 & $-$0.08 & 0.06 & 0.02 & 0.08 \\
        Gaia DR3 604922164840316672 & $-$0.04 & 0.05 & $-$0.10 & 0.05 & $-$0.07 & 0.08 & $-$0.11 & 0.05 & 0.00 & 0.01 \\
        Gaia DR3 604917629355042176 & $-$0.09 & 0.06 & $-$0.14 & 0.03 & $-$0.09 & 0.06 & $-$0.08 & 0.03 & $-$0.07 & 0.06 \\
        Sun & $-$0.04 & 0.04 & 0.00 & 0.01 & $-$0.01 & 0.02 & $-$0.03 & 0.05 & $-$0.04 & 0.02 \\
        \hline
        \hline
    \end{tabular}
    \label{x_fe_reference_1}
\end{table*}

\begin{sidewaystable}[ht]
\caption{Sensitivity table.}
\centering
\resizebox{\textheight}{!}{
\begin{tabular}{c c|c c c c|c c c c|c c c c|c c c c}
\hline
\hline
$\lambda$ [$\AA$] & $\Delta\mathrm{[X/H]}$ & \multicolumn{4}{c}{\textbf{$\mathrm{\Delta \log(g)}$} [dex]} & \multicolumn{4}{c}{\textbf{$\mathrm{\Delta [Fe/H]}$} [dex]} & \multicolumn{4}{c}{\textbf{$\mathrm{\Delta T_{\mathrm{eff}}}$} [K]} & \multicolumn{4}{c}{\textbf{$\mathrm{\Delta \xi}$} [km~s$^{-1}$]} \\
 & $\mathrm{X}$ & $-$0.1 & $-$0.2 & +0.1 & +0.2 & $-$0.05 & $-$0.1 & +0.05 & +0.1 & $-$100 & $-$50 & +100 & +50 & $-$0.2 & $-$0.4 & +0.2 & +0.4 \\
\hline
4874.01 & Ti & 0.111 & 0.070 & 0.256 & 0.291 & 0.280 & 0.338 & 0.162 & 0.105 & 0.192 & 0.207 & 0.252 & 0.237 & 0.300 & 0.375 & 0.154 & 0.095 \\
5105.53 & Cu & $-$0.002 & $-$0.007 & 0.009 & 0.017 & 0.048 & 0.092 & $-$0.034 & $-$0.074 & $-$0.074 & $-$0.037 & 0.095 & 0.046 & 0.033 & 0.063 & $-$0.017 & $-$0.041 \\
5211.53 & Ti & 0.117 & 0.087 & 0.190 & 0.220 & 0.201 & 0.238 & 0.119 & 0.082 & 0.149 & 0.153 & 0.174 & 0.166 & 0.212 & 0.267 & 0.112 & 0.078 \\
5218.20 & Cu & $-$0.012 & 0.001 & $-$0.006 & $-$0.010 & 0.055 & 0.125 & $-$0.072 & $-$0.126 & $-$0.088 & $-$0.043 & 0.062 & 0.027 & $-$0.004 & 0.034 & $-$0.018 & $-$0.014 \\
5219.70 & Ti & 0.150 & 0.151 & 0.161 & 0.156 & 0.167 & 0.168 & 0.150 & 0.129 & 0.099 & 0.137 & 0.177 & 0.172 & 0.164 & 0.163 & 0.157 & 0.155 \\
5220.07 & Cu & $-$0.057 & $-$0.162 & 0.173 & 0.249 & 0.227 & 0.321 & 0.009 & $-$0.165 & $-$0.009 & 0.036 & 0.316 & 0.183 & 0.121 & 0.124 & $-$0.013 & $-$0.075 \\
5260.39 & Ca & $-$0.024 & $-$0.043 & 0.012 & 0.026 & 0.054 & 0.109 & $-$0.080 & $-$0.157 & $-$0.031 & $-$0.026 & 0.019 & 0.010 & $-$0.003 & $-$0.011 & $-$0.010 & $-$0.016 \\
5381.02 & Ti & $-$0.223 & $-$0.239 & $-$0.138 & $-$0.129 & $-$0.135 & $-$0.125 & $-$0.178 & $-$0.200 & $-$0.165 & $-$0.159 & $-$0.143 & $-$0.148 & $-$0.129 & $-$0.113 & $-$0.201 & $-$0.239 \\
... & ... & ... & ... & ... & ... & ... & ... & ... & ... & ... & ... & ... & ... & ... & ... & ... & ... \\
\hline
\hline
\end{tabular}
}
\tablefoot{This table lists the values of $\Delta[X/H]$ for each spectral line, computed as the mean abundance variation over 10 spectra, in response to changes in the atmospheric parameters (top row of the table). These values quantify the sensitivity of each line to the individual parameter variations. The full table is available at the CDS.}
\label{sensitivitytable}
\end{sidewaystable}
\begin{sidewaystable}[ht]
\caption{Data and results for each Cepheid.}
\centering
\resizebox{\textheight}{!}{
\begin{tabular}{ccccccccccccccc}
\hline
\hline
$\mathrm{Star}$ & $\mathrm{Gaia\, ID}$ & $\mathrm{RA}$ & $\mathrm{Dec}$ & $\mathrm{\varpi}$ & $\mathrm{\varpi_{err}}$ & parallax over error & pmra & pmdec & RUWE & N25 & dS23 & d & $\mathrm{d_{err}}$ & ... \\
 & & [deg] & [deg] & [mas] & [mas] & & [$\mathrm{mas\,yr^{-1}}$] & [$\mathrm{mas\,yr^{-1}}$] & & & & [pc] & [pc] & \\
\hline
AA Gem & 3430067092837622272 & 91.646 & 26.329 & 0.275 & 0.018 & 15.548 & $-$0.217 & $-$0.670 & 1.249 & n & y & 3245 & 184 & ... \\
AB Vel & 5354898841641168768 & 154.967 & $-$56.310 & 0.229 & 0.012 & 18.876 & $-$5.640 & 2.616 & 0.907 & y & y & 4430 & 235 & ... \\
AC Mon & 3050050207554658048 & 105.249 & $-$8.709 & 0.355 & 0.019 & 19.010 & $-$2.237 & 1.964 & 1.379 & y & y & 2631 & 126 &... \\
AD Cru & 6057514092119497472 & 183.249 & $-$62.097 & 0.293 & 0.013 & 22.121 & $-$5.338 & 0.473 & 1.017 & y & y & 3220 & 149 & ... \\
AD Pup & 5614312705966204288 & 117.016 & $-$25.578 & 0.233 & 0.017 & 14.088 & $-$2.358 & 3.054 & 1.362 & y & y & 3995 & 276 & ... \\
AE Tau & 3441145084801214464 & 83.653 & 26.202 & 0.274 & 0.020 & 14.044 & 0.071 & $-$1.252 & 1.050 & y & y & 3347 & 231 & ... \\
AE Vel & 5309174967720762496 & 144.214 & $-$53.033 & 0.353 & 0.012 & 29.138 & $-$5.137 & 4.069 & 0.969 & y & y & 2726 & 99 & ... \\
AG Cru & 6059635702888301952 & 190.358 & $-$59.794 & 0.742 & 0.020 & 37.358 & $-$4.422 & $-$1.847 & 1.016 & y & y & 1323 & 32 & ... \\
... & ... & ... & ... & ... & ... & ... & ... & ... & ... & ... & ... & ... & ... & ... \\
\hline
\hline
\end{tabular}
}
\tablefoot{Information for each Cepheid, including kinematic and chemical properties. The Cepheid name and the Gaia DR3 source ID are listed in the first two columns. RA, Dec, parallax ($\mathrm{\varpi}$), parallax error ($\mathrm{\varpi}_{err}$), parallax over error, proper motions in RA and Dec and RUWE are reported in columns (3)-(10). Columns (11) and (12) provide flags indicating whether chemical abundances are presented in this work (N25)  and in \citetalias{da2023oxygen} (dS23), respectively. Columns (13) and (14) list the distance and its associated uncertainty. The full table is available at the CDS.}
\label{ccs_info}
\end{sidewaystable}
\begin{sidewaystable}[ht]
\caption{Data and results for each spectrum.}
\centering
\resizebox{\textheight}{!}{
\begin{tabular}{cccccccccccccccccc}
\hline
\hline
spectrum ID & MJD & instrument & Star & Gaia ID & $\mathrm{{T_{eff}}}$ & $\mathrm{{T_{eff,err}}}$ & $\mathrm{\log g}$ & $\mathrm{(\log g)_{err}}$ & $\mathrm{[Fe/H]}$ &  $\mathrm{[Fe/H]_{err}}$ & $\mathrm{v_{mic}}$ & ... \\
 & & & & & [K] & [K] & [dex] & [dex] & [dex] & [dex] & $\mathrm{[km \,s^{-1}]}$ & \\
\hline
ADP.2014-09-26T16\_50\_57.353 & 56213.9879650 & HARPS & V496 Aql & 4204653587029046400 & 5664 & 140 & 1.02 & 0.15 & $-$0.12 & 0.08 & 3.55 & ... \\
ADP.2014-09-26T16\_54\_13.337 & 56213.9822101 & HARPS & Y Oph & 4175017625462647168 & 5662 & 150 & 1.01 & 0.15 & $-$0.08 & 0.05 & 4.06 & ... \\
ADP.2014-09-26T16\_54\_45.020 & 56239.9935618 & HARPS & V496 Aql & 4204653587029046400 & 5765 & 150 & 1.01 & 0.15 & $-$0.13 & 0.05 & 3.90 & ... \\
ADP.2014-09-26T16\_54\_49.173 & 56212.9902382 & HARPS & Y Sgr & 4096107909387492992 & 5755 & 100 & 1.03 & 0.10 & $-$0.12 & 0.08 & 3.720 & ... \\
ADP.2014-10-01T10\_19\_05.720 & 53034.1506477 & HARPS & $\zeta$ Gem & 3366754155291545344 & 5528 & 90 & 1.25 & 0.10 & 0.05 & 0.08 & 3.08 & ... \\
ADP.2014-10-01T10\_19\_06.007 & 53149.2916913 & HARPS & Y Sgr & 4096107909387492992 & 5564 & 130 & 1.43 & 0.15 & 0.00 & 0.05 & 3.96 & ... \\
ADP.2014-10-01T10\_19\_06.560 & 53021.1981818 & HARPS & $\zeta$ Gem & 3366754155291545344 & 5842 & 130 & 1.01 & 0.15 & $-$0.04 & 0.08 & 3.58 & ... \\
ADP.2014-10-01T10\_19\_12.130 & 53025.1780493 & HARPS & $\beta$ Dor & 4757601523650165120 & 5270 & 100 & 0.99 & 0.10 & $-$0.11 & 0.08 & 3.21 & ... \\
... & ... & ... & ... & ... & ... & ... & ... & ... & ... & ... & ... & ... \\
\hline
\hline
\end{tabular}
}
\tablefoot{Information for each spectrum, including the derived atmospheric parameters and chemical abundances estimates. The spectrum ID, MJD, instrument, Cepheid name, and Gaia DR3 source ID are listed in the first five columns. Columns (6)-(12) report the estimates of $\mathrm{T_{eff}}$, $\mathrm{\log g}$, $\mathrm{ [Fe/H]}$, and $\mathrm{v_{mic}}$ along with their associated uncertainties. The full table is available at the CDS.}
\label{spectra_info}
\end{sidewaystable}

    \section{New distance estimates for two Cepheids}
    \label{distances}
    The targets $\mathrm{ASAS\,\,181024}$-$2049.6$ and $\mathrm{ASAS\,\,SN\,\,J065046.50}$-$085808.7$ have been estimated respectively at a distance of 5.64 kpc (\citealt{andrievsky16}) and of 10.04 kpc (\citealt{trentin23}).

    \citet{andrievsky16} applied the method of the colour excess determination based on the use of the $\mathrm{E(B-V) - EW(DIB)}$ calibrating relation from \citet{friedman11}. Here EW(DIB) is the equivalent width of the 6613~$\AA$ diffuse interstellar band (DIB), which is also seen in Cepheid spectra. Using this relation, they found that the $\mathrm{E(B-V)}$ value for $\mathrm{ASAS 181024}$-$2049.6$ is equal to 1.41. Combining the approximate relation $\mathrm{A_V = E(B-V)/3.2}$, the "absolute magnitude–pulsational period" relation of \citet{gieren98}, and the mean V magnitude, they found a heliocentric distance of 5641 pc, and $\mathrm{R_{GC}= 2.53}$ kpc.
    Given the $\mathrm{parallax\_over\_error\approx 9}$, we used the Gaia parallax to obtain a heliocentric distance of $2.7\pm0.3$ kpc and $\mathrm{R_{GC}=5.5}$~kpc ($\mathrm{external.gaiaedr3\_distance}$ table, \citealt{bailerjones21}). Given this value so different from that obtained by the other authors, we used the W1-band PL relation, calibrated from Galactic CCs by \citet{wang18}, obtaining a distance of $\sim 3.0$ kpc, consistent with the value that is obtained by the parallax method.
    
    \citet{trentin23} found a heliocentric distance of $10.0\pm0.6$~kpc for $\mathrm{ASAS\,\,SN\,\,J065046.50}$-$085808.7$ using PW relations calibrated on Galactic CCs by \citet{ripepi20} and on LMC CCs by \citet{ripepi22}. This measurement is in contrast with that of \citet{skowron19}, who found $7.6\pm0.3$~kpc using mid-IR PL relations. In a coherent manner as described in Sec. \ref{spectral sample}, we found a heliocentric distance of $6.2\pm 0.7$~kpc using the Gaia DR3 parallax, which is consistent with that of \citet{skowron19} within 2$\sigma$.

    \section{Spectral synthesis details}\label{details_synthesis}
   The required steps for computing atmospheric parameters by using PySME are the following:
   
   \begin{enumerate}
       \item Observed spectrum: with a specific spectral resolution.
       \item Atomic linelist: in a specified format, i.e. VALD.
       \item Atmosphere model: we utilized MARCS12 (\citealt{gustafsson2008grid}).
       \item NLTE grids: we used \citet{amarsi2020galah} grids.
       \item Initial chemical abundances: we used those from \citet{asplund09}.
       \item Initial set of atmospheric parameters.
       \item Spectral intervals (masks) around the spectral lines used for parameter 
        estimates with an identification of continuum and line regions.
   \end{enumerate}
     
   \begin{figure}[b]
   \centering
   \includegraphics[width=9cm]{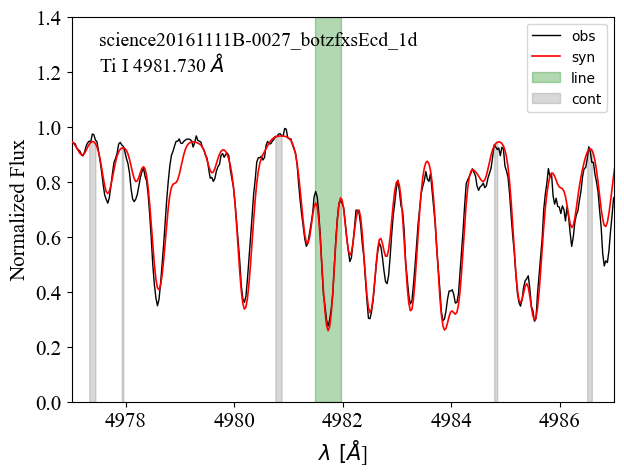}
      \caption{Fit of a Ti~I line for a Cepheid spectrum. The normalized observed spectrum is shown in black, while the synthetic spectrum, derived by estimating the atmospheric parameters through line fitting, is displayed in red. The continuum regions are marked by grey boxes, while the green box highlights the line region where the best fit is performed.}
         \label{sme}
   \end{figure}

\begin{figure}
    \centering
    \includegraphics[width=0.9\linewidth]{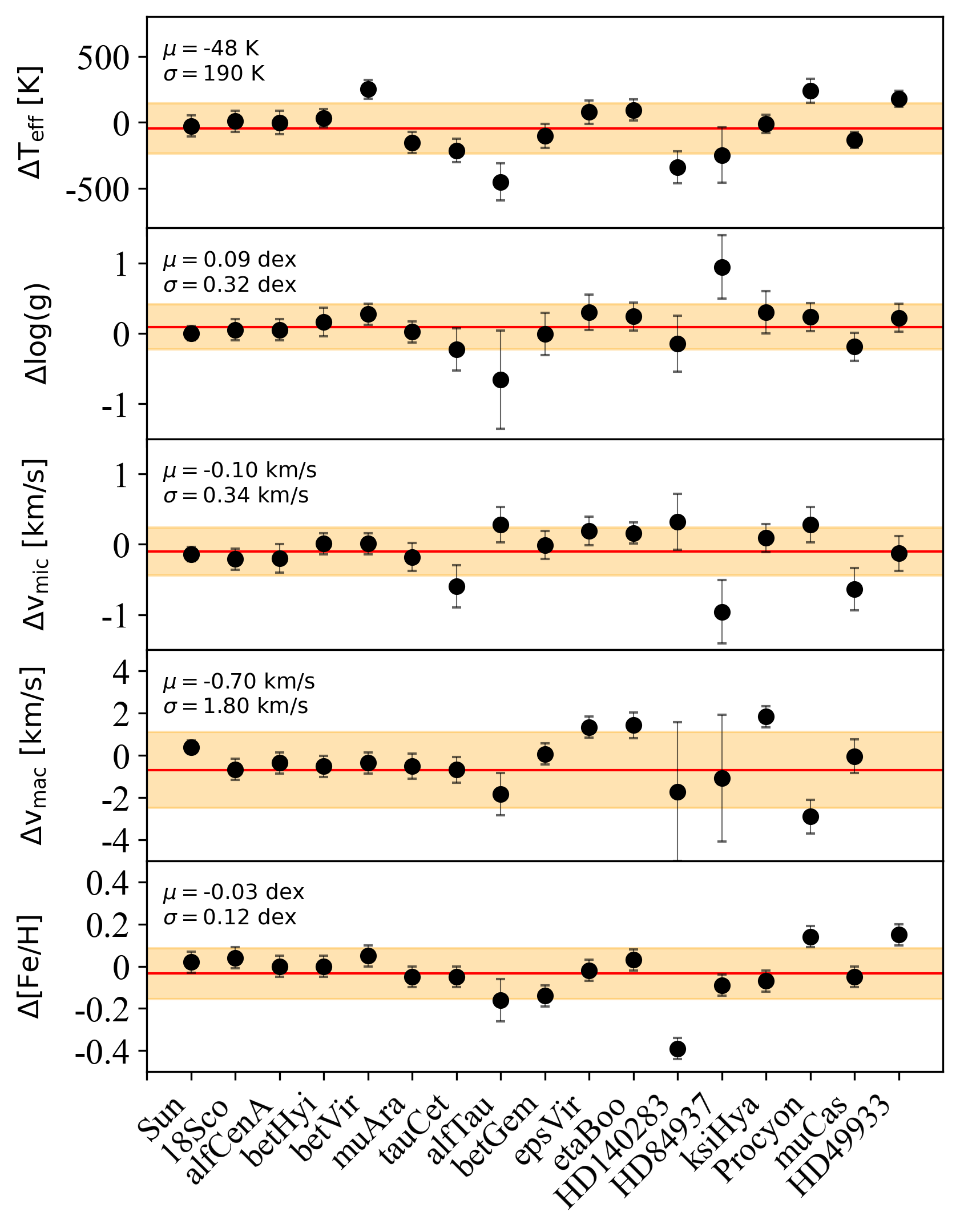}
    \caption{Comparison of the atmospheric parameters derived with our method and the reference values for the Sun and 16 Gaia benchmark stars from \citet{blancocuaresma14}. The black dots represent the differences, with error bars indicating uncertainties. The shaded regions are $\pm1\sigma$ around the mean difference value $\mu$ (red solid line).}
    \label{benchmark}
\end{figure}

   PySME is designed to work in conjunction with the Vienna Atomic Line Database (VALD, \citealt{ryabchikova15}), which provides a comprehensive collection of atomic and molecular transition parameters of astronomical interest. In order to have good estimators we selected spectral lines with accurate atomic parameters.
   We updated the $\mathrm{\log(gf)}$ values for the lines used as estimators by referring to the Gaia-ESO database (\citealt{heiter21}). Gaia-ESO provides two quality parameters for each line: \emph{gfflag}, which assesses the accuracy of $\mathrm{\log(gf)}$, and \emph{synflag}, which evaluates the blending resolution between closely spaced lines. These parameters are rated as "y" (recommended use), "n" (not recommended), or "u" (undecided quality).
    For the line selection we need a relatively large number of estimators, possibly both neutral and ionized lines to be sensitive to the gravity variations. We selected Fe~I, Fe~II, Ti~I, and Ti~II lines with \emph{gfflag} and \emph{synflag} rated as "y" for their high quality. This selection includes 21 Fe~I lines, 2 Fe~II lines, 6 Ti~I lines, and 1 Ti~II line. To expand the Fe~II line list, we also included lines with \emph{gfflag} rated as "u", adding 3 additional lines. Similarly, for the Ti~II line list, we included lines with \emph{synflag} rated as "u", adding 3 more lines. Fig. \ref{sme} displays an example of the fit of a Ti~I line in a Cepheid spectrum. The complete line list is presented in Table \ref{linelist}.
    We compared the atmospheric parameters derived from our method and the reference values for the Sun and 16 benchmark stars in Fig. \ref{benchmark}.

\begin{figure}
   \centering
   \includegraphics[width=8.5cm]{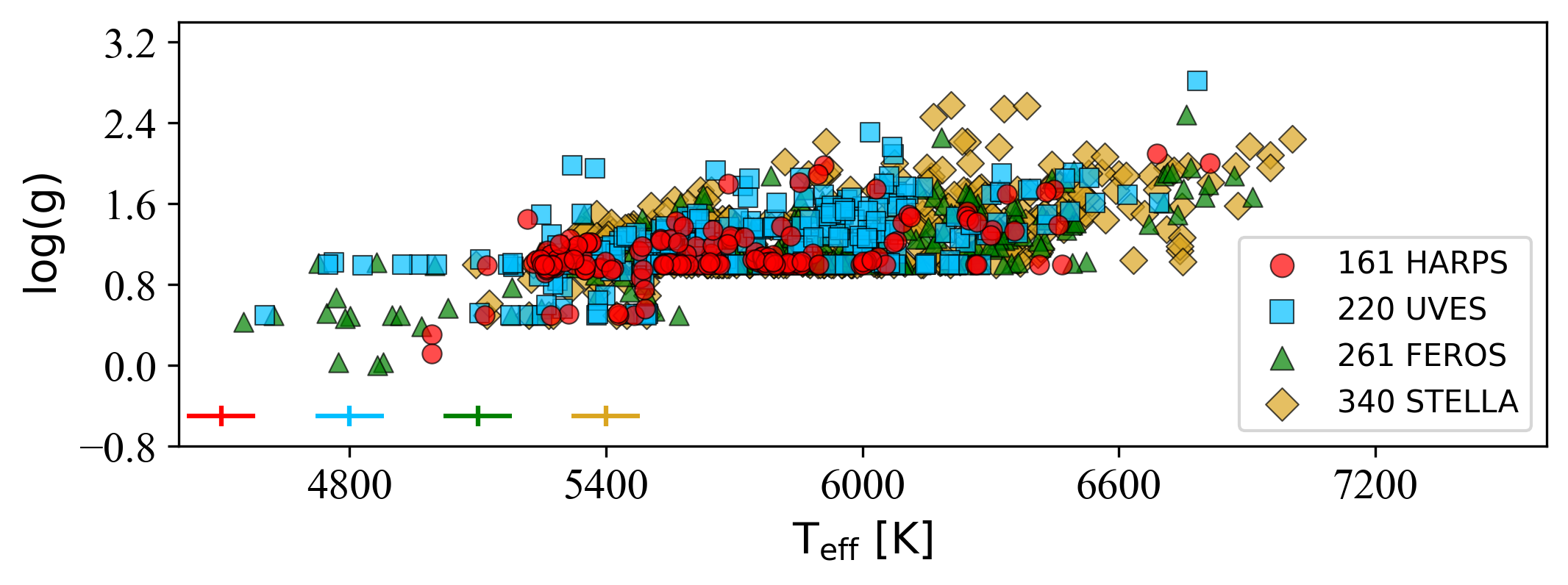}
      \caption{Surface gravity as a function of the effective temperature. The 'staircase structure' shows the limit of the grid of the atmospheric model.}
         \label{grid_limit}
   \end{figure}

\begin{figure}
   \centering
   \includegraphics[width=8.5cm]{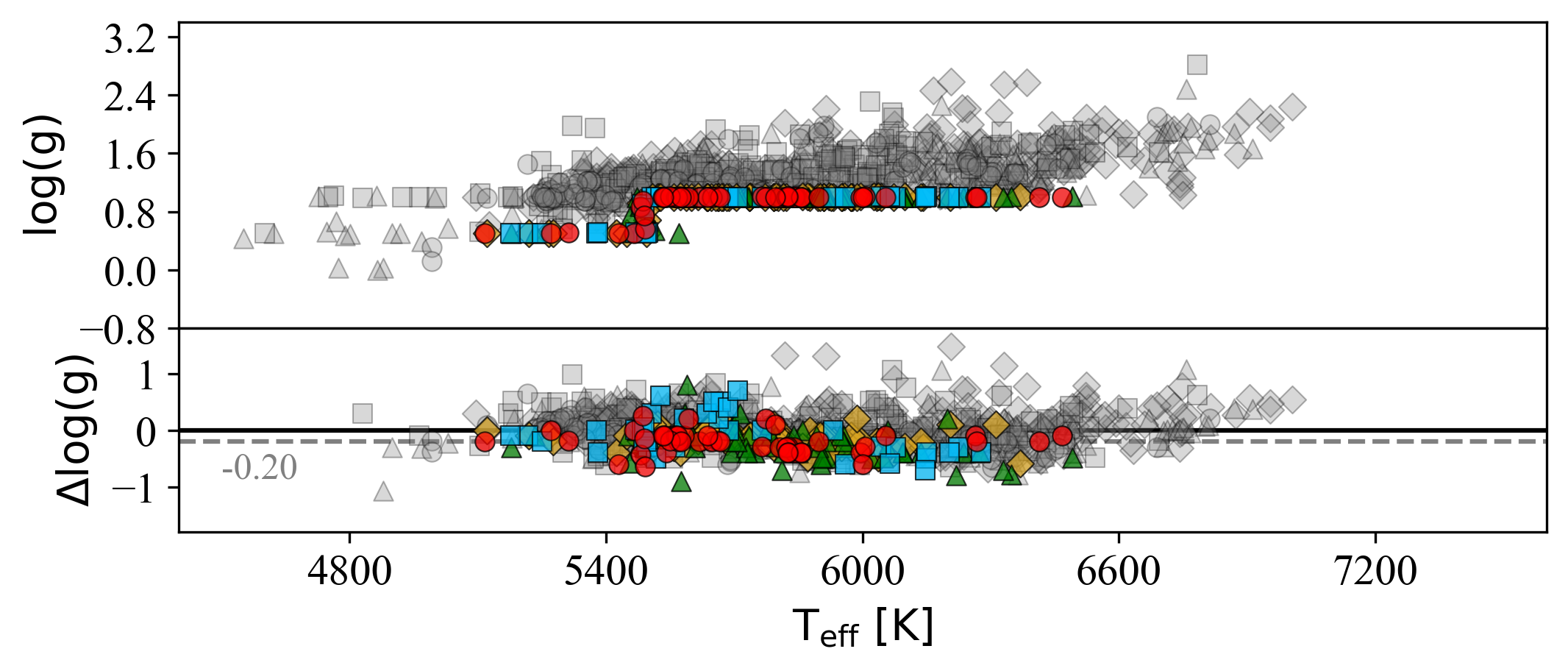}
      \caption{The top panel shows the same distribution of Fig.~\ref{grid_limit} in grey. Values at the limit of the grid are shown with the same colour-code of Fig. \ref{grid_limit}. The bottom panel highlights the difference between this study and \citetalias{da2023oxygen} values. The grey dotted-line highlights $\mathrm{\Delta \log g=}-0.20$ dex.}
         \label{grid_limit_bis}
   \end{figure}

We point out a limitation given by the atmosphere model used, since the atmospheric parameters are estimated at the edge of the grid for 22\% of the spectra, as shown in Figure \ref{grid_limit}, where a "staircase structure" artifact is in place. This implies that for the objects at the edge of the grid the surface gravity is in general overestimated, and/or the effective temperature is underestimated. As shown in Fig. \ref{grid_limit_bis}, we compared our values of $\mathrm{\log(g)}$ and $\mathrm{T_{eff}}$ with those of \citetalias{da2023oxygen} to quantify this effect. We found an average difference of $-0.20$~dex in $\mathrm{\log(g)}$ and 5~K in $\mathrm{T_{eff}}$. While for the temperature the underestimation is completely negligible, for the surface gravity the average difference is comparable with the errors. For this reason, our final estimations of the surface gravity uncertainties take into account this effect.
    
\begin{figure}
   \centering
   \includegraphics[width=8.5cm]{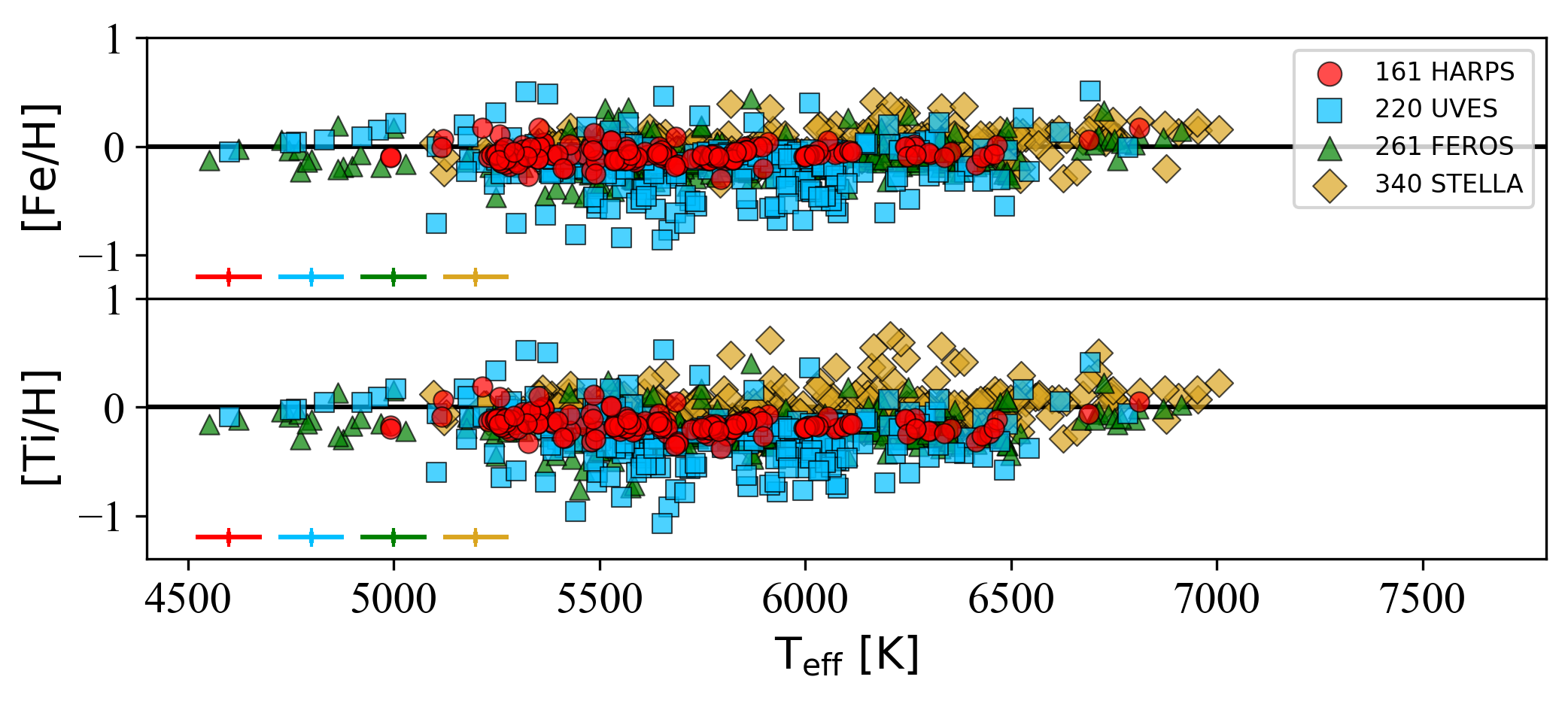}
   \includegraphics[width=8.5cm]{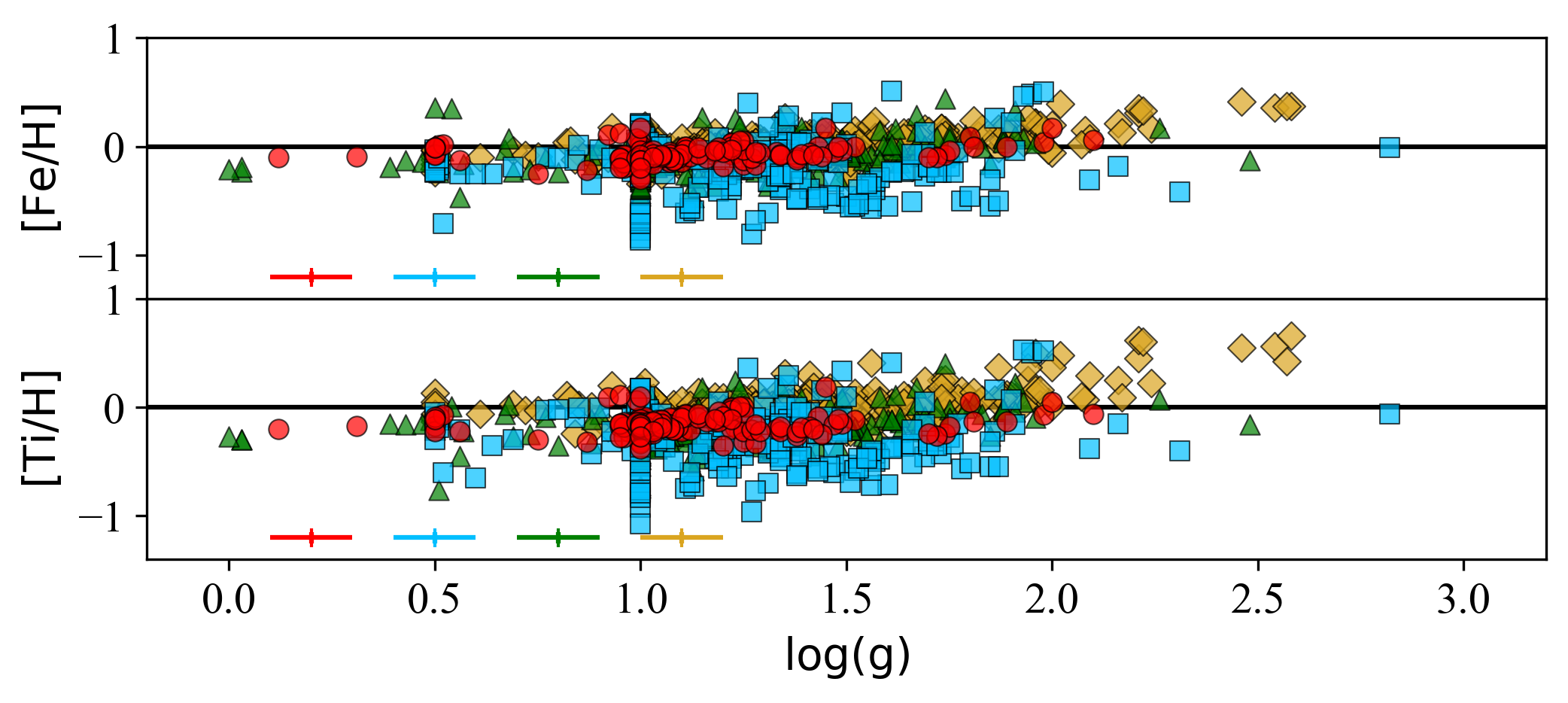}
   \includegraphics[width=8.5cm]{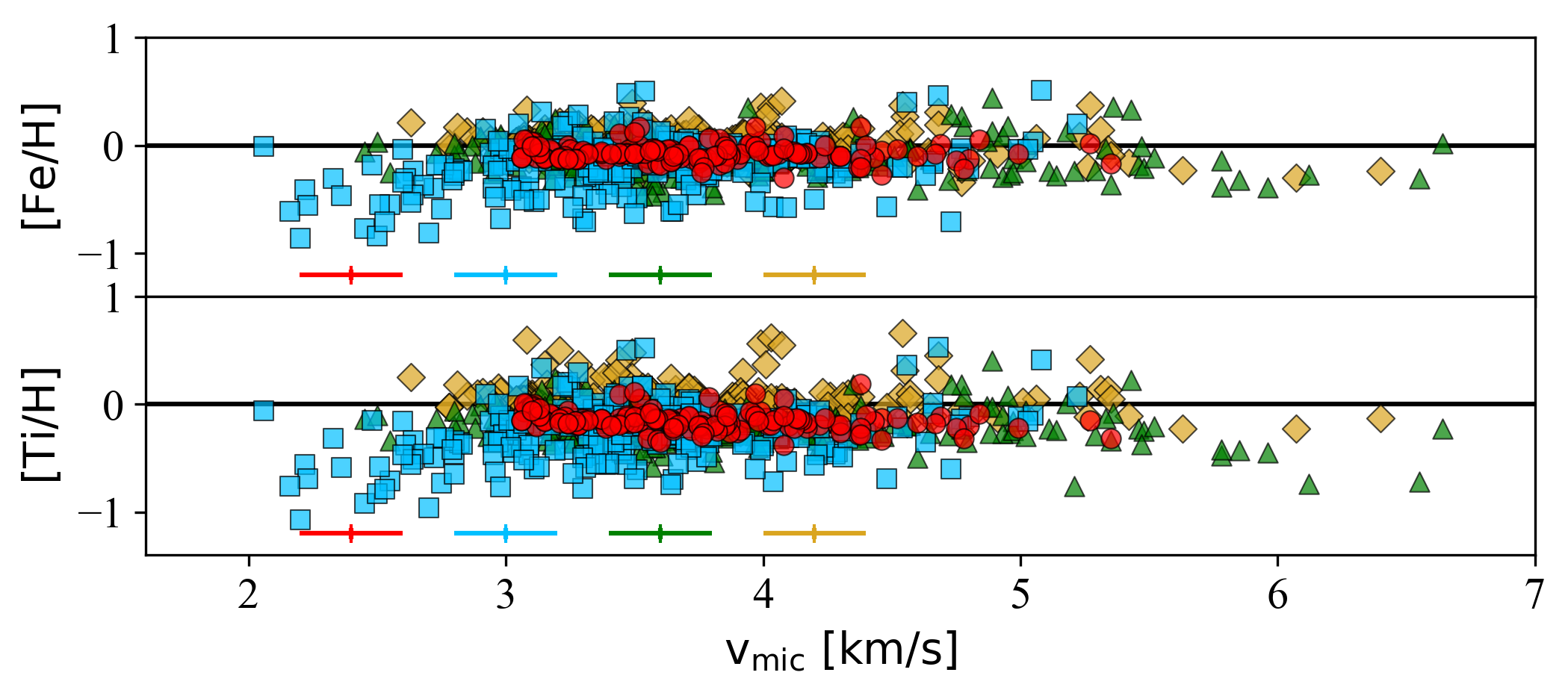}
   \includegraphics[width=8.5cm]{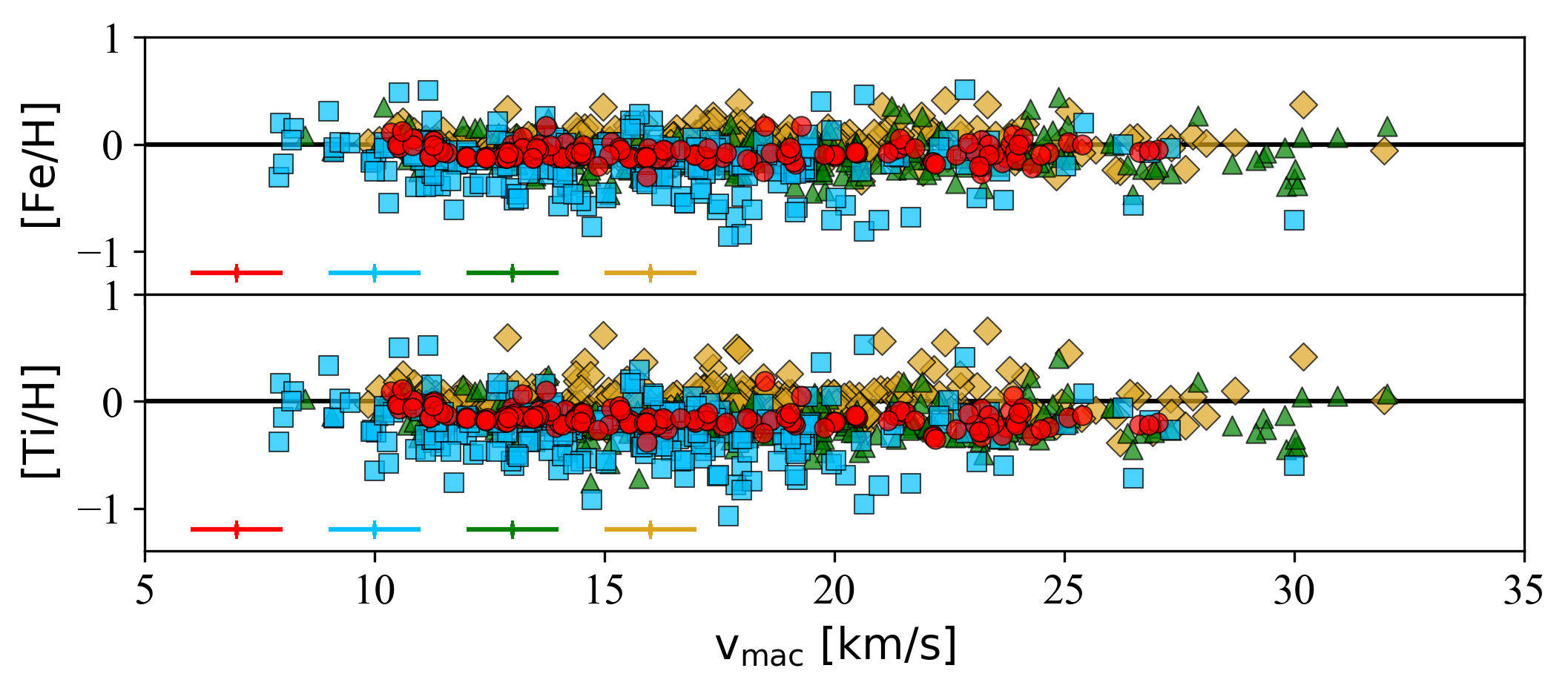}
      \caption{Iron and titanium abundances as a function of the atmospheric parameters. Colours and symbols are the same of Fig.~\ref{comp_ronaldo}.}
         \label{trends_fe_ti}
   \end{figure}

A double check on possible trends among the different parameters has been performed, and in Figs. \ref{trends_fe_ti} the iron and titanium abundances are shown as a function of the atmospheric parameters.
    
Furthermore, to provide a more quantitative analysis of the differences between the NLTE and LTE approaches, we made use of the large number of HR spectra collected by our group for 20 calibrating Cepheids \citep{da2022new}. For these objects, the HR spectra cover the entire pulsation cycle, making them an excellent benchmark for validating spectroscopic approaches. The reason is twofold. 
a) Classical Cepheids with moderate to large amplitudes show variations along the pulsation cycle of approximately 1000 K in $\mathrm{T_{eff}}$, about half a dex in $\mathrm{\log g}$, and a factor of two in $\mathrm{v_{mic}}$ (from 2 to 4 $\mathrm{km\, s^{-1}}$). However, the chemical abundances for each object are the same. This makes them a robust laboratory for constraining the physical assumptions involved when transitioning from LTE to NLTE analyses. 
b) Classical Cepheids trace a loop in both the colour-magnitude diagram and the spectroscopic colour-magnitude diagram (or Kiel diagram; $\mathrm{T_{eff}}$ vs.\ $\mathrm{\log g}$; see \citealt{kudritzki20}, \citeyear{kudritzki24}). As a consequence, during the pulsation cycle a Cepheid can reach, at a fixed surface gravity, two different effective temperatures: one on the rising branch and one on the declining branch. This provides a unique opportunity to test the impact of $\mathrm{T_{eff}}$ on NLTE analysis while keeping both surface gravity and chemical composition fixed. 

\begin{figure}[h]
    \centering
    \includegraphics[width=1.\linewidth]{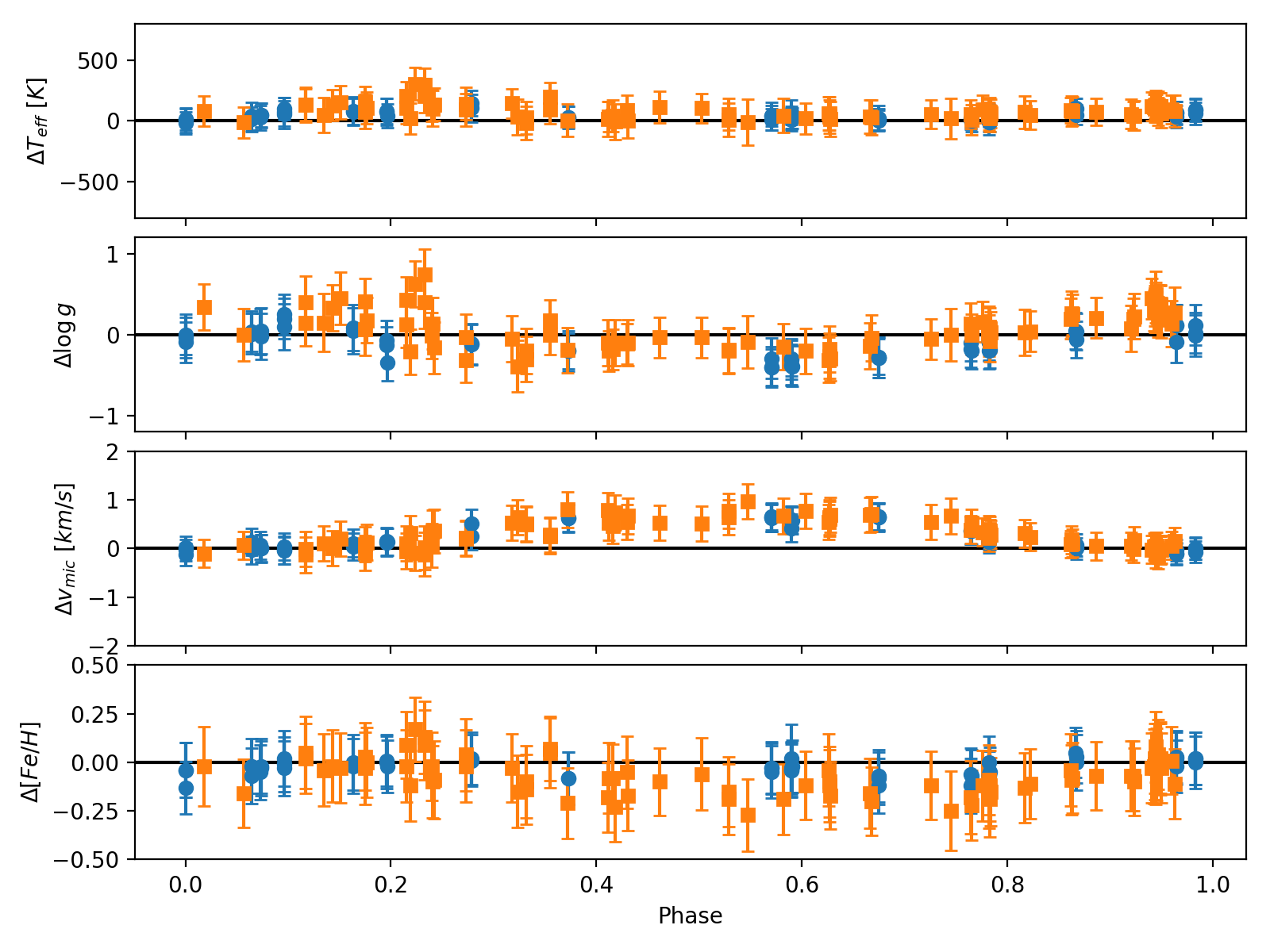}
    \includegraphics[width=1.\linewidth]{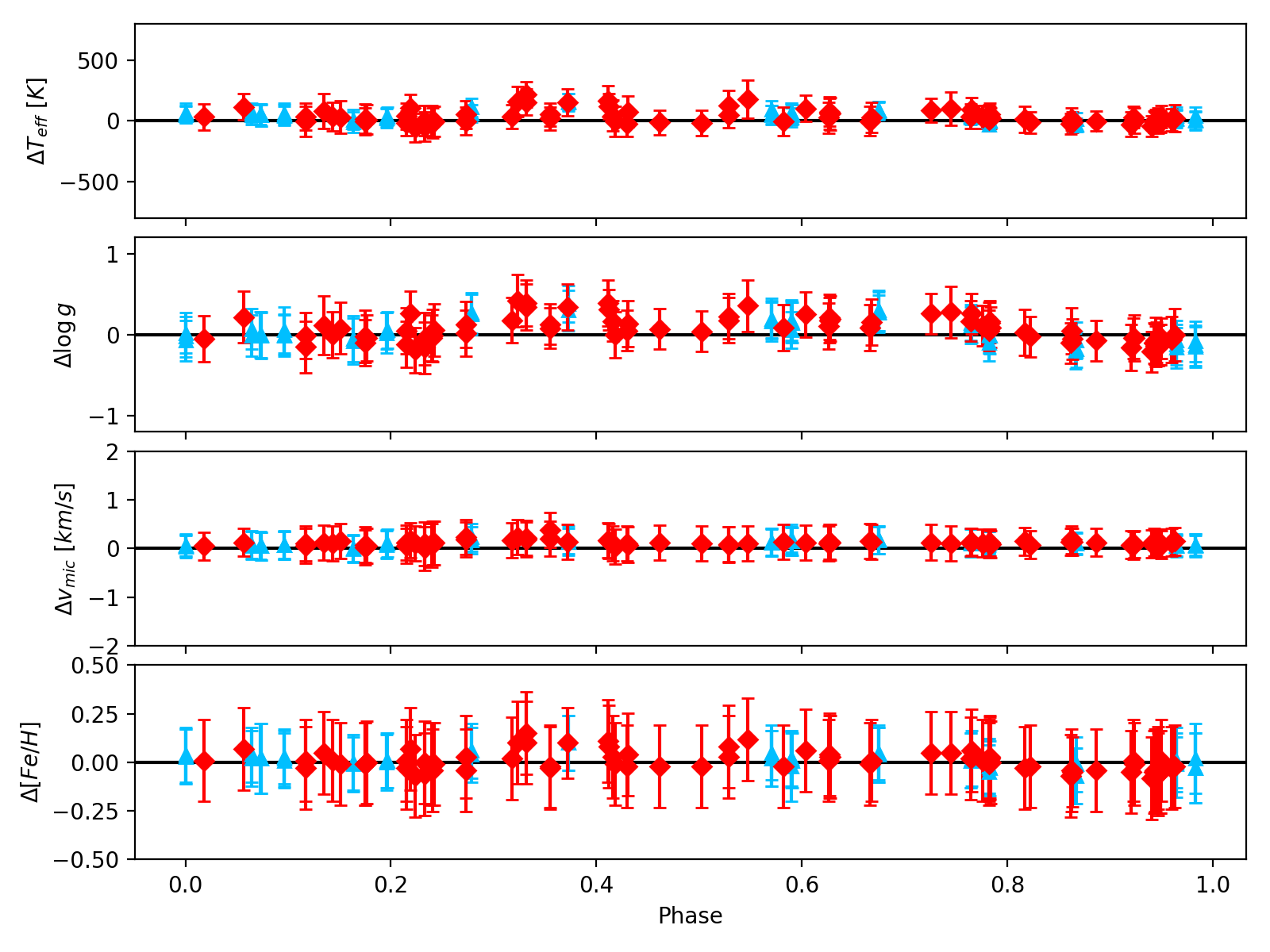}
    \caption{Top four panels -- Difference in atmospheric parameters between the current analysis and \citet{da2022new} as a function of pulsation phase for the calibrating Cepheid $\mathrm{\zeta \,\, Gem}$.  Dark symbols indicate STELLA spectra, whereas orange symbols mark HARPS spectra. Black solid lines display zero difference.
    Bottom four panels -- same as the top four panels, but the difference is between NLTE and LTE estimates of the atmospheric parameters. Light blue symbols indicate STELLA spectra, whereas red symbols mark HARPS spectra.}
    \label{calib}
\end{figure}

\begin{figure}[h]
    \centering
    \includegraphics[width=0.84\linewidth]{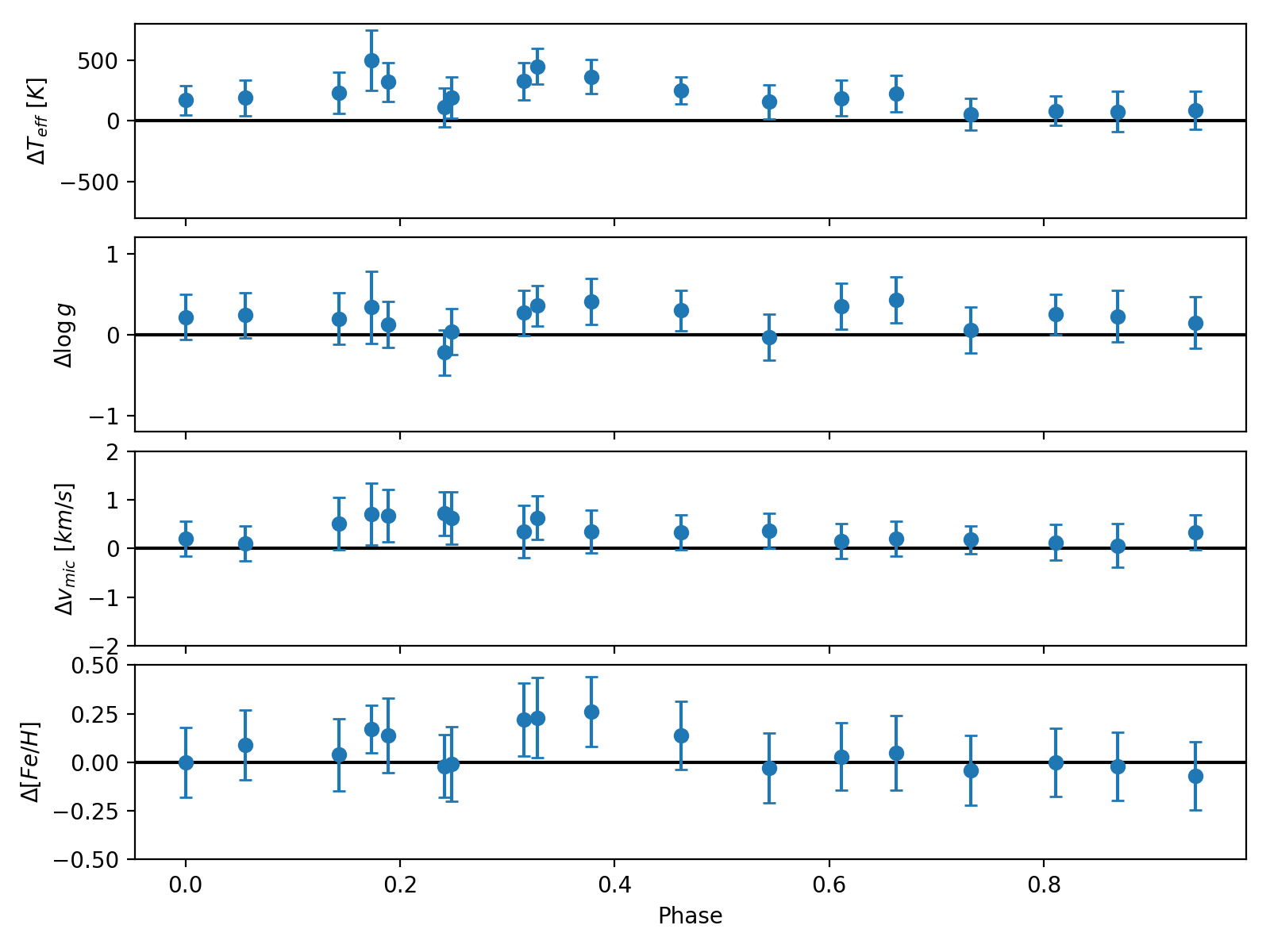}
    \includegraphics[width=0.84\linewidth]{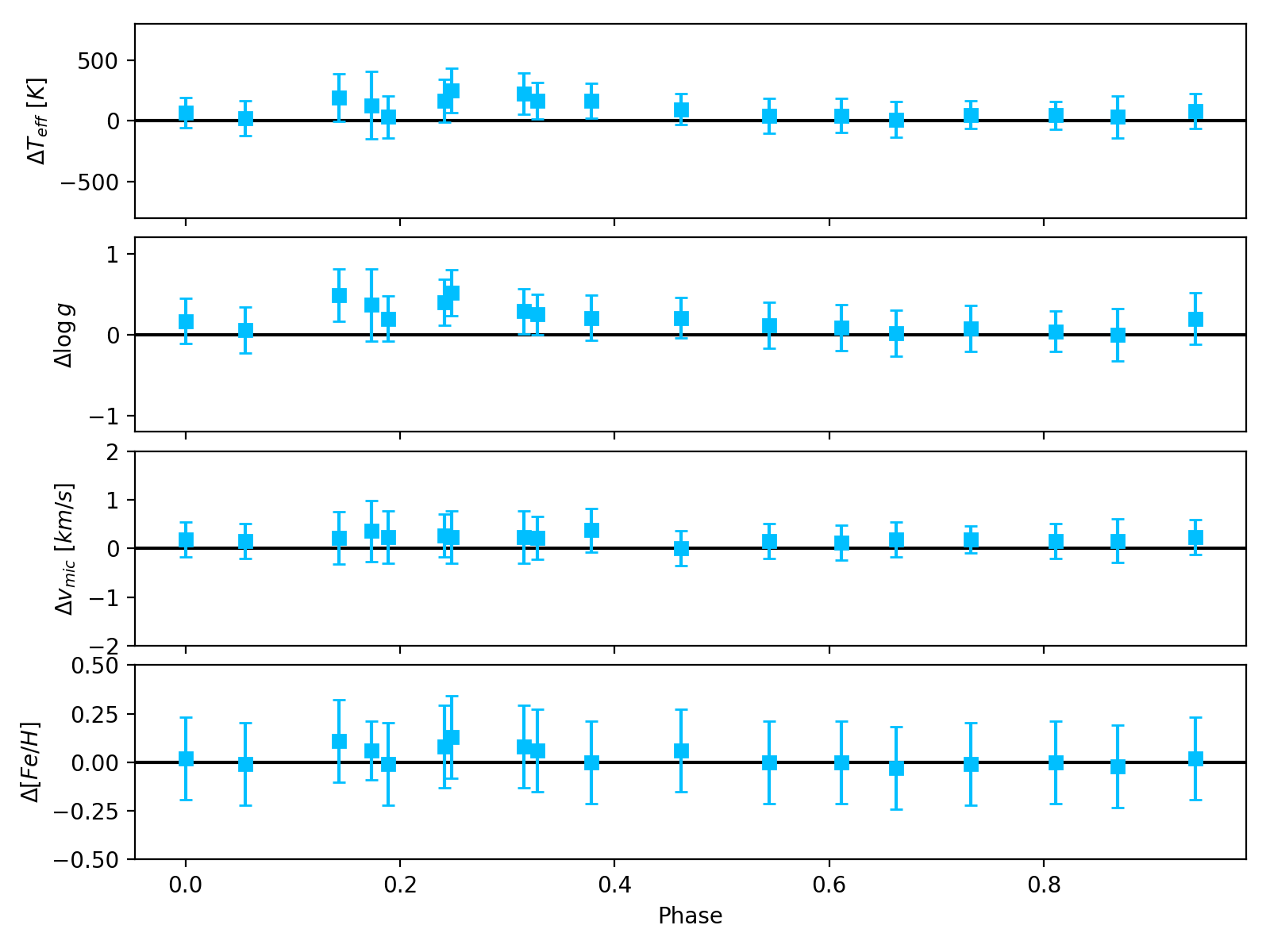}
    \caption{Same as Fig.~\ref{calib}, but for $\mathrm{\delta \,\, Cep}$.}
    \label{calib2}
\end{figure}

\begin{figure}[h]
    \centering
    \includegraphics[width=0.84\linewidth]{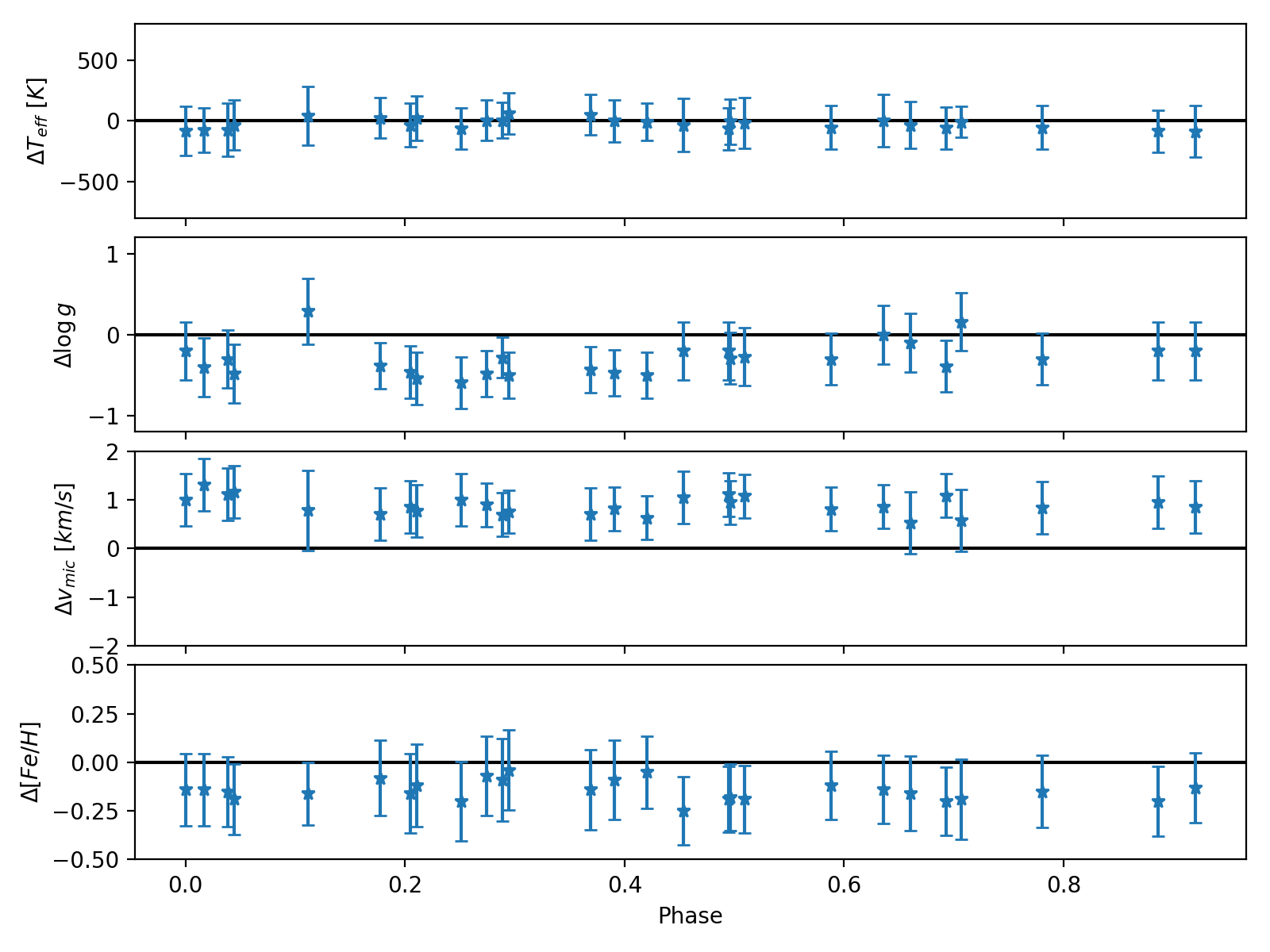}
    \includegraphics[width=0.84\linewidth]{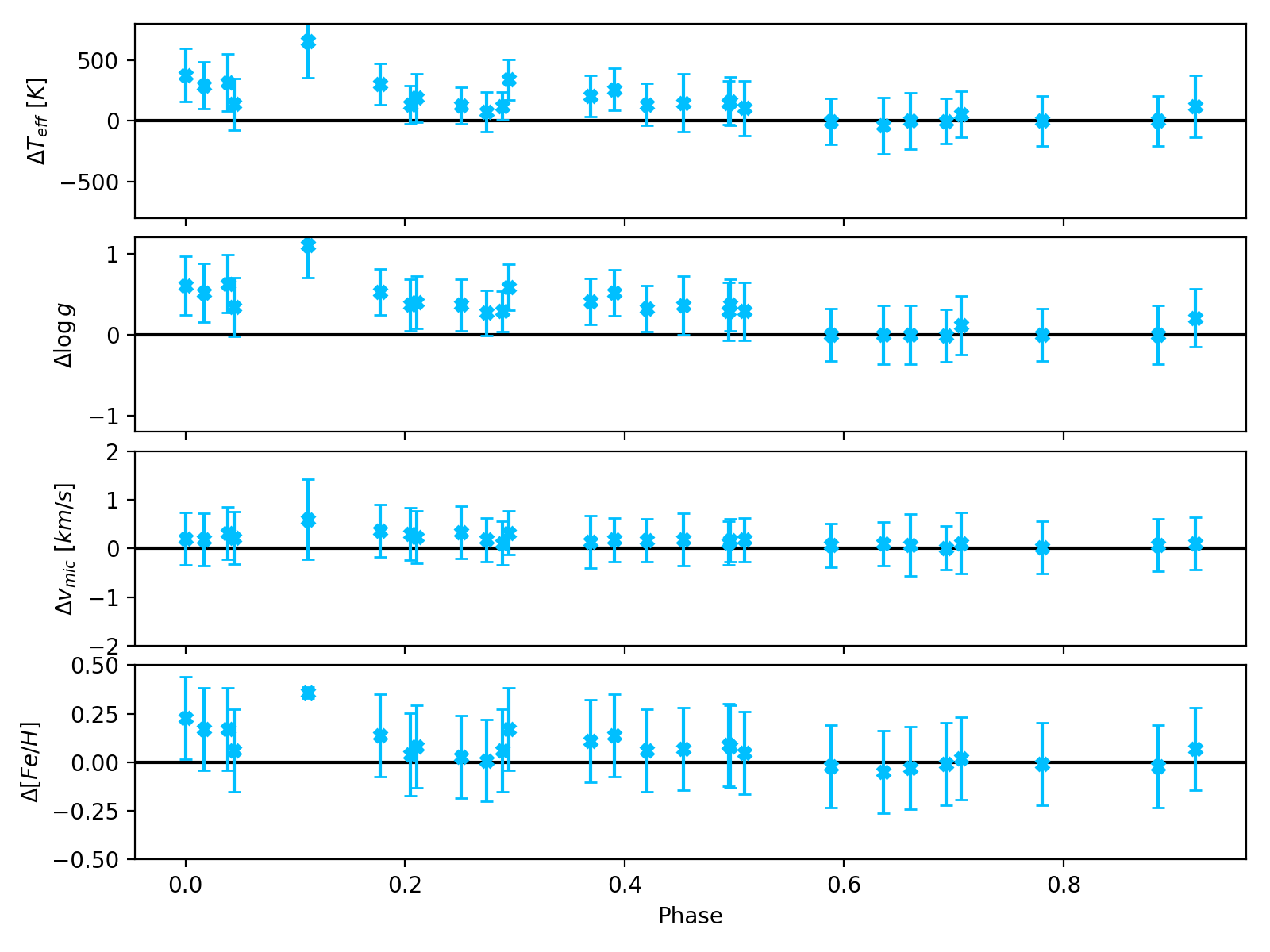}
    \caption{Same as Fig.~\ref{calib}, but for $\mathrm{FF\,\,Aql}$.}
    \label{calib3}
\end{figure}

We focused our analysis on three Cepheids for which we have very detailed phase coverage: $\mathrm{\zeta\,\,Gem}$, (fundamental mode; Fig.~\ref{calib}), $\mathrm{\delta\,\,Cep}$ (fundamental mode; Fig.~\ref{calib2}), and $\mathrm{FF\,\,Aql}$ (first overtone; Fig.~\ref{calib3}). The bottom four panels of these figures show the differences in atmospheric parameters and iron abundances between the NLTE and LTE analyses. These panels reveal overall consistency between the two approaches, as no significant trends or systematic offsets are observed. The same conclusion holds when comparing the LTE results from \citet{da2022new} with our current NLTE analysis (top four panels of the same figures).


\section{Validation of the NLTE parameters and abundances}
\label{test_nlte}
\begin{figure}
    \centering
    \includegraphics[width=0.8\linewidth]{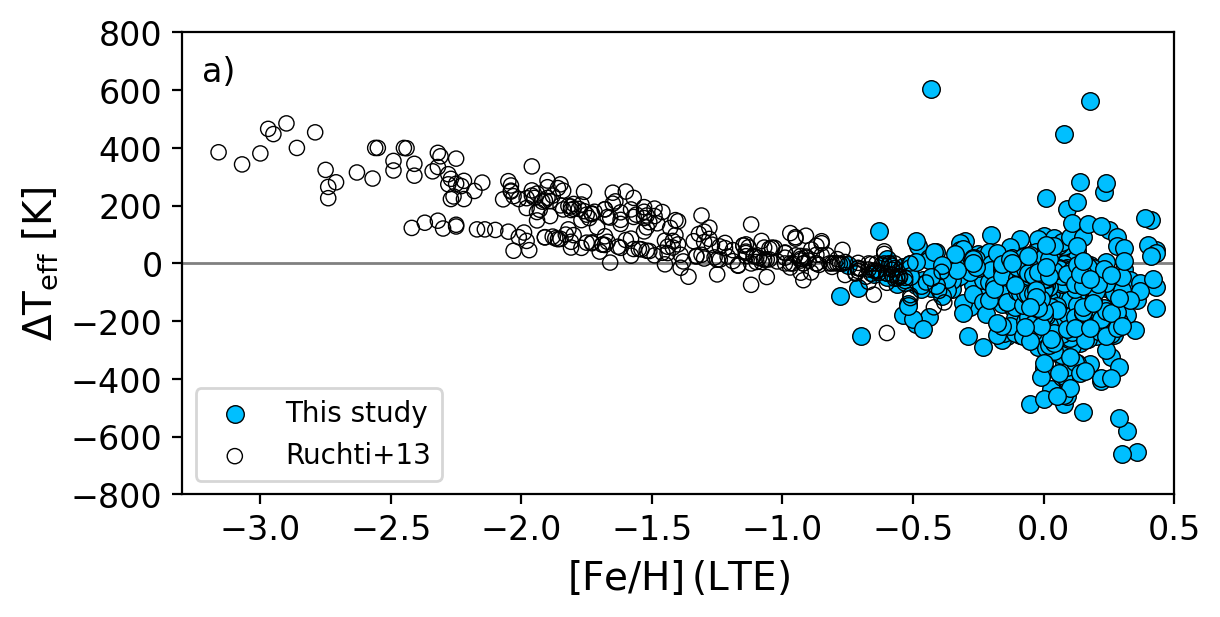}
    \includegraphics[width=0.8\linewidth]{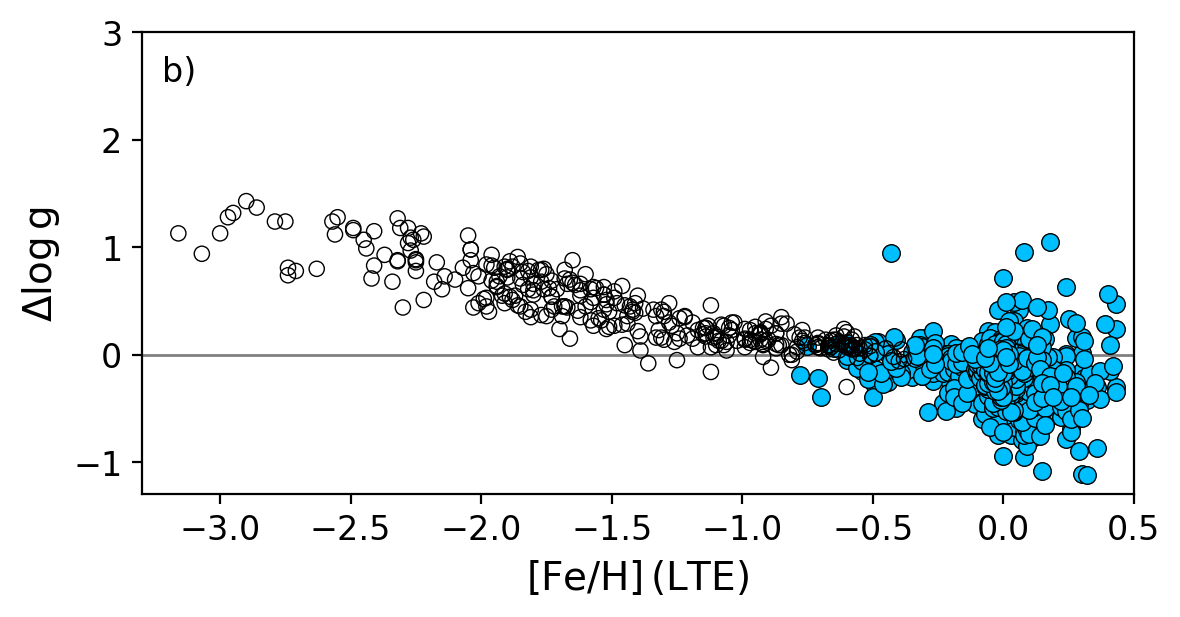}
    \includegraphics[width=0.8\linewidth]{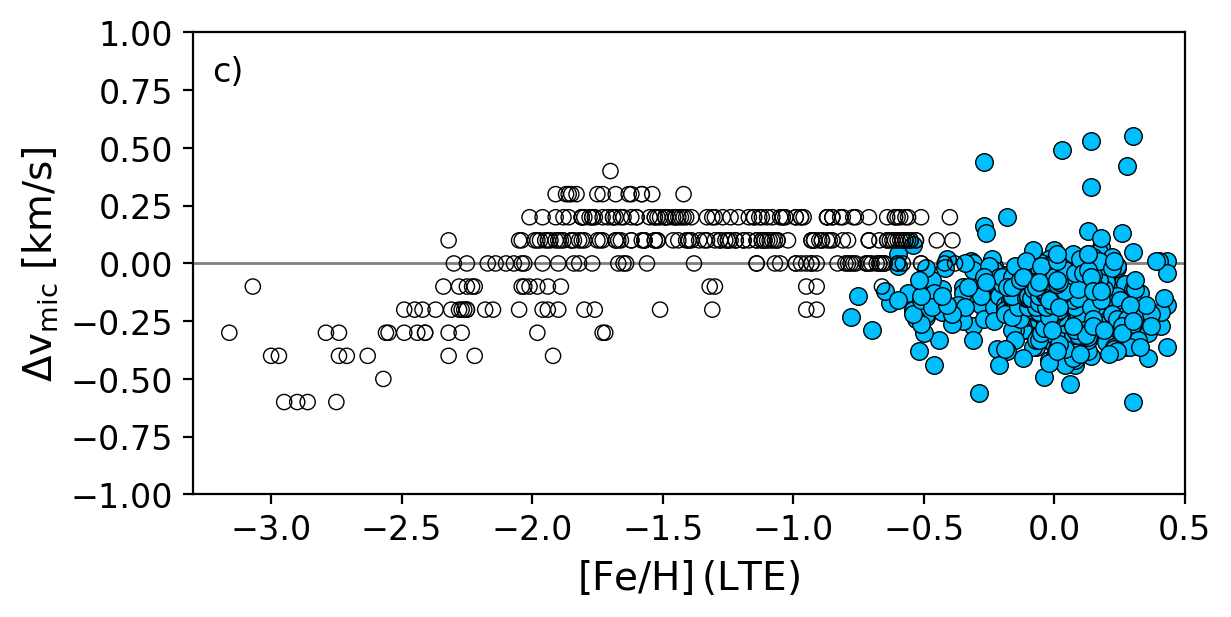}
    \includegraphics[width=0.8\linewidth]{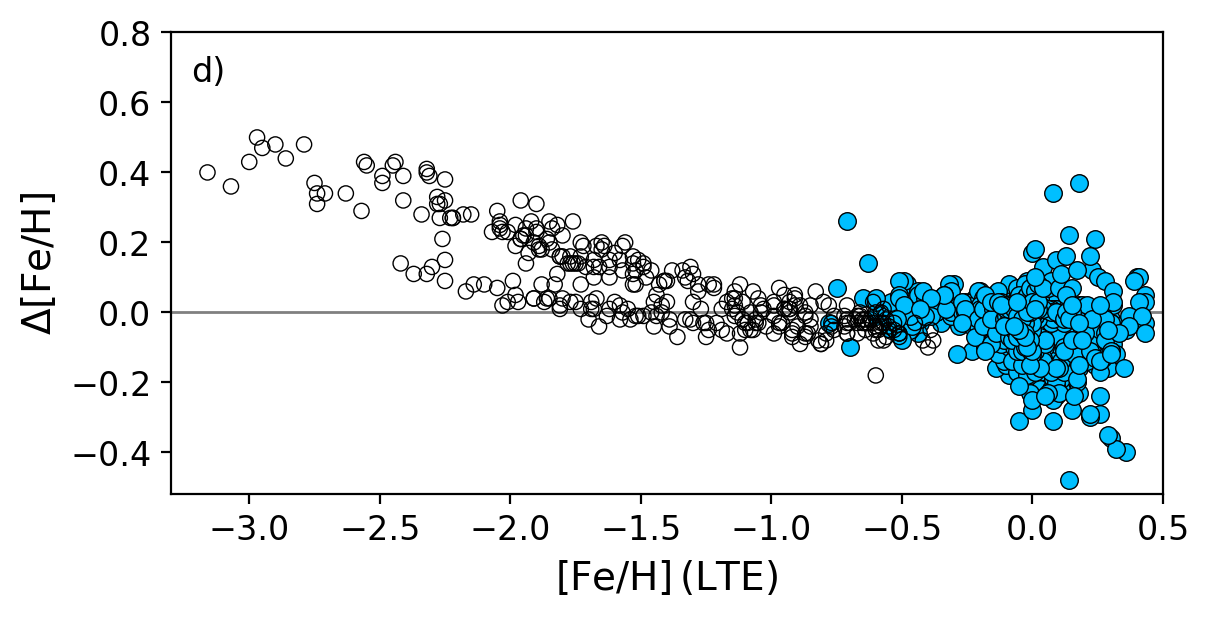}
    
    \caption{Differences between NLTE and LTE estimates of $\mathrm{T_{eff}}$ (panel a), $\mathrm{\log g}$ (panel b), $\mathrm{v_{mic}}$ (panel c) and $\mathrm{[Fe/H]}$ as a function of LTE-metallicity. \citet{ruchti13} results are shown with black circles, while this work's results are highlighted in blue.}
    \label{comp_ruchti}
\end{figure}

We compared our NLTE atmospheric parameters and chemical abundances with previous studies that investigated the impact of NLTE analysis, focusing in particular on the behaviour of NLTE--LTE differences as a function of metallicity. \citet{ruchti13} introduced an optimised methodology, termed NLTE-Opt, specifically designed to mitigate the systematic biases typical of standard 1D LTE analyses for FGK-type stars. In this approach, $\mathrm{T_{eff}}$ is determined from Balmer lines profiles, while $\mathrm{\log g}$ and [Fe/H] are derived from the NLTE ionization balance of Fe.
Fig.~\ref{comp_ruchti} shows the systematic offsets in stellar parameters as a function of metallicity based on standard LTE analysis when compared with NLTE-Opt estimates. These trends are generally attributed to the inability of the LTE Fe~I excitation balance to provide an accurate $\mathrm{T_{eff}}$ scale. As shown by the black circles representing the results of \citet{ruchti13}, these differences decrease moving toward higher metallicities (see also \citealt{kovalev19}). Our Cepheids (in blue) follow the same general behaviour, displaying a good agreement with \citet{ruchti13}, particularly in $\mathrm{\Delta T_{eff}}$ (panel a), $\mathrm{\Delta \log g}$ (panel b) and $\mathrm{\Delta [Fe/H]}$ (panel d). However, the panel c) of the same figure shows the difference in microturbulence differences, revealing an offset of 0.20--0.30 $\mathrm{km\, s^{-1}}$. Since Cepheids typically exhibit higher micro- and macroturbulent velocities, both the observed trends and the intrinsic scatter should be interpreted with caution. We also note that $\mathrm{v_{mic}}$ is the only parameter for which we find a statistically significant offset between the NLTE and LTE determinations ($-0.15\pm0.11\, \mathrm{km\, s^{-1}}$, see Fig.~\ref{lte_plots}). Finally, our fitting method simultaneously optimizes all atmospheric parameters, which means that the contribution of $\mathrm{v_{mic}}$ significantly affects the determination of the others. 

\begin{figure}
    \centering
    \includegraphics[width=0.8\linewidth]{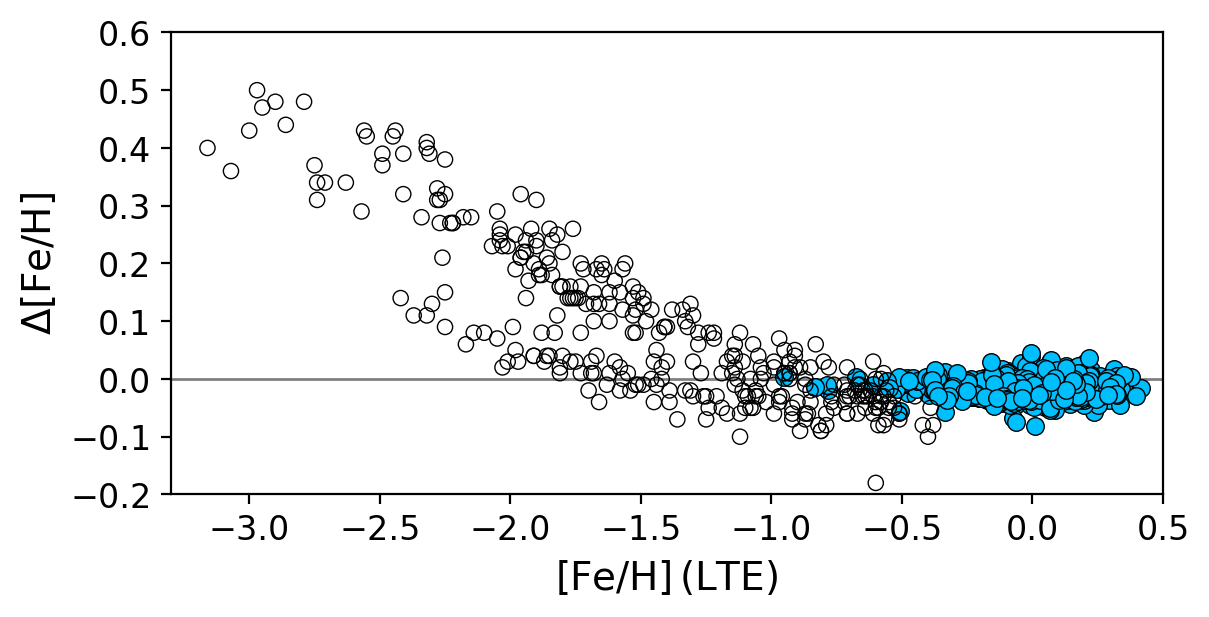}
    \caption{Differences between NLTE and LTE estimates of $\mathrm{[Fe/H]}$ as a function of LTE-metallicity, keeping fixed $\mathrm{T_{eff}}$, $\mathrm{\log g}$, and $\mathrm{v_{mic}}$ to \citetalias{da2023oxygen} estimates. \citet{ruchti13} results are shown with black circles, while this work's results are highlighted in blue.}
    \label{deltafe_additional}
\end{figure}

These results were obtained by varying all parameters simultaneously, performing the analysis once in NLTE and once in LTE. We further carried out an additional test to quantify the specific contribution of the NLTE versus LTE treatment to $\mathrm{[Fe/H]}$. In this test, we measured the NLTE and LTE metallicity of the \citetalias{da2023oxygen} sample while fixing $\mathrm{T_{eff}}$, $\mathrm{\log g}$, and $\mathrm{v_{mic}}$ to their estimates. Fig.~\ref{deltafe_additional} shows the $\mathrm{\Delta[Fe/H]}$ (NLTE--LTE) as a function of the LTE iron abundance, compared with the results of \citet{ruchti13}. Our estimates closely follow the general behaviour reported by \citet{ruchti13} and remain well within 0.1 dex. These results further demonstrate that the offsets and trends observed in the comparison of metallicities with the literature (Figs.~\ref{comp_ronaldo}~and~\ref{comp_luck_trentin}) are not primarily driven by the NLTE versus LTE treatment, but instead mainly reflect methodological differences, i.e. the adopted spectral analysis techniques and line lists.

\begin{figure}
    \centering
    \includegraphics[width=0.49\linewidth]{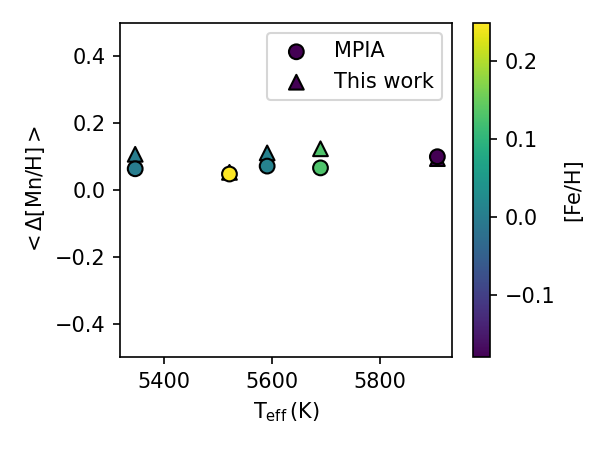}
    \includegraphics[width=0.49\linewidth]{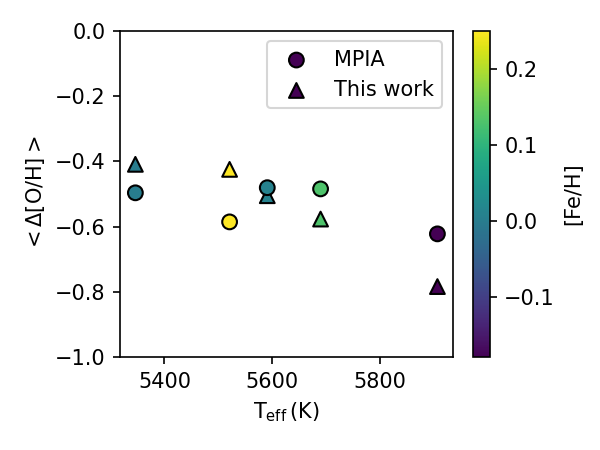}
    \caption{Comparison of the mean NLTE abundance corrections provided by the MPIA database (\url{https://nlte.mpia.de/index.php}) and this study's results ($\mathrm{\Delta[X/H] = [X/H]_{NLTE}-[X/H]_{LTE}}$) for manganese (left panel) and oxygen (right panel) as a function of $\mathrm{T_{eff}}$ and colour-coded by $\mathrm{[Fe/H]}$.}
    \label{mpia}
\end{figure}

We independently verified our NLTE abundances by applying the line-by-line NLTE abundance corrections provided by the MPIA database\footnote{https://nlte.mpia.de/} (\citealt{kovalev18}). Specifically, we focused on Mn and O. The atomic model for Mn is the same as that used in our study from \cite{bergemann19}, while for O, we used the atomic model from \cite{amarsi2018}, while the MPIA database relies on \cite{bergemann21}. We performed this comparison for five sample spectra whose atmospheric parameters are representative of the whole population. We downloaded the NLTE abundance correction from the MPIA database for Mn (5394.68 $\AA$, 5420.33 $\AA$ and 6021.77 $\AA$) and O (7771.94 $\AA$, 7774.17 $\AA$ and 7775.39 $\AA$). As shown in Fig.~\ref{mpia}, the comparison indicates good agreement for all Cepheids, especially for Mn. For oxygen the agreement is also good (differences within 0.1 dex) for three Cepheids, while the remaining two show discrepancies of approximately 0.2 dex. This is well within our observational uncertainties. Finally, the decreasing trend observed in the plane $\mathrm{<\Delta[O/H]>}$ vs $\mathrm{T_{eff}}$ has also been obtained in \cite{vasilyev19} for the 777.4 nm line with both 1D and 2D models, and previously observed in Cepheids in \cite{luck13}.

\end{appendix}
\end{document}